\documentclass[a4paper,fleqn,usenatbib]{mnras}

\usepackage{amssymb} 
\usepackage{amsmath} 
\usepackage{graphicx}

\title[Scaled-up jets from IM protostars]{[Fe~{\sc ii}] jets from intermediate-mass protostars in Carina} 

\author[Reiter et al.]{
Megan Reiter,$^{1}$\thanks{email: mreiter@umich.edu (MR)}, 
Nathan Smith,$^{2}$, and John Bally$^{3}$ \\
$^{1}$University of Michigan, Ann Arbor, MI 48109, USA \\
$^{2}$Steward Observatory, University of Arizona, Tucson, AZ 85721, USA \\
$^{3}$Center for Astrophysics and Space Astronomy, University of Colorado, 389 UCB, Boulder, CO 80309, USA}

\date{Accepted XXX. Received YYY; in original form ZZZ}

\pubyear{2016} 

\begin{document}

\label{firstpage} 
\pagerange{\pageref{firstpage}--\pageref{lastpage}} 
\maketitle


\begin{abstract}
We present new HST/WFC3-IR narrowband [Fe~{\sc ii}] images of protostellar jets in the Carina Nebula. 
Combined with 5 previously published sources, we have a sample of 18 jets and 2 HH objects. 
All of the jets we targeted with WFC3 show bright infrared [Fe~{\sc ii}] emission, and a few H$\alpha$ candidate jets are confirmed as collimated outflows based on the morphology of their [Fe~{\sc ii}] emission. 
Continuum-subtracted images clearly separate jet emission from the adjacent ionization front, providing a better tracer of the collimated jet than H$\alpha$ and allowing us to connect these jets with their embedded driving sources. 
The [Fe~{\sc ii}] 1.64 $\mu$m/H$\alpha$ flux ratio measured in the jets is $\gtrsim 5$ times larger than in the adjacent ionization fronts. 
The low-ionization jet core requires high densities to shield Fe$^+$ against further ionization by the FUV radiation from O-type stars in the H~{\sc ii} region. 
High jet densities imply high mass-loss rates, consistent with the intermediate-mass driving sources we identify for 13 jets. 
The remaining jets emerge from opaque globules that obscure emission from the protostar. 
In many respects, the HH jets in Carina look like a scaled-up version of the jets driven by low-mass protostars. 
Altogether, these observations suggest that [Fe~{\sc ii}] emission is a reliable tracer of dense, irradiated jets driven by intermediate-mass protostars. 
We argue that highly collimated outflows are common to more massive protostars, and that they suggest the outflow physics inferred for low-mass stars formation scales up to at least $\sim 8$ M$_{\odot}$. 
\end{abstract}

\begin{keywords}
stars: formation --- jets --- outflows 
\end{keywords}


\section{Introduction}\label{s:ir_synth_intro}

Understanding accretion and outflow in young stars is key to constraining the physics that govern their formation and early evolution. 
Accretion and outflow will shape the circumstellar environment around young stars where planet formation may already be ongoing, fostering or inhibiting the genesis of sub-stellar companions. 
Abundant low-mass sources in the solar neighborhood allow for detailed, multi-wavelength studies of the evolution of protostars from deeply embedded cores that are only accessible at long wavelengths \citep[e.g.][]{eno08,jor09} to the IR excess and strong optical emission lines characteristic of more evolved T~Tauri stars \citep[e.g.][]{eva03,harvey07,eva09}. 
For the nearby sources, high angular resolution observations can resolve the circumstellar geometry, clearly illustrating the importance of accretion disks, jets, and outflows for the formation of stars \citep[e.g.][]{bur96,kri98,mcc98,pad99}.

In low-mass stars, disk material accretes along stellar magnetic field lines, ultimately splashing onto the stellar surface at high latitudes \citep[see, e.g. review by][]{bou07}. 
This process of magnetospheric accretion requires strong, large-scale stellar magnetic fields with a predominately dipolar topology to lift material from the inner edge of the disk and carry it along field lines to the star. 
Strong magnetic fields have been found in many T Tauri stars, supporting the magnetospheric accretion paradigm \citep[e.g.][]{joh99a,joh99b,joh04,joh13}. 
Whether a scaled-up version of this formation scenario applies to intermediate- and high-mass stars remains unclear. 
In particular, it is not settled whether intermediate- and high-mass stars generate magnetic fields of sufficient strength to support magnetospheric accretion. 
Surveys of Herbig Ae/Be stars find a low magnetic incidence of $\leq 10$\% \citep[see, e.g.][]{wad07}. 
Derived upper limits on the magnetic field strength are smaller than the minimum field strength required for magnetospheric accretion in both Herbig Ae and Be stars \citep[derived from the models of][]{joh99b}. 
More recent surveys with higher sensitivity to weaker fields further constrain the average field strength of intermediate-mass protostars to be an order of magnitude less than typically observed in T Tauri stars \citep[e.g.][]{ale13,hub15}.

Evidence for circumstellar disks around intermediate- and high-mass protostars also argues for massive stars forming via a scaled-up version of low-mass star formation \citep[e.g.][]{tan03}.
Direct detection is difficult, however, given the large median distances, high optical depths, and short timescales involved \citep[e.g.][]{beu09,kra10,pre11b,car12,joh15}. 
For this reason, indirect accretion indicators provide an important avenue to understand the evolution of intermediate-mass protostars. 
Jets are one such signpost as they require active disk accretion. 
Observed similarities in the physical properties of outflows from low- and high-mass stars suggest a common production mechanism regardless of protostellar mass \citep{ric00}. 
Alternate formation pathways, for example the coalescence of lower-mass cores, are unlikely to form collimated outflows \citep{bal05}. Thus, the detection of collimated jets provides compelling, though indirect, evidence of circumstellar disks \citep[e.g.][]{guz12}.

Protostellar \textit{outflows} appear to be a ubiquitous feature of star formation \citep[e.g.][]{arc07}, and thus provide an avenue to identify accreting systems even at distances where the accretion geometry is not directly resolved. 
A well-collimated \textit{jet} launched near the protostar requires energy from disk accretion to produce a collimated stream of high-velocity gas \citep[e.g.][]{fer06}. 
An underlying jet, a wide-angle disk wind, or a combination of the two \citep[e.g.][]{bac95,san99,arc01b,yba06} may power the outflows typically studied at millimeter wavelengths. 
Outflows tend to have lower velocities and broader morphologies than the collimated jets that may be obscured inside the optically thick outflow lobe. 
This is especially true for high-mass protostars as they tend to form out of dense clumps that will birth a cluster of stars that have large columns of gas and dust that thoroughly enshroud the earliest evolutionary stages.

Comparatively little attention has been paid to intermediate mass ($\sim 2-8$ M$_{\odot}$) stars, even though they sample the changes in stellar structure that may be related to a transition in accretion mechanisms between low- and high-mass stars. 
Unlike the highest mass sources, modest optical depths obscure intermediate-mass stars, permitting multi-wavelength observations of a variety of spatial and temporal scales. 
Employing similar observational techniques as low-mass star formation makes it easier to directly compare the results \citep[e.g.][]{cal04,vin05,men12}. 
This potential to link low- and high-mass star formation has led to growing interest in intermediate-mass stars. 
However, studies typically target only a few jets or outflows from intermediate-mass stars, providing a heterogeneous sample of objects observed in different regions and with different techniques 
\citep[e.g.][]{she03,ell13}. 
Ideally, multiple sources in a single region would be observed with the same technique to develop a coherent picture of accretion and outflow throughout the intermediate-mass range.

\begin{table*}
\caption[New WFC3-IR observations]{Observations}
\vspace{5pt}
\centering
\vspace{3pt}
\begin{tabular}{lllllll}
\hline\hline
\vspace{5pt}
Target & $\alpha_{\mathrm{J2000}}$ & $\delta_{\mathrm{J2000}}$ & Date & 
Exp. time (s) & Comment \\ 
\hline 
HH~666 &  10:43:51.3  &  -59:55:21  &  2009 Jul 24-29  &  2397/2797  &  no cont.; see also \citet{rei13}  \\ 
HH~901  &  10:44:03.5  &  -59:31:02  &  2010 Feb 1-2  &  2797/3197  &  no cont.; see also \citet{rei13}  \\
HH~902  &  10:44:01.7  &  -59:30:32  &  2010 Feb 1-2  &  2797/3197  &  no cont.; see also \citet{rei13}  \\
HH~1066 &  10:44:05.4  &  -59:29:40  &  2010 Feb 1-2  &  2797/3197  &  no cont.; see also \citet{rei13}  \\
\hline
HH~900 &  10:45:19.3  &  -59:44:23  &  2013 Dec 28  &  2397  &  see also \citet{rei15a}\\ 
HH~903 &  10:45:56.6  &  -60:06:08  &  2014 Apr 17  &  2397  &  \\ 
HH~1004 &  10:46:44.8  &  -60:10:20  &  2014 Sep 28  &  2397  &  \\ 
HH~1005 &  10:46:44.2  &  -60:10:35  &  2014 Sep 28  &  2397  &  \\ 
HH~1006 &  10:46:33.0  &  -60:03:54  &  2014 Jun 05  &  2397  &  \\ 
HH~1007 &  10:44:29.5  &  -60:23:05  &  2014 Apr 17  &  2397  &  \\ 
HH~1010 &  10:41:48.7  &  -59:43:38  &  2014 Apr 18  &  2397  &  \\ 
HH~1014 &  10:45:45.9  &  -59:41:06  &  2015 Jan 16  &  2397  &  \\ 
HH~1015 &  10:44:27.9  &  -60:22:57  &  2014 Apr 17  &  2397  &  \\ 
HH~c-3 &  10:45:04.6  &  -60:03:02  &  2014 Feb 20  &  2397  &  \\ 
HH~1159 &  10:45:08.3  &  -60:02:31  &  2014 Feb 20  &  2397  &  HH~c-4 in \citet{smi10} \\ 
HH~1160 &  10:45:09.3  &  -60:01:59  &  2014 Feb 20  &  2397  &  HH~c-5 in \citet{smi10} \\ 
HH~1161 &  10:45:09.3  &  -60:02:26  &  2014 Feb 20  &  2397  &  HH~c-6 in \citet{smi10} \\ 
HH~1162 &  10:45:13.4  &  -60:02:55  &  2014 Feb 20  &  2397  &  HH~c-7 in \citet{smi10} \\ 
HH~1163 &  10:45:12.2  &  -60:03:09  &  2014 Feb 20  &  2397  &  HH~c-8 in \citet{smi10} \\ 
HH~1164 &  10:45:10.5  &  -60:02:42  &  2014 Feb 20  &  2397  &  new jet identified in this work  \\ 
HH~c-10 &  10:45:56.6  &  -60:06:08  &  2014 Apr 17  &  2397  &  \\ 
HH~1156 &  10:45:45.9  &  -59:41:06  &  2015 Jan 16  &  2397  &  HH~c-14 in \citet{smi10} \\ 
\hline
\end{tabular} 
\label{t:ir_synth_obs}
\end{table*}

H~{\sc ii} regions offer one environment where jets from low- and intermediate-mass stars can be studied with similar techniques. 
Feedback from massive stars will have cleared much of the original molecular cloud, allowing UV radiation from nearby O-type stars to illuminate the jet body after it breaks free from its natal cloud. 
External irradiation lights up otherwise invisible components of the jet, including cold material that has not been excited in shocks and would therefore remain unseen in a quiescent region \citep[see, e.g.][]{rei98,bal01}. 
This more complete view allows for a better census of the mass in the jet. 
The physical properties of the jets can be calculated using the theory of photoionized gas, rather than complicated and time-dependent shock models \citep[e.g.][]{bal06}. 
These jets are bright in many of the same lines (e.g., H$\alpha$, [S~{\sc ii}]) as shock-excited Herbig-Haro (HH) objects that are now known to be associated with protostellar outflows \citep{her50,her51,har52,har53}, and are therefore called HH jets. 
Many HH jets have been seen emanating from low-mass stars in Orion \citep{rei98,bal00,bal01,bal06}, but few have been observed from intermediate-mass stars \citep[e.g.][]{ell13}.

\citet{smi10} discovered 39 jets and candidate jets in the Carina Nebula in an H$\alpha$ imaging survey with \emph{HST}/ACS that imaged many of the brighter regions in the nebula with indications of ongoing star formation. 
Jet mass-loss rates, estimated from the H$\alpha$ emission measure (EM), are higher than those measured the same way for the jets in Orion, suggesting that the driving sources are intermediate-mass protostars. 
The high-luminosities of protostars identified along the jet axes support this hypothesis \citep[see][]{smi04,ohl12,rei13}.

\citet{rei13} showed that bright [Fe~{\sc ii}] emission from these jets arises in high density, low ionization (or neutral) regions of the jets that are not bright in H$\alpha$ emission. 
While [Fe~{\sc ii}] is often assumed to be shock-excited, this is not necessarily the case in regions with significant photoexcitation. 
Regardless of the excitation mechanism, however, the survival of Fe$^+$ emission in the harsh UV environment created by $\sim 70$ O-type stars in the Carina Nebula requires a large column of neutral material to shield Fe$^+$ from further ionization by photons with energy $\geq$16.2~eV \citep{rei13}.  
Less energetic photons will penetrate deeper into the jet, producing [Fe~{\sc ii}] emission deeper in the jet core (the first ionization potential of Fe is 7.9~eV). 
Accounting for the neutral material in the jets increases the estimated mass-loss rate by as much as an order of magnitude, compared to that derived from the H$\alpha$ EM. This points to a distinct class of powerful outflows from intermediate-mass protostars.

Several other factors point to near-IR [Fe~{\sc ii}] emission as a better tracer of irradiated, high mass-loss rate jets in H~{\sc ii} regions. 
The irradiated pillars from which many of the jets emerge are themselves bright in H$\alpha$, making it difficult to distinguish faint jet features from filamentary structures associated with the photoevaporative flow off the pillar. 
[Fe~{\sc ii}] emission from the jet provides better contrast between the jet and environment \citep[e.g.][]{smi04}, especially when offline-continuum images exist to subtract PDR emission that might otherwise obscure faint jet features \citep[see, e.g.][]{rei15a}.
Near-IR [Fe~{\sc ii}] emission helps penetrate the dusty birthplaces of the jet-driving protostars, connecting the larger-scale H$\alpha$ outflow to the embedded IR source that drives it \citep[see, e.g.][]{smi04,rei13}. 
In addition, H$\alpha$ and [Fe~{\sc ii}] emission trace different morphologies and kinematics near the protostar in some jets with embedded driving sources \citep{rei15a,rei15b}. 
When the two emission lines trace different outflow components, [Fe~{\sc ii}] emission appears to be a better tracer of the protostellar \textit{jet}, while the broader morphology and slower kinematics of H$\alpha$ resemble those observed in entrained \textit{outflows}.

In this paper, we present near-IR [Fe~{\sc ii}] images obtained with \emph{HST}/WFC3-IR of 18 jets, 2 HH objects, and one candidate jet in the Carina Nebula. 
We targeted 14 of the HH jets discovered by \citet{smi10} with a candidate driving source identified along the jet axis. 
Three additional candidate jets from \citet{smi10} and one new jet serendipitously fell within the imaged area. 
We confirm two of these as collimated jets based on their [Fe~{\sc ii}] morphology. 
Combined with earlier [Fe~{\sc ii}] imaging of four of the most powerful HH jets in Carina \citep{rei13}, this provides a sample of 18 jets driven by sources throughout the $\sim 2-8$ M$_{\odot}$ intermediate-mass range.


\section{Observations}\label{s:ir_synth_obs}
\begin{figure*}
\centering
\includegraphics[trim=25mm 15mm 25mm 15mm,angle=0,scale=0.725]{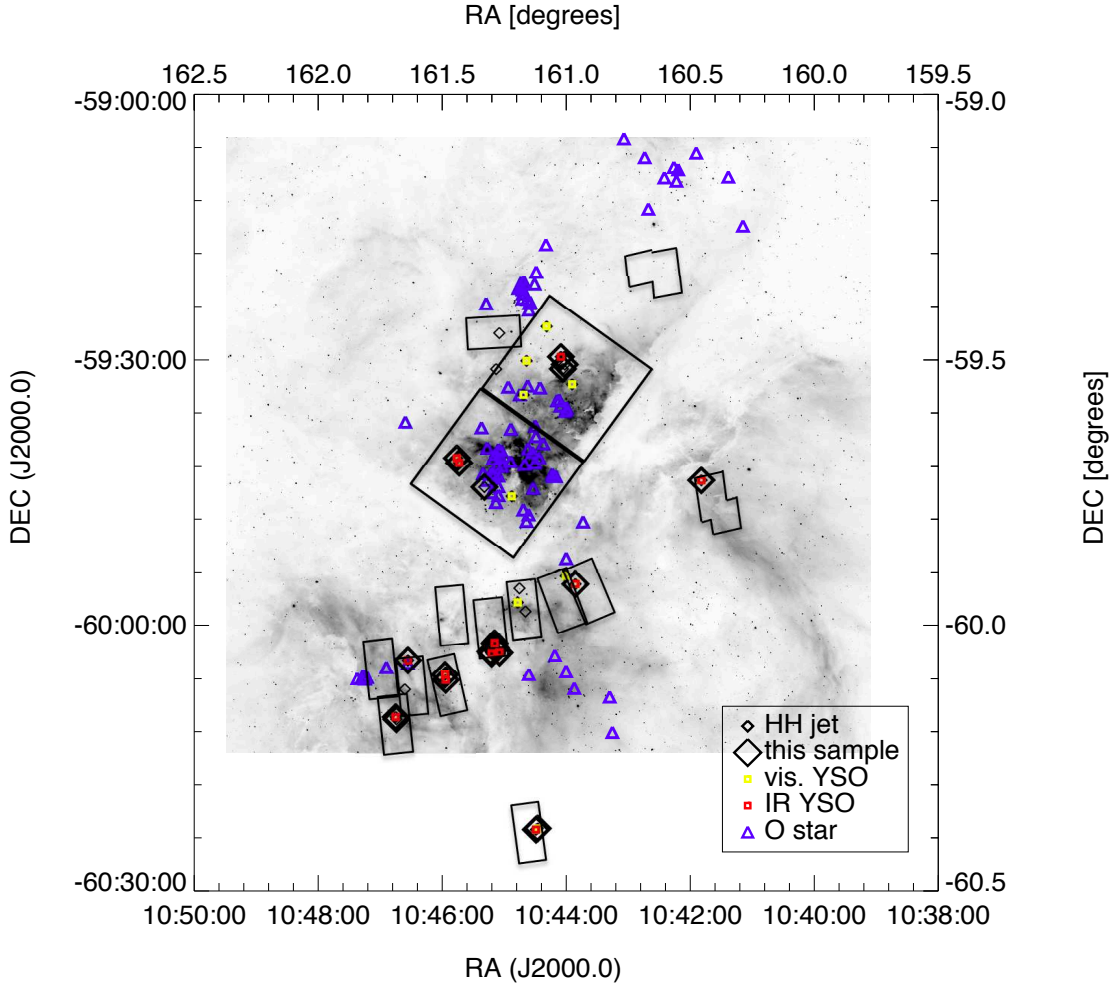} 
\caption[Jet and O-type star distribution in Carina]{A ground-based H$\alpha$ image of Carina with black diamonds that show the location of the HH jets discovered by \citet{smi10}. Larger, thicker diamonds denote the subset of jets that we observed in [Fe~{\sc ii}] with WFC3-IR. Red and yellow squares indicate jets with a driving source identified in the IR or optical, respectively. Blue triangles show the O-type stars in Carina cataloged by \citet{smi06a}. Rectangles show the footprint of the original \emph{HST}/ACS survey \citep[also see Figure~1 in][]{smi10}. }\label{fig:jet_ostar_distro} 
\end{figure*}


\subsection{New [Fe~{\sc ii}] images from \textit{HST}/WFC3-IR} 
Table~\ref{t:ir_synth_obs} lists jet positions, observation dates, and integration times of the new \emph{HST}/WFC3-IR images presented in this paper obtained under program GO-13391. 
Regions targeted with WFC3-IR contain one or more jets with a candidate driving source identified near the outflow axis. These candidate driving sources come from the Pan-Carina YSO Catalog (PCYC) identified and modeled by \citet{pov11}. 
Using a total of 8 pointings with \emph{HST}, we obtained near-IR, narrowband [Fe~{\sc ii}] images and complementary off-line continuum images for 16 HH jets in Carina. 

Observing these near-IR lines with \emph{HST} allows us to obtain images free of the OH lines that contaminate narrowband images obtained from the ground. 
Photon shot-noise dominates the signal-to-noise ratio in ground-based data, severely limiting sensitivity to low-surface brightness features. 
In addition, sky lines are variable, reducing the quality of images that isolate [Fe~{\sc ii}] emission by subtracting an image of the off-line continuum. 
The sensitivity and stability of \emph{HST} are key to detailed study of jet emission, even though similar angular resolution is now possible with AO-fed 8-m telescopes on the ground. 

For each pointing, we obtained images with a total integration time of 2397 s in the F126N, F130N, F164N, and F167N filters. 
Observations employ the same setup and integration times used to obtain the [Fe~{\sc ii}] images of the HH~666, HH~901, HH~902, and HH~1066 jets in Carina presented by \citet{rei13}. 
We include those observations in this analysis. 
Relevant details of those images are listed in Table~\ref{t:ir_synth_obs}, and described in depth in \citet{rei13}. 
Together with new [Fe~{\sc ii}] observations, this dataset provides a sample with nearly uniform sensitivity to [Fe~{\sc ii}] that allows us to detect faint jet features. 
Unlike the [Fe~{\sc ii}] observations presented in \citet{rei13}\footnote{We analyzed archival narrowband [Fe~{\sc ii}] images obtained with WFC3-IR after SM4 and to commemorate the 20$^{th}$ anniversary of \emph{HST}.}, the new observations presented in this paper also include simultaneous offline continuum images (in the F130N and F167N filters to complement F126N and F164N, respectively). 
Using the same integration time in all four filters allows us to subtract continuum emission at the same signal-to-noise. 
All images were obtained using a box-dither pattern to avoid dead pixel artifacts and provide modest resolution enhancement.

\section{Results}\label{s:ir_synth_results}

We detect near-IR [Fe~{\sc ii}] emission from every HH jet that we targeted in the Carina Nebula targeted for follow-up with \emph{HST}/WFC3-IR. 
Figure~\ref{fig:jet_ostar_distro} shows the distribution of the known O-type stars and HH jets in the Carina Nebula. 
Jets studied in this paper are denoted with thick black diamonds. 
We targeted a subset of jets with an embedded YSO identified near the jet axis. 
Candidate jet-driving sources were selected from YSOs identified and modeled in the PCYC by \citet{pov11}. 
In many cases, new [Fe~{\sc ii}] observations confirm this source as the embedded protostar that drives the jet; these are marked with red squares in Figure~\ref{fig:jet_ostar_distro}. 
Only for HH~900 do [Fe~{\sc ii}] observations definitively rule out the candidate driving sources identified near the outflow axis \citep[see Section~\ref{ss:new_w_cont} and][]{rei15a}. 
In the following sections, we briefly describe each jet.

\begin{figure}
\centering
$\begin{array}{c}
\includegraphics[trim=15mm 0mm 0mm 0mm,angle=0,scale=0.325]{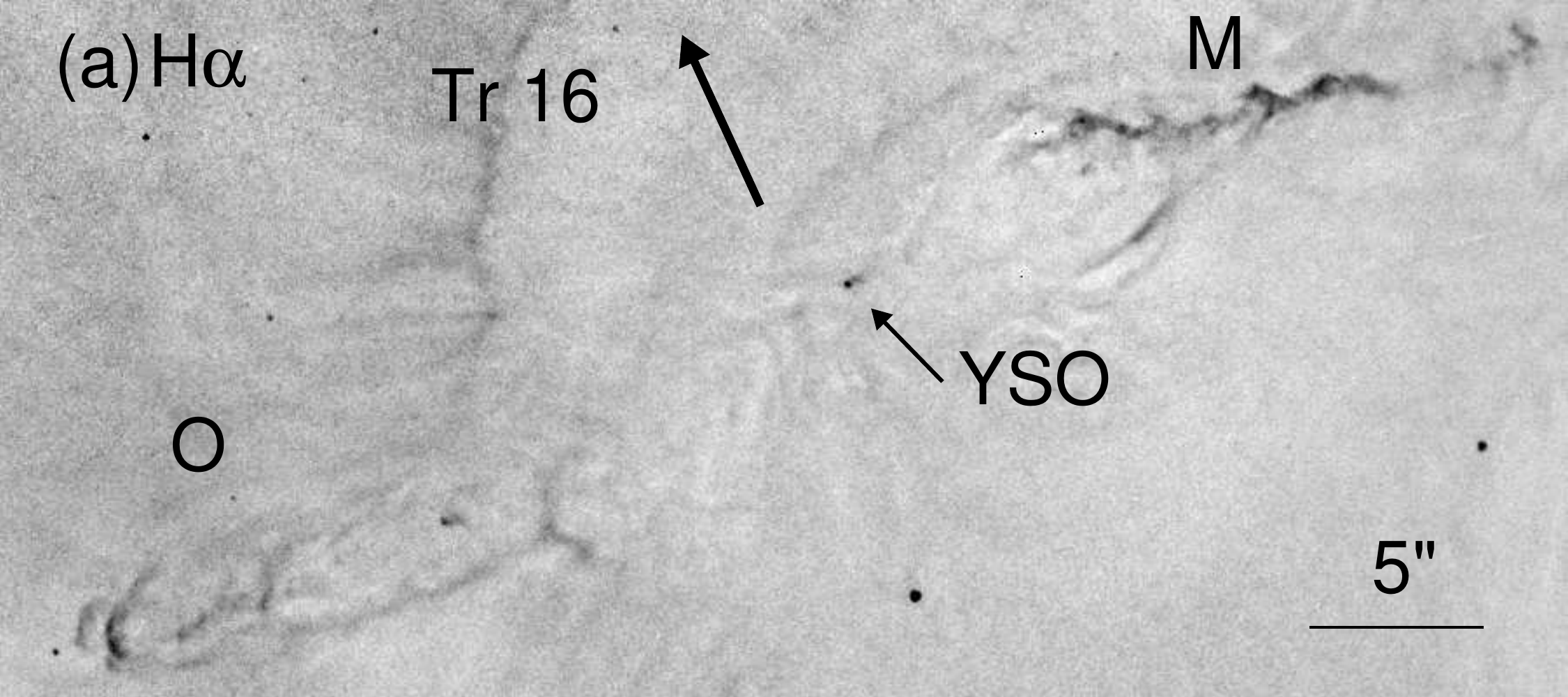} \\
\includegraphics[angle=0,scale=0.305]{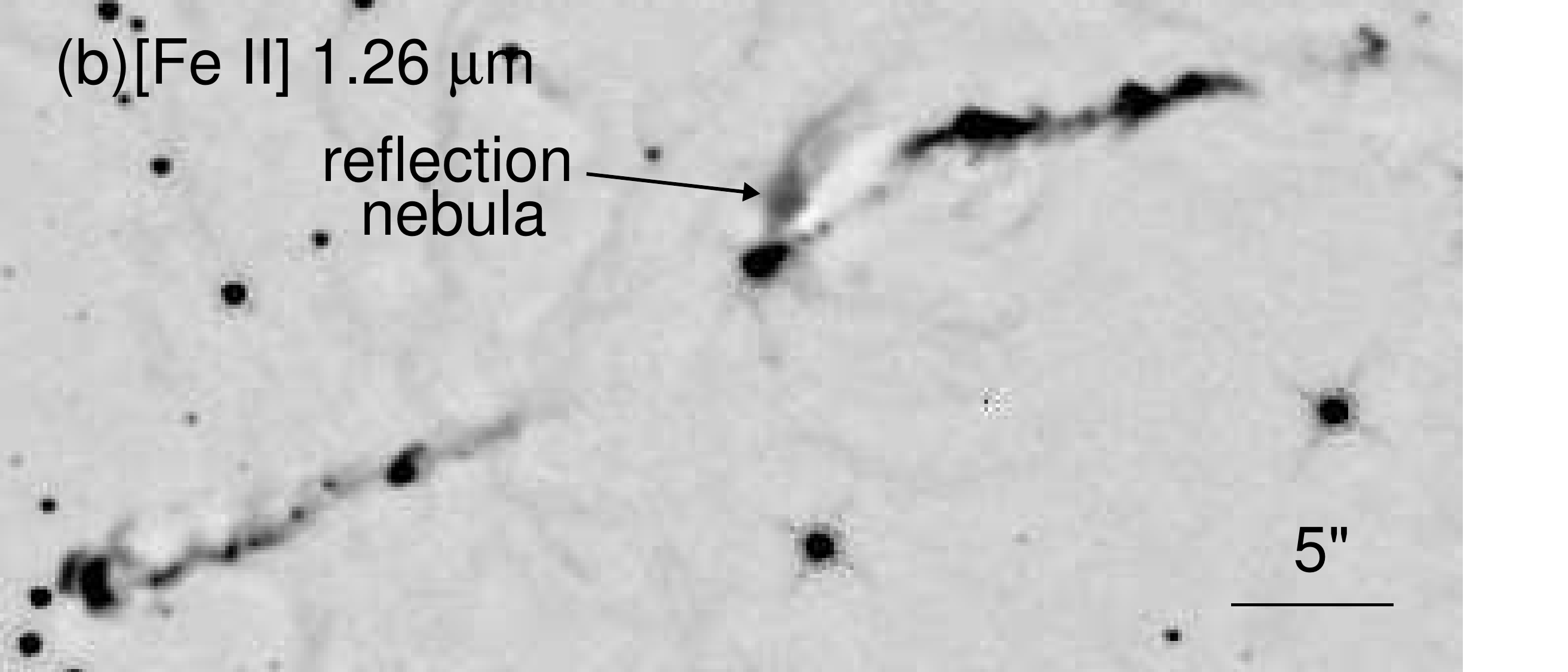} \\
\includegraphics[angle=0,scale=0.305]{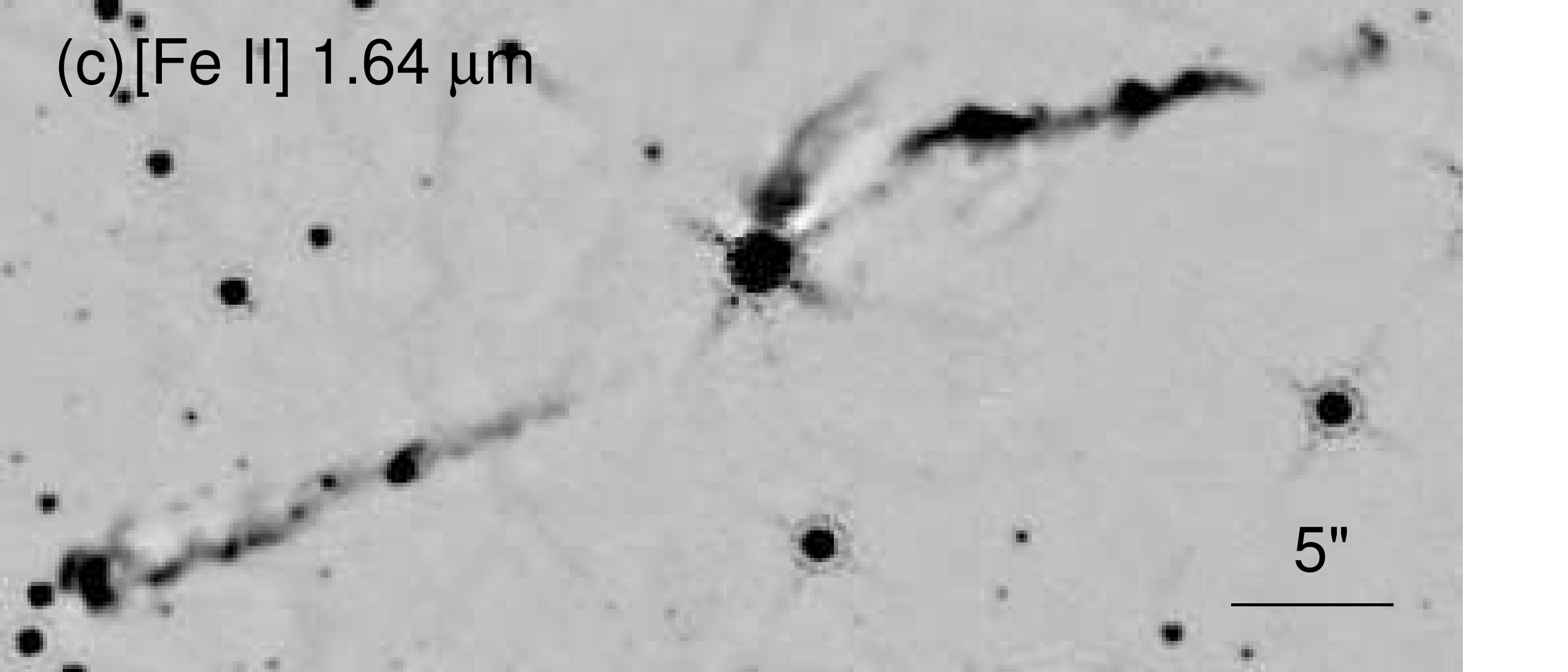} \\
\includegraphics[trim=25mm 0mm 0mm 0mm,angle=0,scale=0.450]{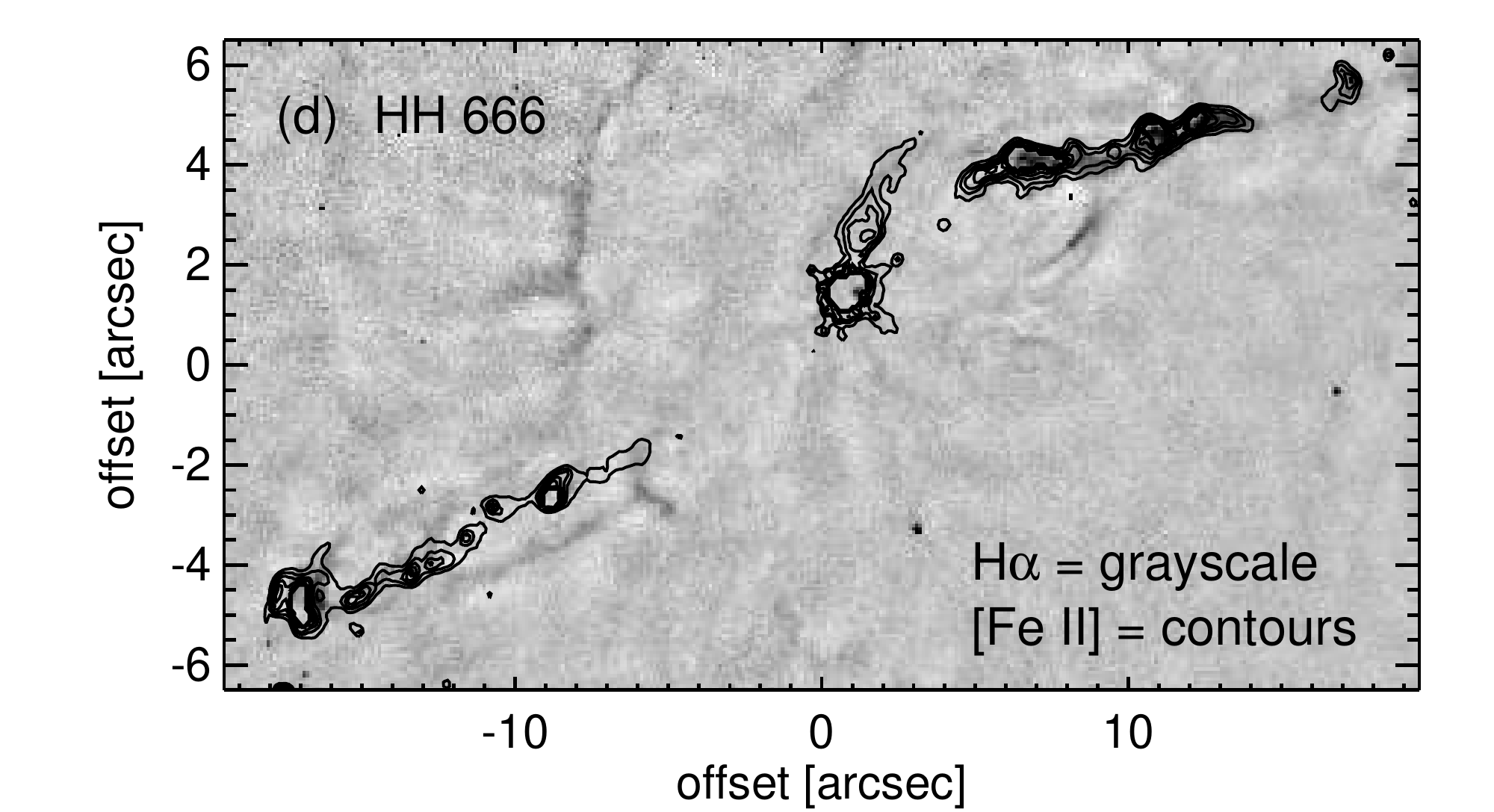} \\
\end{array}$
\caption{H$\alpha$ (a) and [Fe~{\sc ii}] (b,c) images of HH~666 from \citet{rei13}. These archival [Fe~{\sc ii}] data were not obtained with simultaneous off-line continuum images. Each image has been median-filtered to improve contrast with emission from the irradiated dust pillar. 
Panel (d) shows [Fe~{\sc ii}] contours (1.26 \micron\ $+$ 1.64 \micron) on the \emph{HST}/ACS H$\alpha$ image. 
HH~666 is an example of a two-component jet, with wide-angle H$\alpha$ emission that surrounds the collimated [Fe~{\sc ii}] jet. 
Straight near-IR [Fe~{\sc ii}] emission inside the pillar clearly connects the larger scale jet with the driving protostar. 
}\label{fig:hh666_feii} 
\end{figure}

\subsection{Jets without simultaneous off-line continuum images (archival images)}\label{ss:old_no_cont}

\citet{rei13} presented narrowband WFC3-IR [Fe~{\sc ii}] images of four HH jets in Carina and demonstrated that bright [Fe~{\sc ii}] emission traces low-ionization material in the shielded jet core. 
Two of the four jets studied by \citet{rei13} have an IR-bright candidate protostar identified with \emph{Spitzer} that lies near the jet axis. 
In both cases, [Fe~{\sc ii}] emission traces the jet inside the natal globule, clearly connecting the larger scale H$\alpha$ outflow with its embedded driving source. 
No emission from the jet is observed inside the globule from objects that do not have a point source detected near the jet axis with \emph{Spitzer}. 
Instead, [Fe~{\sc ii}] emission appears to be offset $\sim 1$\arcsec\ ($\sim 2000$~AU or $\sim 0.01$~pc at the distance of Carina) from the globule edge. 
Below, we summarize key results for these four jets. 


\textit{HH~666:} The first jet detected in Carina was seen to have bright H$\alpha$ and [Fe~{\sc ii}] emission in ground-based images \citep{smi04}. 
Subsequent \emph{HST} imaging showed that the two emission lines are offset laterally from one another. 
Bright [Fe~{\sc ii}] emission traces a collimated jet that is surrounded by a cocoon of H$\alpha$ emission. 
Taken together, the two lines show that HH~666 has the same morphology as a jet-driven outflow \citep[see Figure~\ref{fig:hh666_feii} and][]{smi10,rei13}. 
Different kinematics traced by Doppler velocities in spectra of the two lines also argue for a two-component jet-outflow system. 
Fast, steady [Fe~{\sc ii}] radial velocities more typical of a jet are surrounded by slower H$\alpha$ emission that accelerates away from the driving source, similar to the Hubble flows seen in entrained molecular outflows \citep{smi04,rei15b}.

\begin{figure}
\centering
$\begin{array}{c}
\includegraphics[angle=0,scale=0.275]{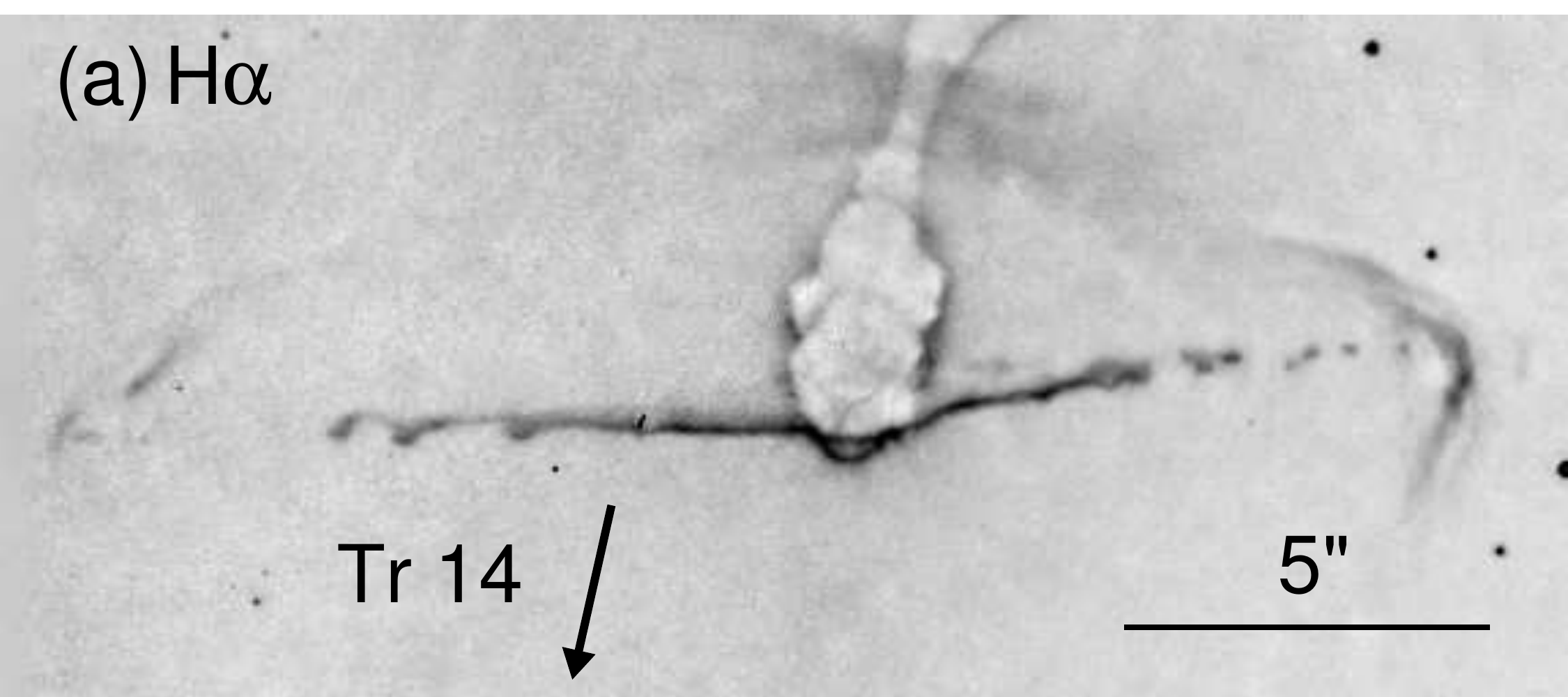} \\
\includegraphics[angle=0,scale=0.385]{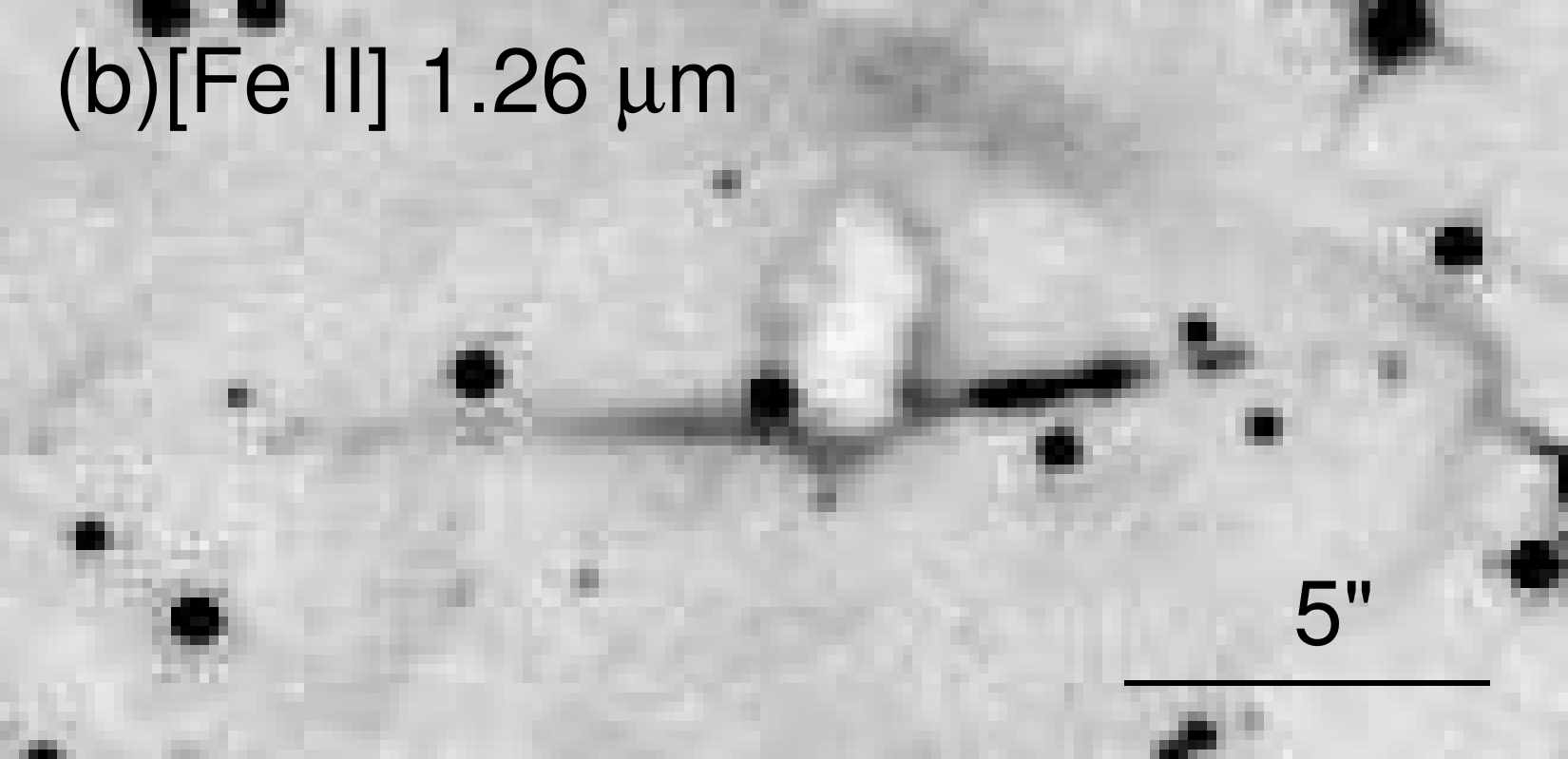} \\
\includegraphics[angle=0,scale=0.385]{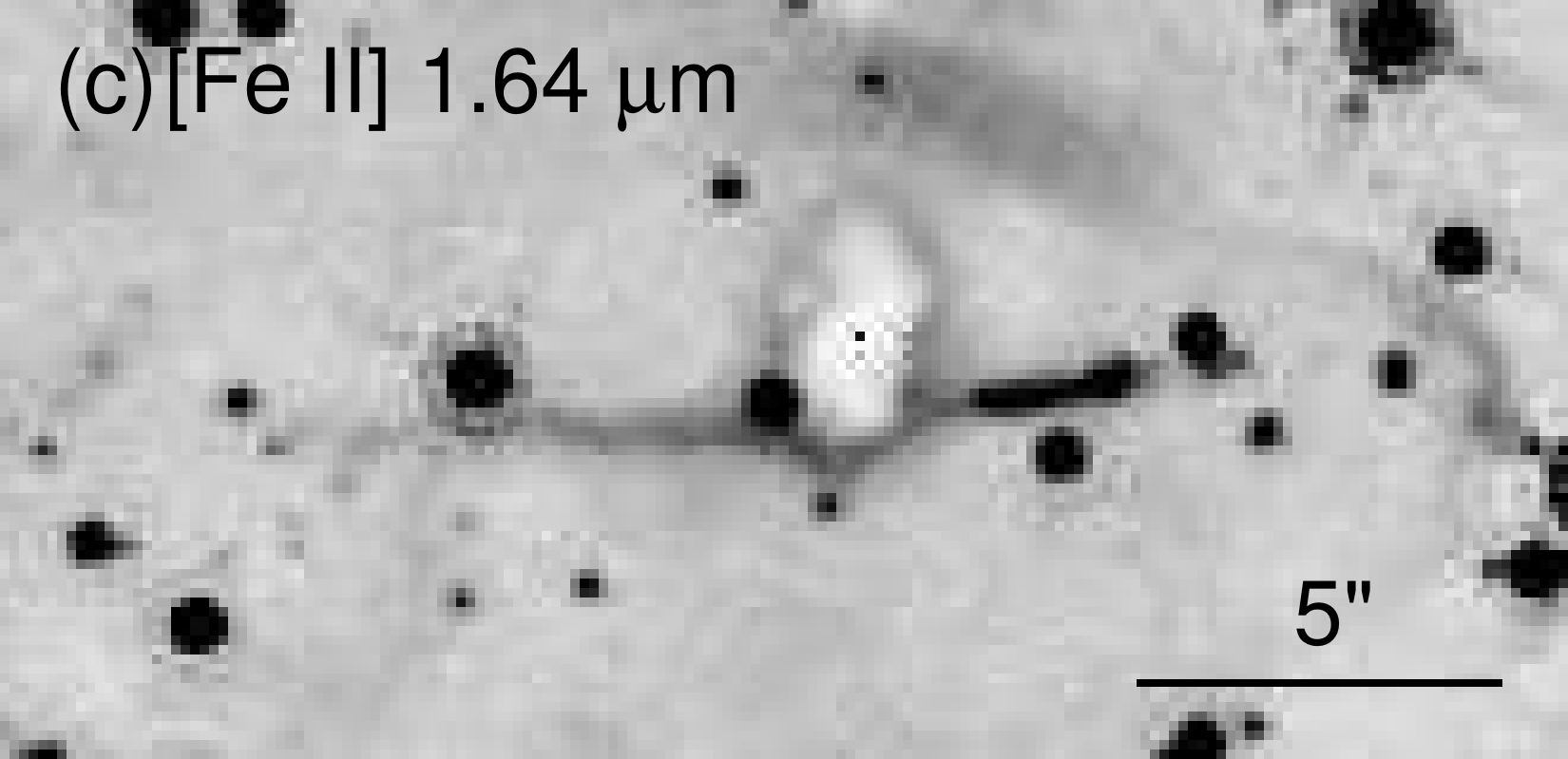} \\
\includegraphics[trim=10mm 0mm 0mm 0mm,angle=0,scale=0.40]{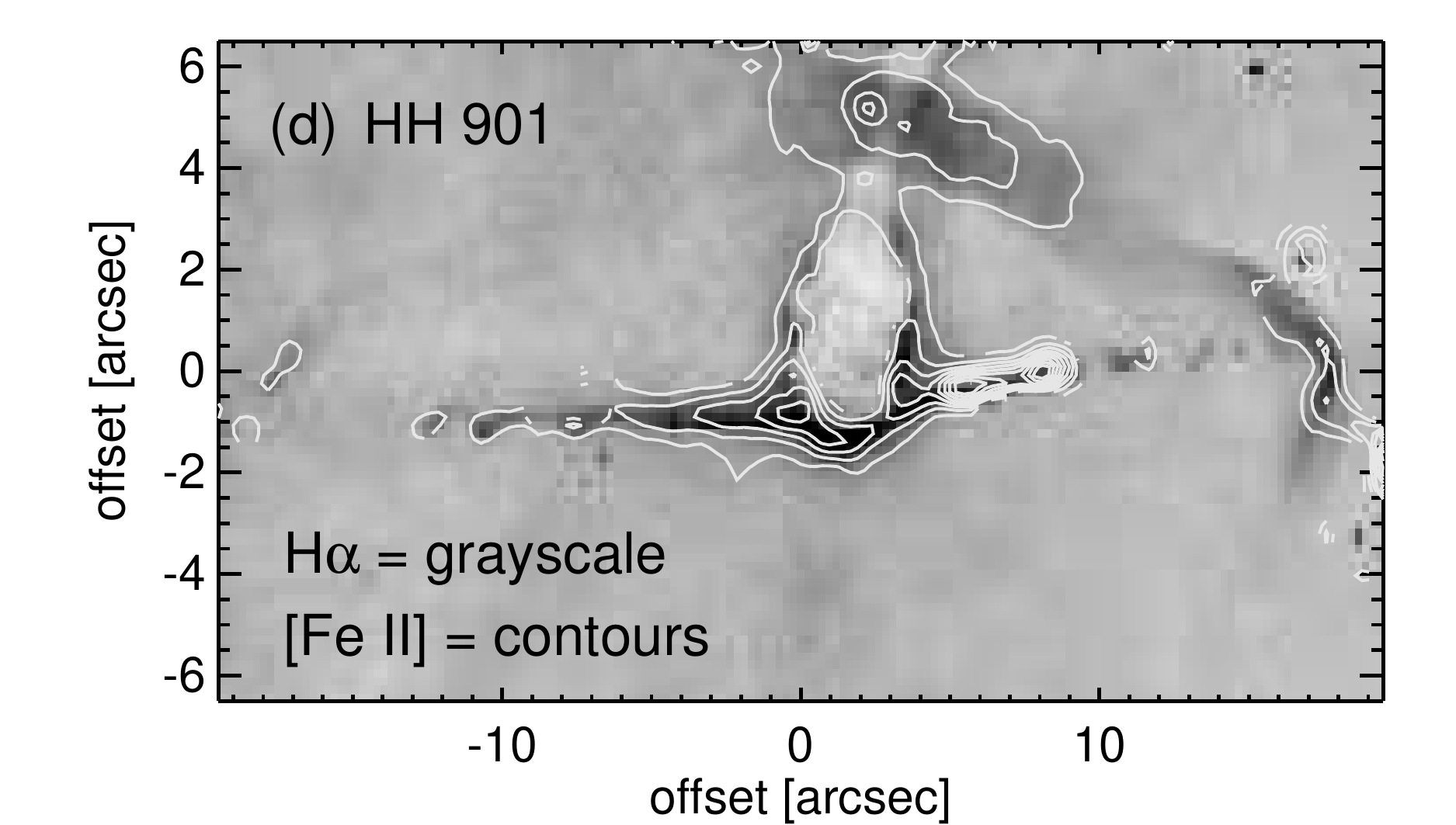} \\
\end{array}$
\caption{H$\alpha$ (a) and [Fe~{\sc ii}] (b,c) images of HH~901 from \citet{rei13}. 
Panel (d) shows contours [Fe~{\sc ii}] emission on an H$\alpha$ image. 
The bright star seen in the near-IR images at the eastern edge of the pillar is an interloper, as it lies too far above the [Fe~{\sc ii}] jet axis to be the driving source. 
Kinematics require a driving source embedded in the head of the globule, although no protostar has been detected. 
}\label{fig:hh901_feii} 
\end{figure}
\begin{figure}
\centering
$\begin{array}{c}
\includegraphics[angle=0,scale=0.545]{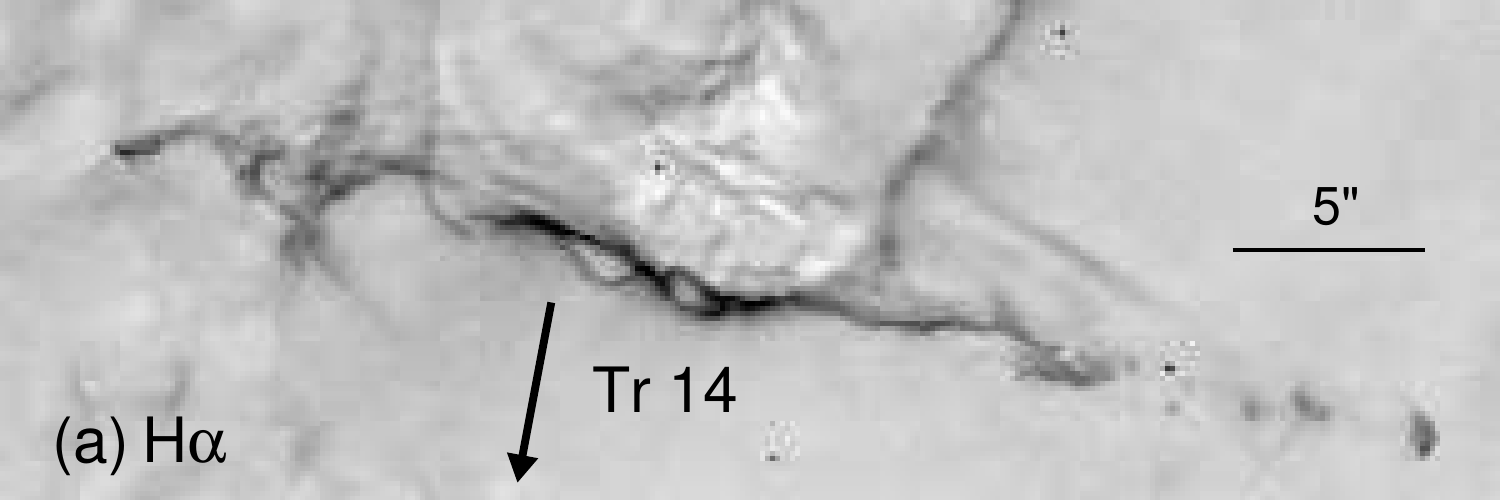} \\
\includegraphics[angle=0,scale=0.545]{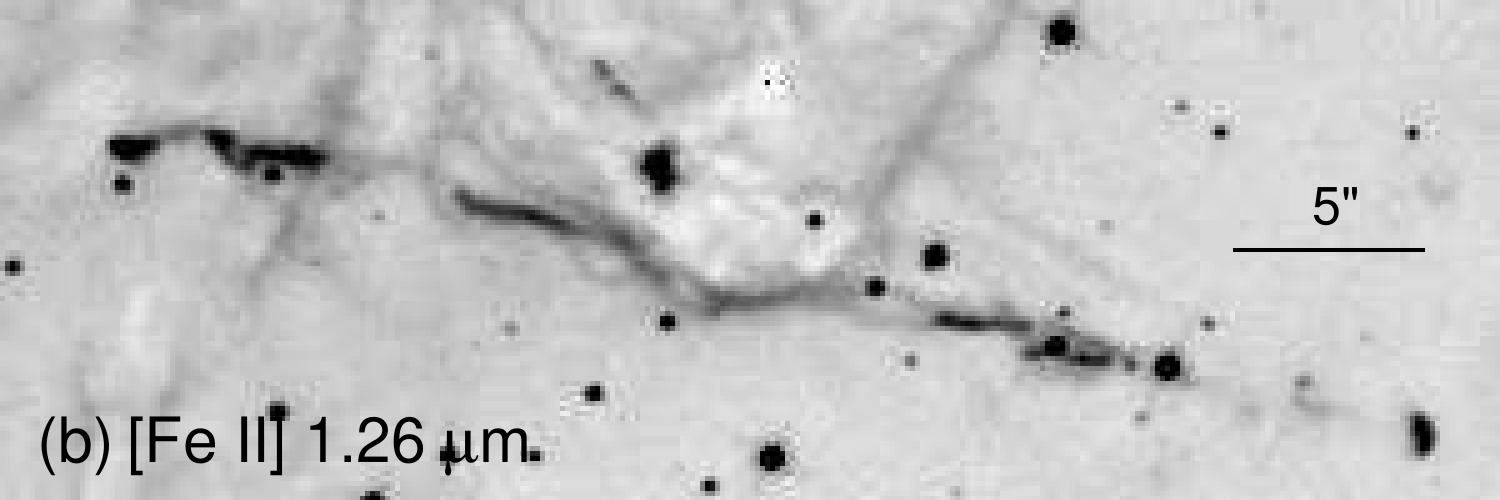} \\
\includegraphics[angle=0,scale=0.545]{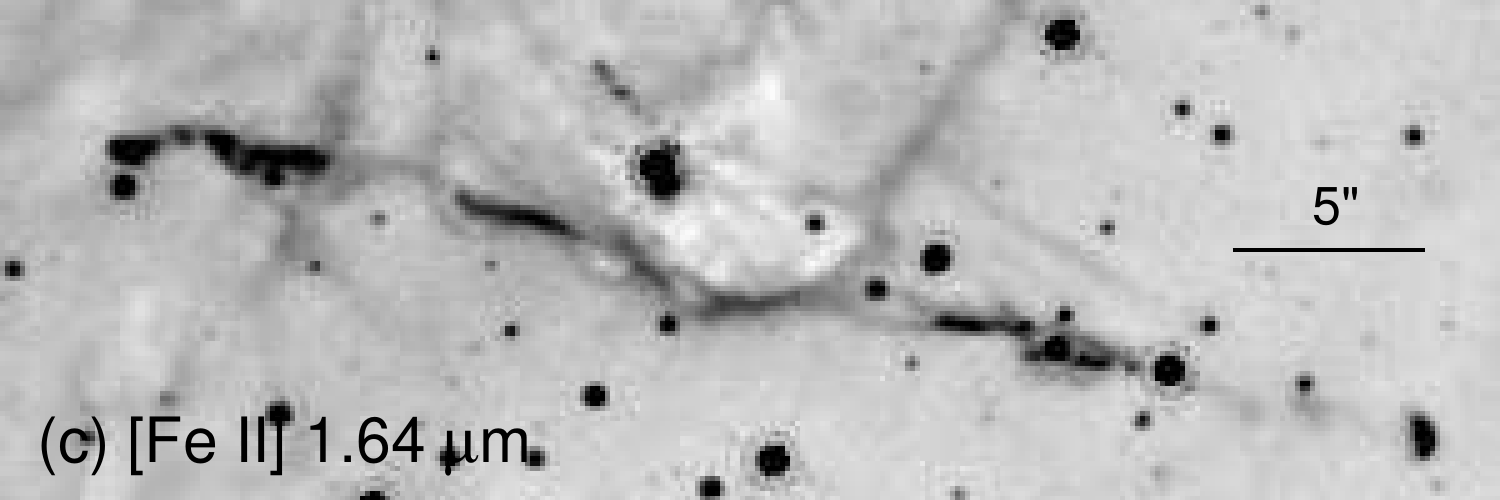} \\
\includegraphics[trim=10mm 0mm 0mm 0mm,angle=0,scale=0.545]{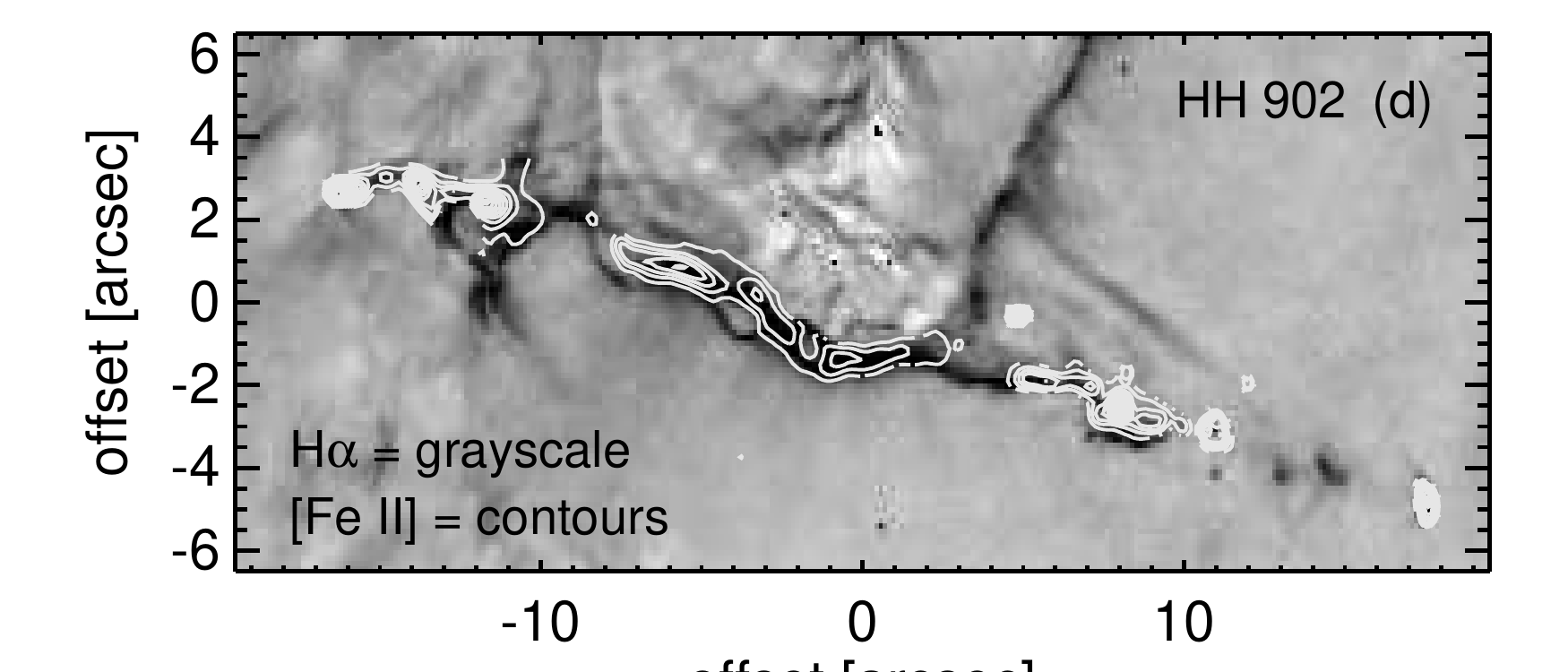} \\
\end{array}$
\caption{H$\alpha$ (a) and [Fe~{\sc ii}] (b,c) images of HH~902 from \citet{rei13}, with a comparison of the two in panel (d). 
Bright [Fe~{\sc ii}] emission traces the body of the jet along the edge of the broad natal pillar. 
Like HH~901, no driving source has been detected but kinematics require a protostar just inside the apex of the pillar. 
}\label{fig:hh902_feii} 
\end{figure}

\textit{HH~901 and HH~902:} \citet{smi10} hypothesized that HH~901 and HH~902 are both high density jets with an unseen neutral core based on sharp H$\alpha$ emission that traces the ionization front along the length of the jets. 
Near-IR [Fe~{\sc ii}] observations presented in \citet{rei13} demonstrate that the ionized gas traced by H$\alpha$ does not sample the full width of the jet. 
Bright [Fe~{\sc ii}] emission peaks \textit{behind} the H$\alpha$, further away from the ionizing source, consistent with an ionized skin (H$\alpha$) that shields the neutral jet core ([Fe~{\sc ii}], see Figures~\ref{fig:hh901_feii}~and~\ref{fig:hh902_feii}). 
Neither HH~901 nor HH~902 has a driving source detected in the globule, although proper motions require a source located near the head of the pillar in both cases \citep{rei14}. 

\begin{figure}
\centering
$\begin{array}{cc}
\includegraphics[trim=-10mm 0mm 0mm 0mm,angle=0,scale=0.40]{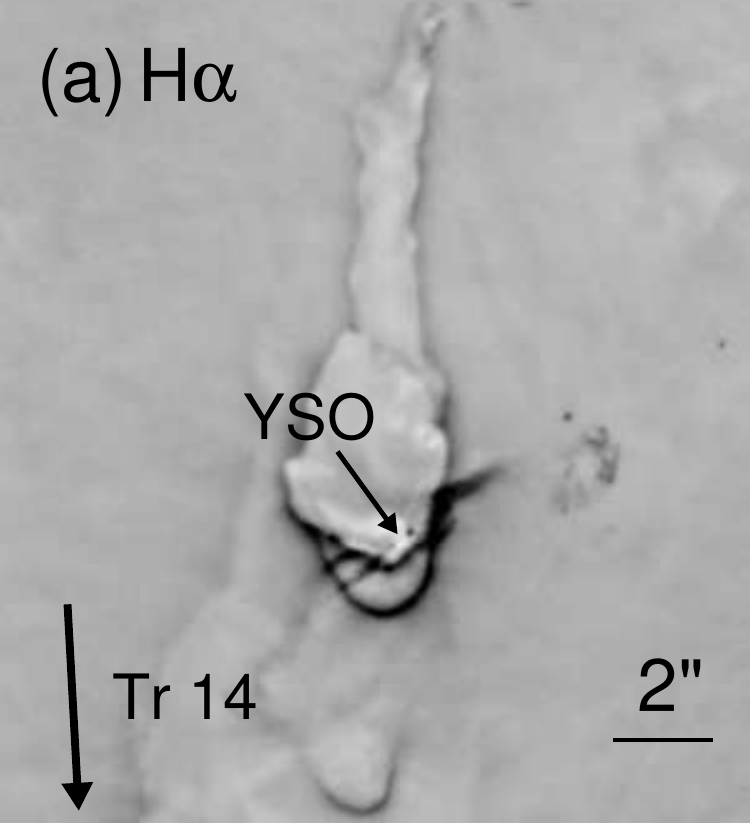} & 
\includegraphics[angle=0,scale=0.325]{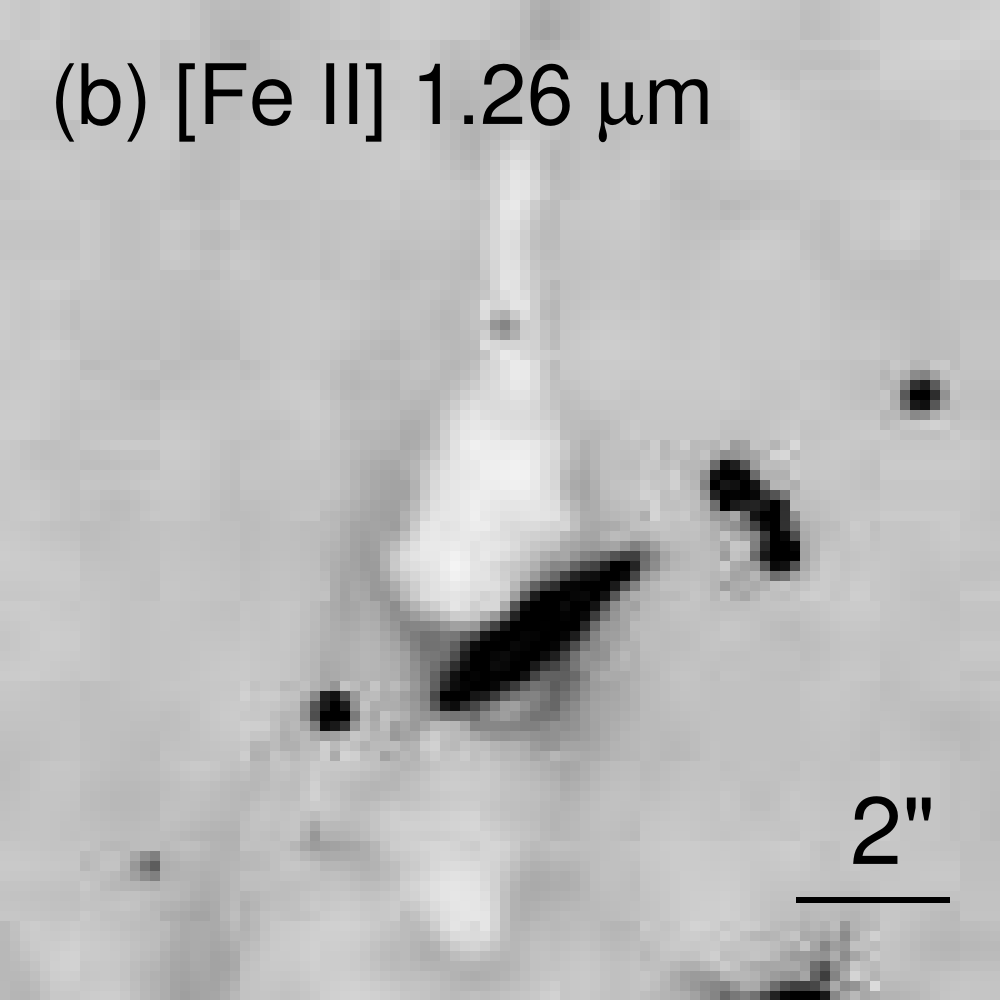} \\
\includegraphics[trim=0mm 0mm 0mm -10mm,angle=0,scale=0.325]{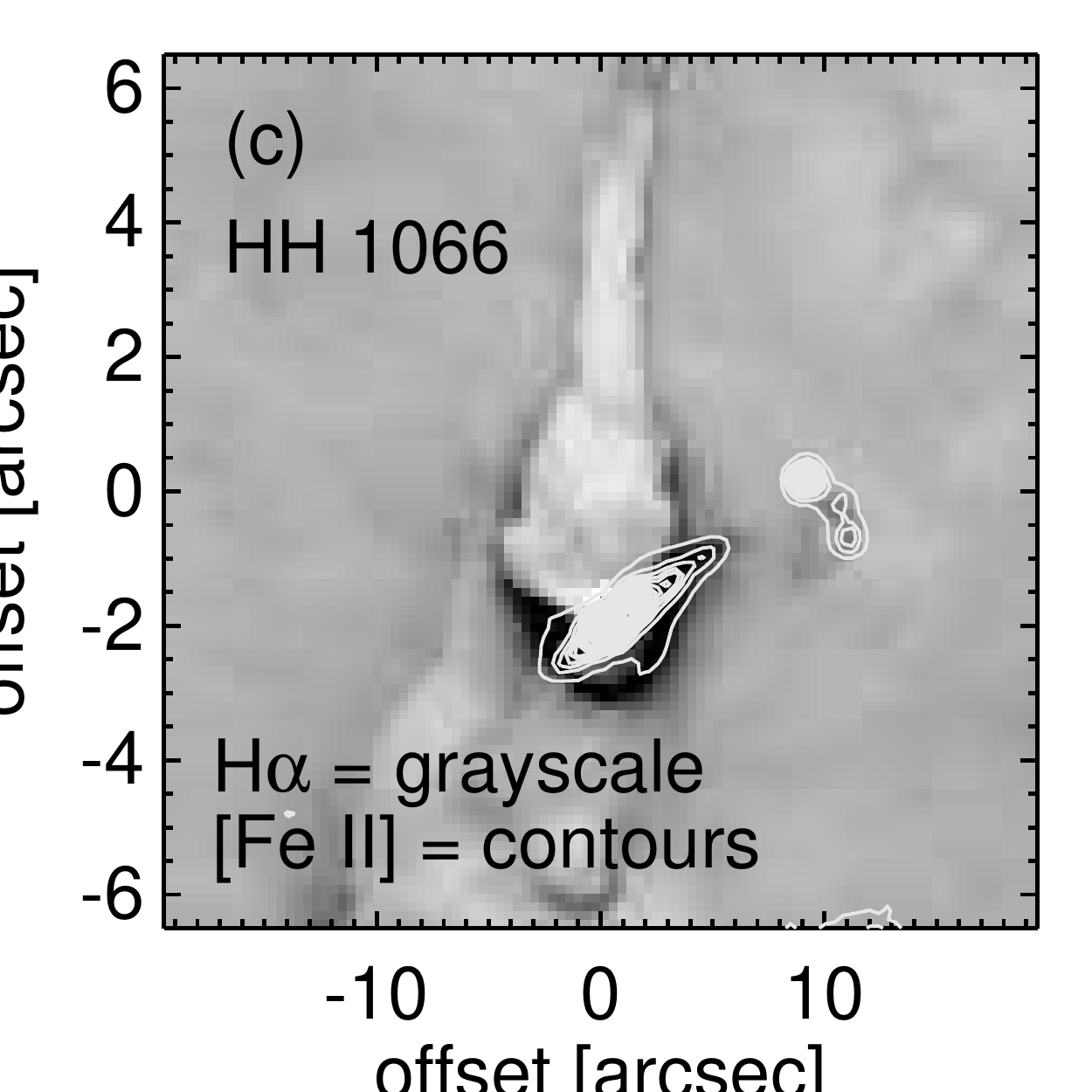} & 
\includegraphics[trim=0mm -20mm 0mm 0mm,angle=0,scale=0.325]{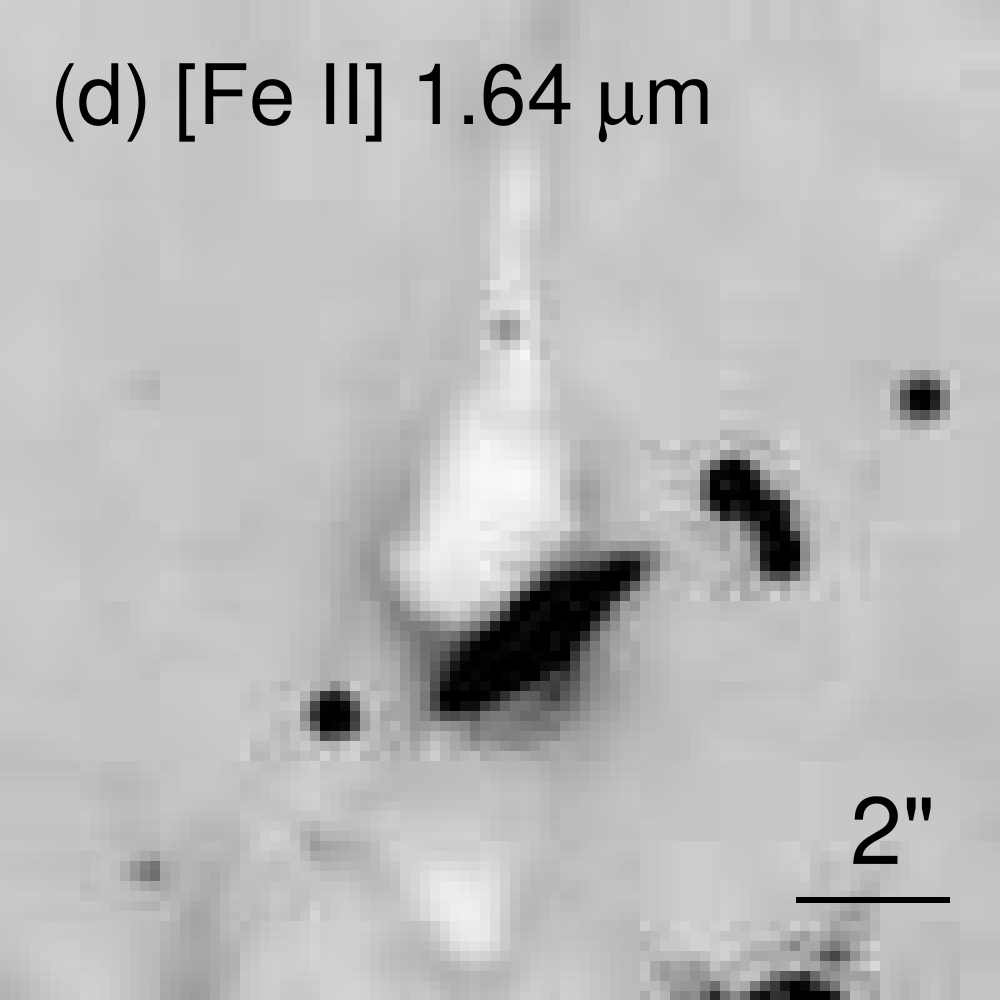} \\
\end{array}$
\caption{H$\alpha$ (a) and [Fe~{\sc ii}] (b,c) images of HH~1066 from \citet{rei13}. 
Strong, collimated [Fe~{\sc ii}] emission confirmed the jet-like nature of HH~1066 (see panel d). 
Two spots of H$\alpha$ emission trace where the jet breaks out of the pillar (see panel a), tracing a bipolar outflow from a young, \emph{Spitzer}-identified protostar located at the tip of the globule. 
}\label{fig:hh1066_feii} 
\end{figure}

\textit{HH~1066:} 
\citet{smi10} identified HH~1066 as a candidate jet based on streamers of H$\alpha$ emission coming off the globule that point toward a bow shock. 
No unambiguous jet body could be identified in H$\alpha$ images alone. 
However, bright [Fe~{\sc ii}] emission clearly traces a collimated jet body that bisects the \emph{Spitzer}-identified YSO \citep[see Figure~\ref{fig:hh1066_feii} and][]{rei13}. 
The ratio of the two [Fe~{\sc ii}] lines shows that the reddening increases toward the inner jet, reaching a maximum at the point source, likely tracing the edge-on optically thick circumstellar disk. 

\subsection{[Fe~{\sc ii}] observations with simultaneous offline-continuum subtracted}\label{ss:new_w_cont}

\begin{figure}
\centering
$\begin{array}{c}
\includegraphics[angle=0,scale=0.35]{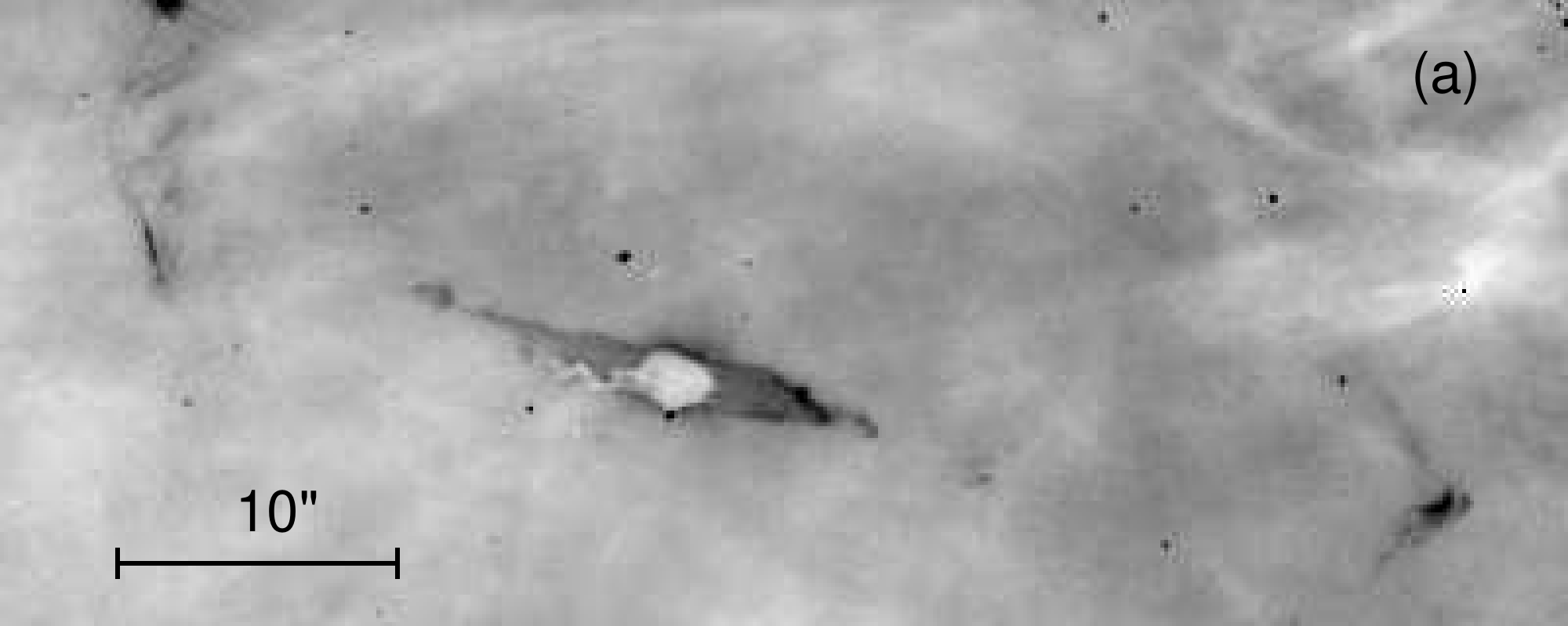} \\ 
\includegraphics[angle=0,scale=0.35]{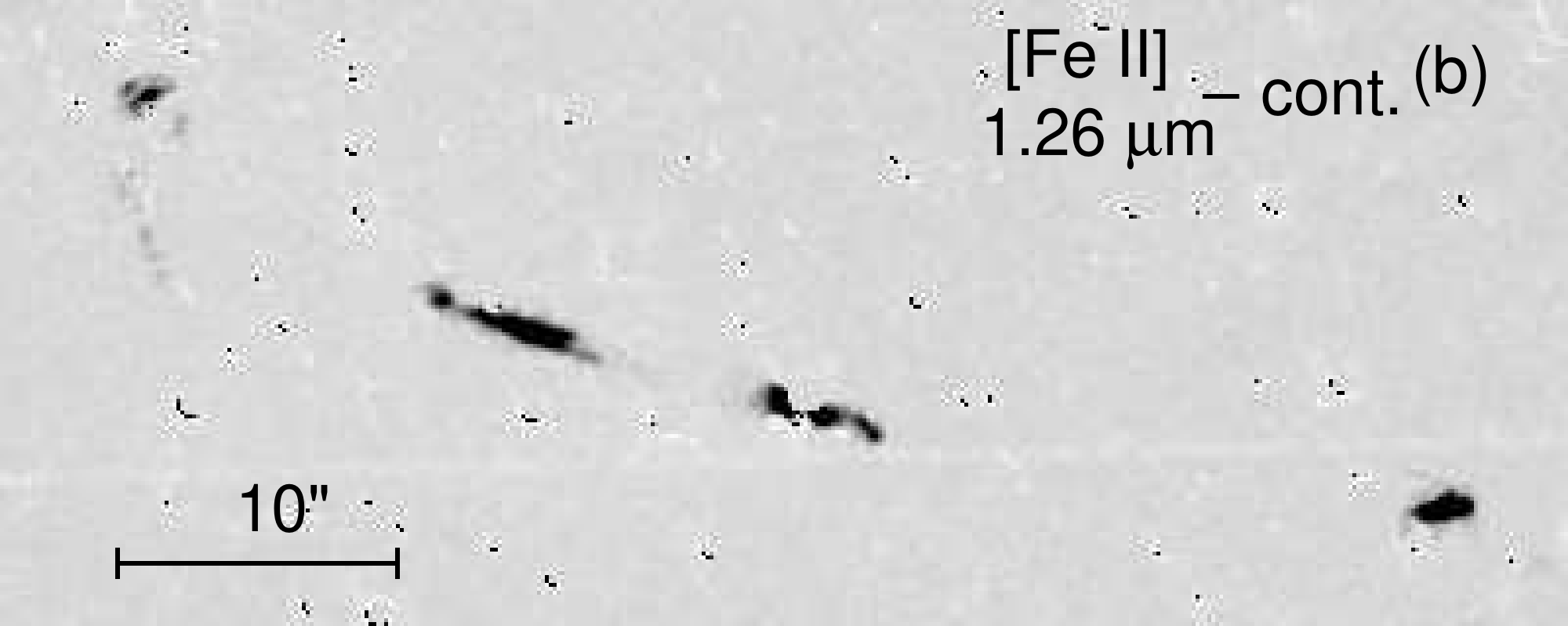} \\ 
\includegraphics[angle=0,scale=0.35]{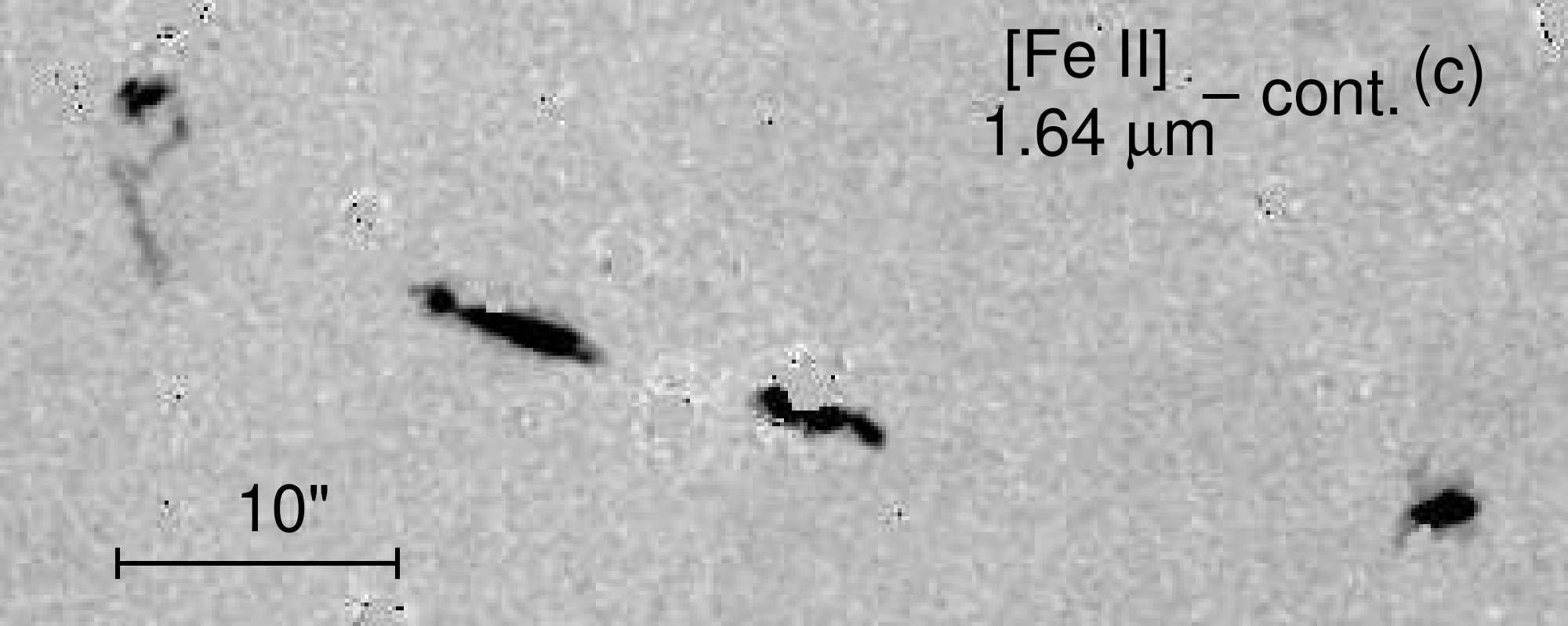} \\ 
\includegraphics[angle=0,scale=0.475]{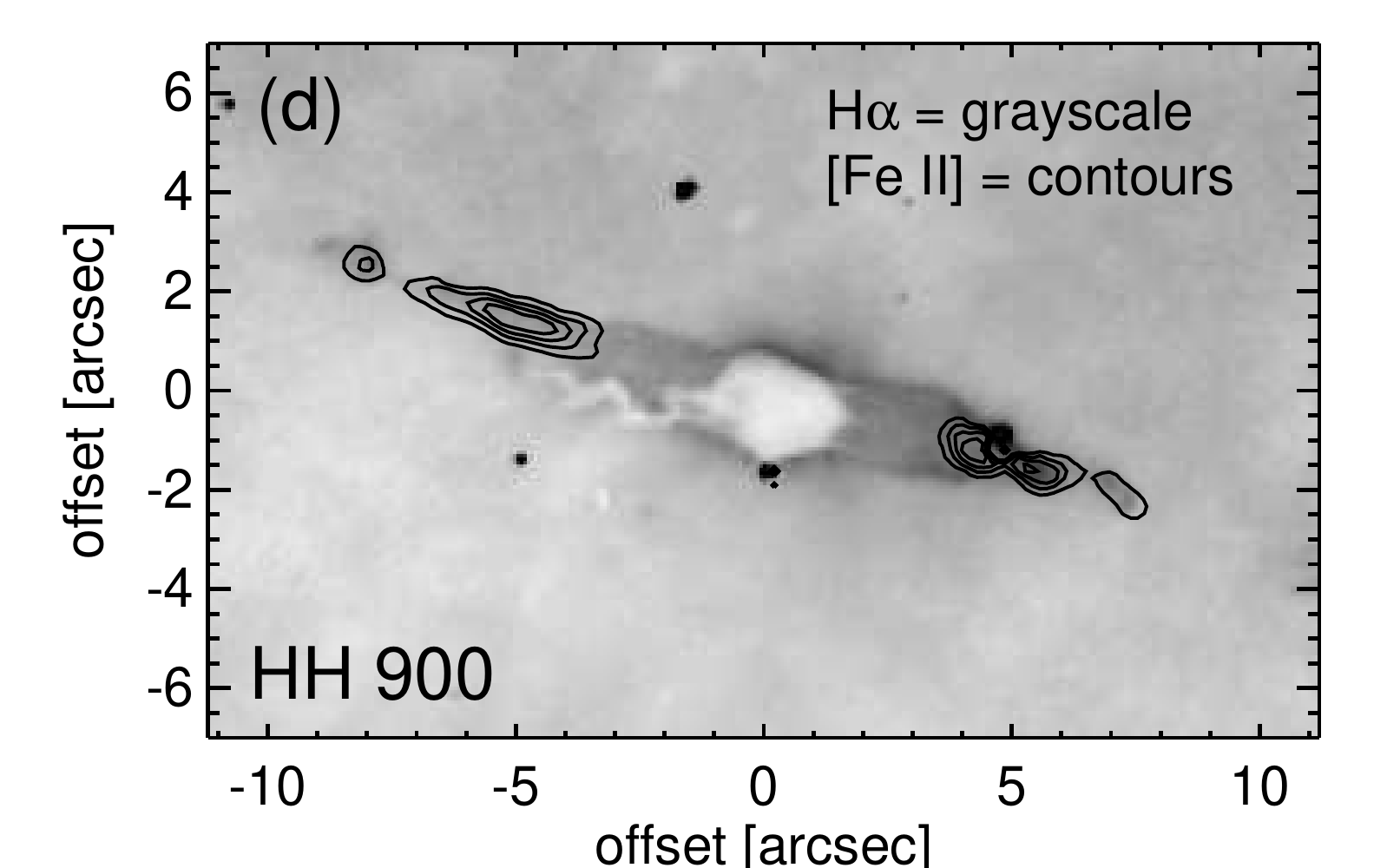} \\ 
\end{array}$ 
\caption{H$\alpha$ image (a) of HH~900 from \citet{smi10} and continuum-subtracted [Fe~{\sc ii}] images (b,c) previously published in \citet{rei15a}. 
HH~900 is one of the two-component jets seen in Carina where H$\alpha$ emission traces the wide-angle, lower velocity outflow and [Fe~{\sc ii}] traces the faster, collimated jet (d). 
}\label{fig:hh900_feii} 
\end{figure}

\textit{HH~900 (previously published):} As in HH~666, HH~900 shows a bright, bipolar [Fe~{\sc ii}] jet with fast, steady velocities moving away from the globule. 
The narrow jet traced by near-IR [Fe~{\sc ii}] emission plows through a broader, slower H$\alpha$ outflow \citep[see Figure~\ref{fig:hh900_feii} and][]{rei15a}. 
H$\alpha$ (and H$_2$) emission from the outflow extend continuously from the edge of the globule. 
In contrast, continuum-subtracted [Fe~{\sc ii}] images clearly show that [Fe~{\sc ii}] emission from the jet is offset from the edge of the globule by $\sim 1$\arcsec\ (and not simply confused with bright emission from the photodissociation region, PDR, along the surface of the globule). 
The jet axis defined by the [Fe~{\sc ii}] emission bisects the globule and H$\alpha$ proper motions require a driving source embedded in the dark, tadpole-shaped globule. 
However, no point source has been detected inside the globule in available IR images.

\begin{figure*}
\centering
$\begin{array}{c}
\includegraphics[angle=0,scale=0.65]{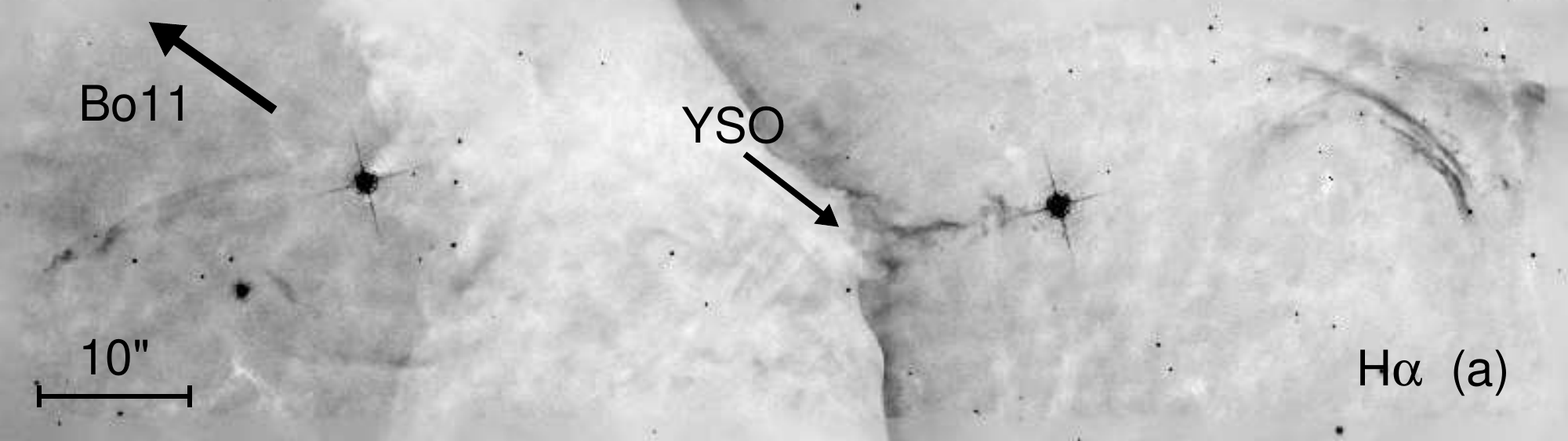} \\ 
\includegraphics[angle=0,scale=0.65]{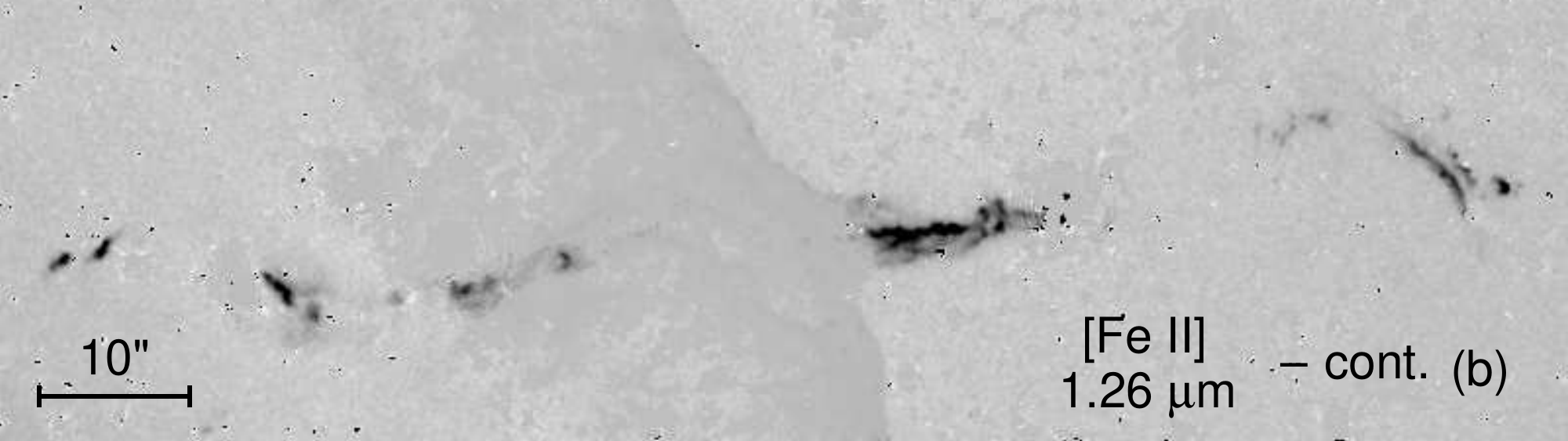} \\ 
\includegraphics[angle=0,scale=0.65]{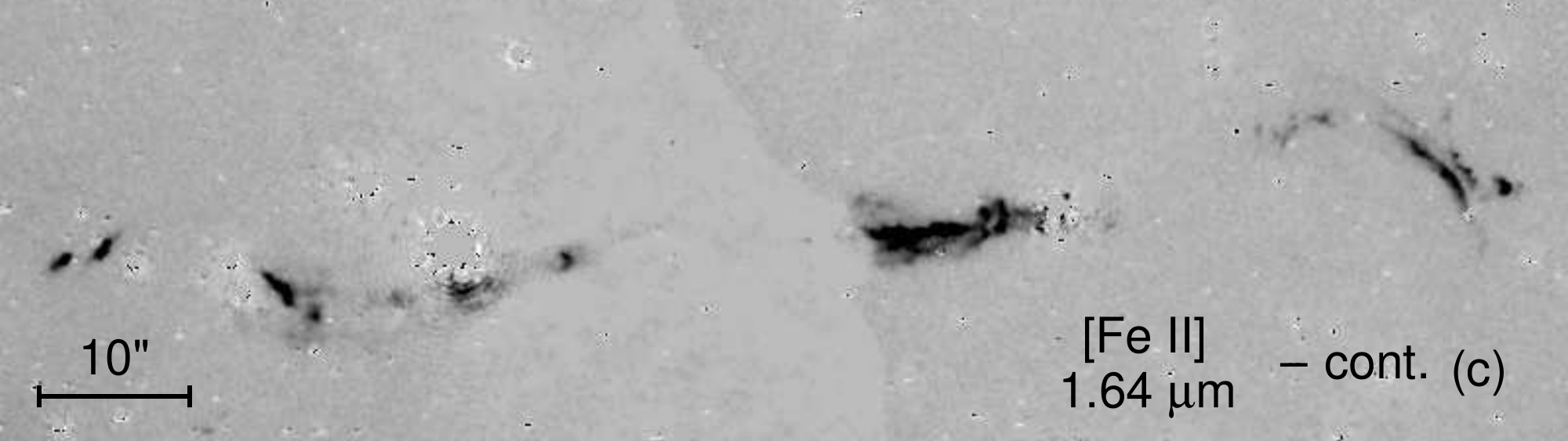} \\ 
\includegraphics[angle=0,scale=0.65]{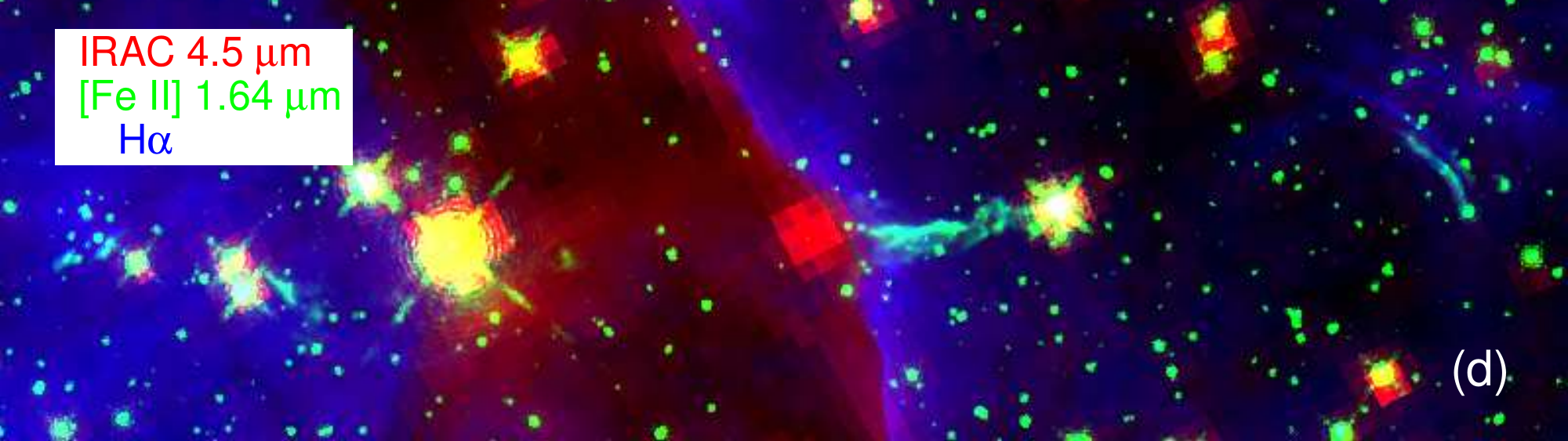} \\ 
\includegraphics[trim=25mm 0mm 0mm 0mm,angle=0,scale=0.65]{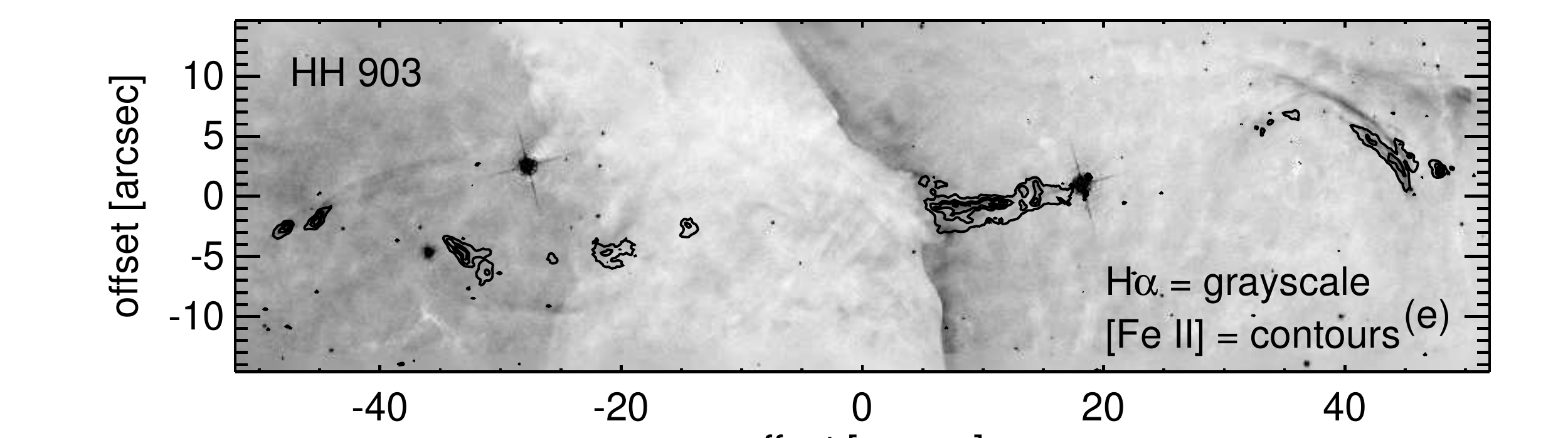} \\ 
\end{array}$ 
\caption{The H$\alpha$ image (a) from \citet{smi10} is shown alongside new, continuum-subtracted near-IR [Fe~{\sc ii}] images (b,c) of HH~903. 
[Fe~{\sc ii}] emission connects the large-scale jet to the IR-bright protostar that lies just inside the pillar ionization front as shown in panel (d). 
Contours of [Fe~{\sc ii}] emission on an H$\alpha$ image (e) highlight portions of the jet inside the pillar that are only seen in the IR. 
}\label{fig:hh903_feii} 
\end{figure*}

\textit{HH~903:} Figure~\ref{fig:hh903_feii} shows HH~903 bursting from the western edge of a dust pillar in the actively star-forming South Pillars \citep{smi10}. 
H$\alpha$ emission from the jet extends on either side of the pillar, reaching a total projected length of $\sim 2$ pc. 
The collimated jet identified in H$\alpha$ images propagates west, delineating a jet axis perpendicular to the major axis of the pillar. 
To the east, H$\alpha$ emission is much less collimated, with tenuous streams of emission that trace a broader sheath like those seen in HH~900 and HH~666~O. 

In new WFC3-IR images, [Fe~{\sc ii}] emission traces the collimated western jet limb seen in H$\alpha$ and delineates the narrow stream of the eastern jet.  
Unlike H$\alpha$, [Fe~{\sc ii}] from the inner jet reaches inside the dust pillar, connecting the larger jet to an embedded IR source. 
The putative driving source lies immediately inside the ionization front along the western edge of pillar. 
The point source can clearly be seen in \emph{Spitzer}/IRAC images with $\lambda \geq 4.5 \mu$m, but is not detected at enough wavelengths to be included among the high-probability YSOs cataloged in the PCYC by \citet{pov11}. 
However, \citet{ohl12} model the mid-IR emission and estimate a protostellar mass of $\sim 4$ M$_{\odot}$ (see Table~\ref{t:jets_ysos}). 
As the jet breaks into the H~{\sc ii} region, it disrupts the bright ionization front along the pillar edge, similar to HH~1016 \citep[see][]{smi10}. 

To the east, [Fe~{\sc ii}] emission traces a single jet axis that bends into an ``S'' shape as it propagates away from the driving source. 
Bright knots of [Fe~{\sc ii}] bisect more diffuse H$\alpha$ emission, similar to HH~666~O \citep[see][]{rei15b}. 
This [Fe~{\sc ii}]-jet and H$\alpha$-cocoon morphology has been seen in other HH jets in Carina \citep[e.g. HH~900 and HH~666, see][]{rei15a,rei15b}. 
Further east, two bright [Fe~{\sc ii}] knots may be part of the larger bow-shock structure seen in H$\alpha$. 

\textit{HH~1004:} HH~1004 emerges from the head of a broad dust pillar. 
Like HH~903 and HH~1016, H$\alpha$ emission along the pillar edge is clumpy, illustrating the disruption of the ionization front by the passage of the jet. 
To the southwest, a diffuse bow shock structure seen in H$\alpha$ lies in front of the pillar, tracing the blueshifted limb of the jet.  

\begin{figure}
\centering
$\begin{array}{c}
\includegraphics[angle=0,scale=0.375]{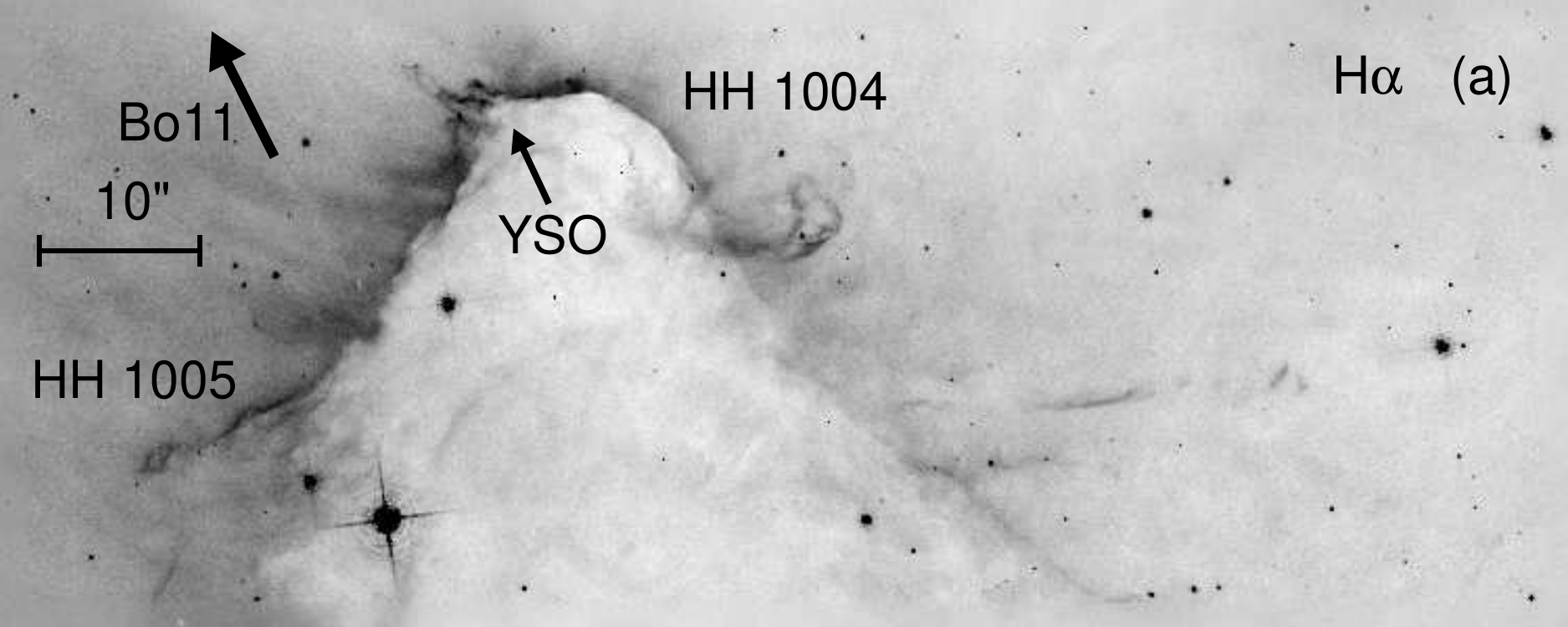} \\ 
\includegraphics[angle=0,scale=0.375]{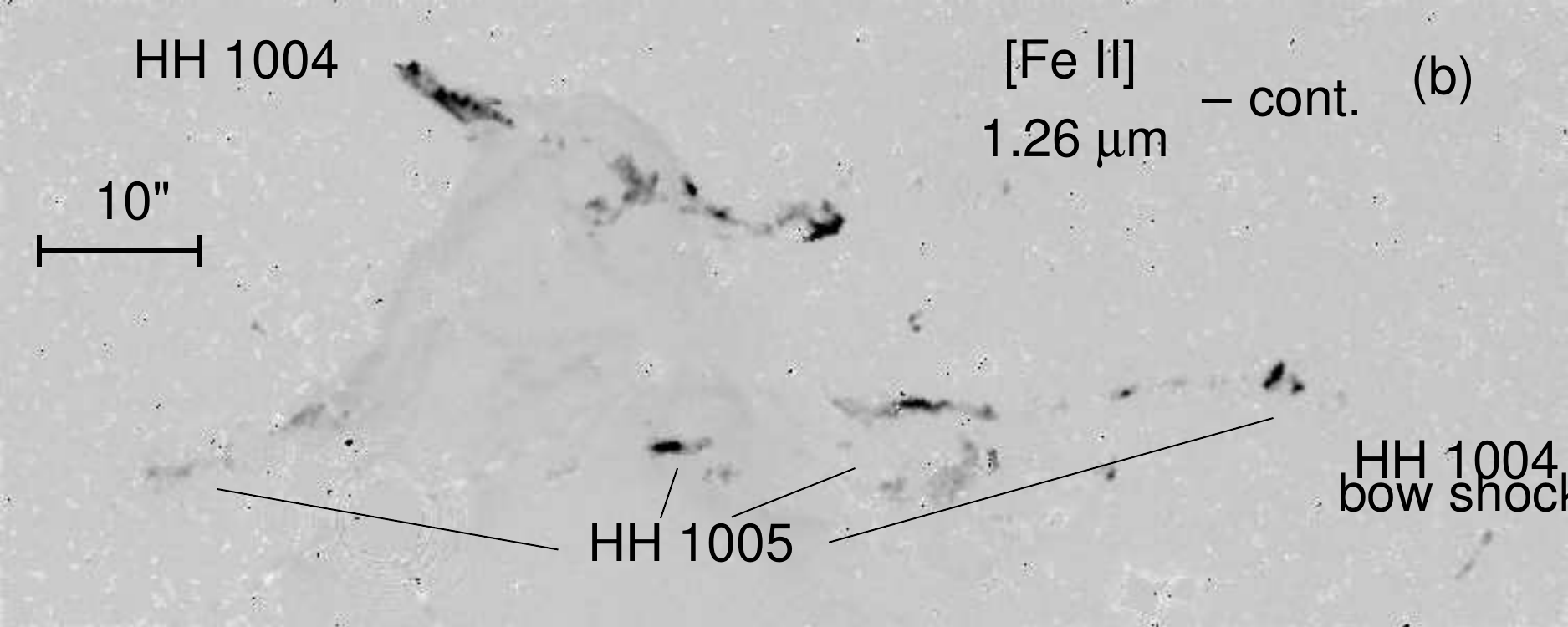} \\ 
\includegraphics[angle=0,scale=0.375]{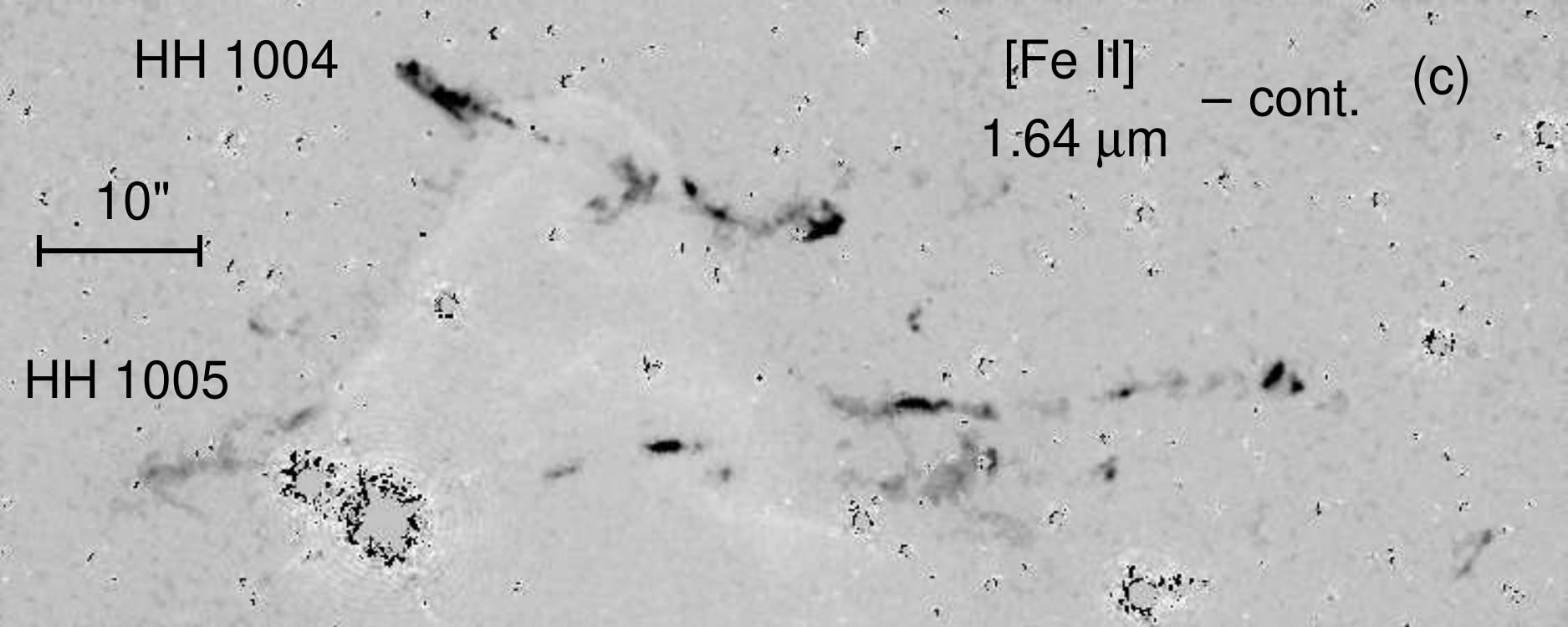} \\ 
\includegraphics[trim=25mm 0mm 0mm 0mm,angle=0,scale=0.375]{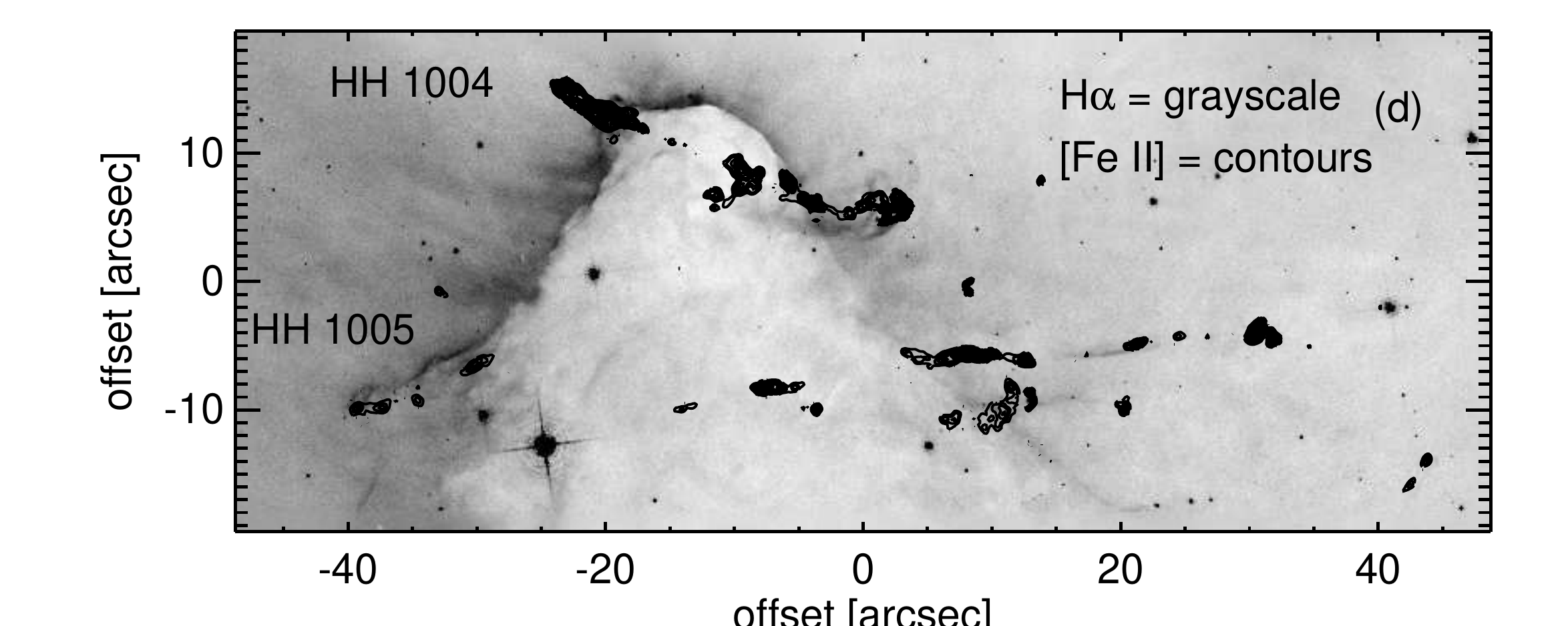} \\ 
\end{array}$ 
\caption{H$\alpha$ image (a) from \citet{smi10} shows two jets, HH~1004 and HH~1005, emerging from a dust pillar. 
New, continuum-subtracted [Fe~{\sc ii}] images (b,c) show the collimated, bipolar body of HH~1004. 
Several [Fe~{\sc ii}] knots trace out HH~1005 as it extends through the pillar (d) and into the H~{\sc ii} region on either side. Unlike HH~1004, the morphology of the [Fe~{\sc ii}] emission in HH~1005 does not clearly identify the driving protostar. 
}\label{fig:hh1004_feii} 
\end{figure}

Bright [Fe~{\sc ii}] traces the collimated jet to the northeast and points to the apex of the bow shock to the southwest (see Figure~\ref{fig:hh1004_feii}). 
A continuous stream of [Fe~{\sc ii}] emission from the northeast connects the jet to the \emph{Spitzer}-detected driving source \citep[PCYC~1198, see Table~\ref{t:jets_ysos} and][]{pov11}. 
Inside the pillar, to the southwest of the driving source, an arc of [Fe~{\sc ii}] emission may trace the shock created as the counterjet emerges from the pillar. 
Outside this point, [Fe~{\sc ii}] traces a narrow jet body that hooks upward, terminating in the center of the bow shock. 
[Fe~{\sc ii}] emission is brightest at the head of the bow shock while H$\alpha$ emission traces the larger structure including the extended bow shock wings. 
As in HH~666 and HH~903, H$\alpha$ appears to trace a cocoon enveloping the [Fe~{\sc ii}] jet. 

\textit{HH~1005:} 
Further south in the same pillar as HH~1004 lies HH~1005. 
\citet{smi10} identified a bright, collimated H$\alpha$ jet that points toward the southeast. 
No obvious bow shocks indicate more distant portions of the outflow. 
\citet{smi10} propose that, like HH~901 and HH~902, bright H$\alpha$ emission in HH~1005 traces an ionization front in a high density jet. 
Indeed, new WFC3-IR images reveal [Fe~{\sc ii}] emission behind the H$\alpha$ stream, tracing the neutral jet core that must be shielded by a large column of material. 
However, H$\alpha$ and [Fe~{\sc ii}] emission are clearly offset ($\sim 1$\arcsec) from each other, similar to the distinct H$\alpha$ and [Fe~{\sc ii}] features seen in HH~900 and HH~666 \citep[see][]{rei15a,rei15b}. 
The position angle of the jet axis inferred from [Fe~{\sc ii}] emission differs from that measured in H$\alpha$ images by \citet{smi10} by $\sim 17^{\circ}$.  
The H$\alpha$ emission resembles an arm extending off the pillar edge while [Fe~{\sc ii}] emission is more closely aligned with a chain of [Fe~{\sc ii}] knots that bisect the width of the pillar and extend past the western edge. 

Several discrete [Fe~{\sc ii}] knots not seen in H$\alpha$ trace the jet axis through the pillar. 
Most knots fall along a single, smooth stream. 
This is not the case where the jet emerges from the western edge of the pillar. 
Two streams separated by $\sim 5$\arcsec\ lie above and below the jet axis defined by the rest of the [Fe~{\sc ii}] knots. 
Where the jet emerges from the pillar, the opening angle between the two streams is $\sim 20^{\circ}$, however this narrows as the jet moves further into the H~{\sc ii} region. 
H$\alpha$ emission from HH~900 shows a similar narrowing of the outflow morphology with increasing distance from the source \citep[see Figure~\ref{fig:hh900_feii} and][]{rei15a}. 
To the west, outside the pillar, [Fe~{\sc ii}] knots coincide with two smooth wisps of H$\alpha$ separated by $\sim 3.5$\arcsec\ that \citet{smi10} identified as part of the HH~1004. 
However, these features lie along the HH~1005 jet axis defined by [Fe~{\sc ii}]. 
Unlike other jets with broad H$\alpha$, the [Fe~{\sc ii}] emission does not resemble a collimated jet body surrounded by an H$\alpha$ cocoon. 
Instead, [Fe~{\sc ii}] knots trace an arc between the two wisps of H$\alpha$ emission. 

Curiously, neither HH~1004 or HH~1005 point toward the H$\alpha$ bow shock that \citet{smi10} identified east of the pillar. 
A separate arc of [Fe~{\sc ii}] emission lies to the southwest of both jets, but not clearly along the axis of either. 
Proper motions are required to determine if either feature is physically associated with these jets.


\textit{HH~1006:} Like HH~900, HH~1006 was first identified in ground-based images as a candidate proplyd by \citet{smi03}. 
H$\alpha$ images from \emph{HST} clearly reveal a collimated bipolar jet emerging from a small cometary cloud \citep{smi10}. 
The jet axis of HH~1006 is aligned parallel to the major axis of the cloud, providing a counterexample to the many jets that are seen perpendicular to the long axis of their natal pillars \citep[see, e.g.][]{rag10}. 
\citet{sah12} estimate a globule mass of 0.35 M$_{\odot}$ from molecular line observations (with $8-19$\arcsec\ resolution, depending on the tracer) and identify the IR-bright driving source embedded in the globule.

\begin{figure}
\centering
$\begin{array}{cc}
\includegraphics[angle=0,scale=0.45]{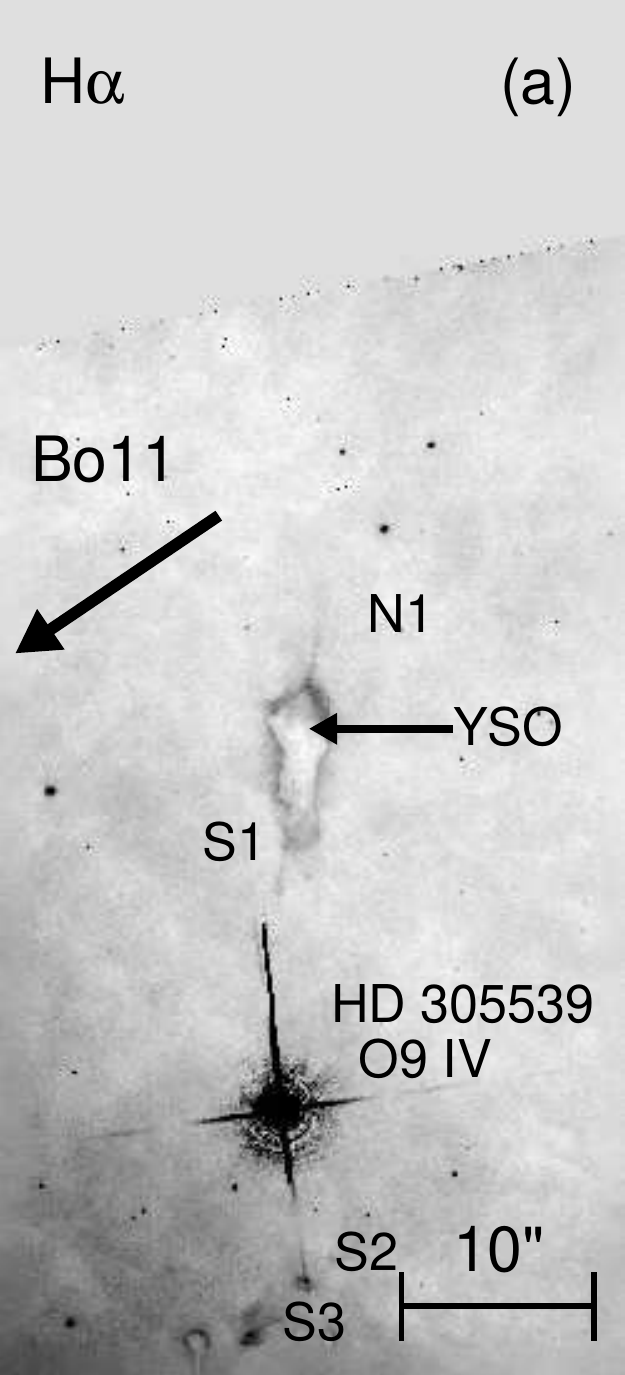} &
\includegraphics[trim=0mm 0mm 0mm 0mm,angle=0,scale=0.45]{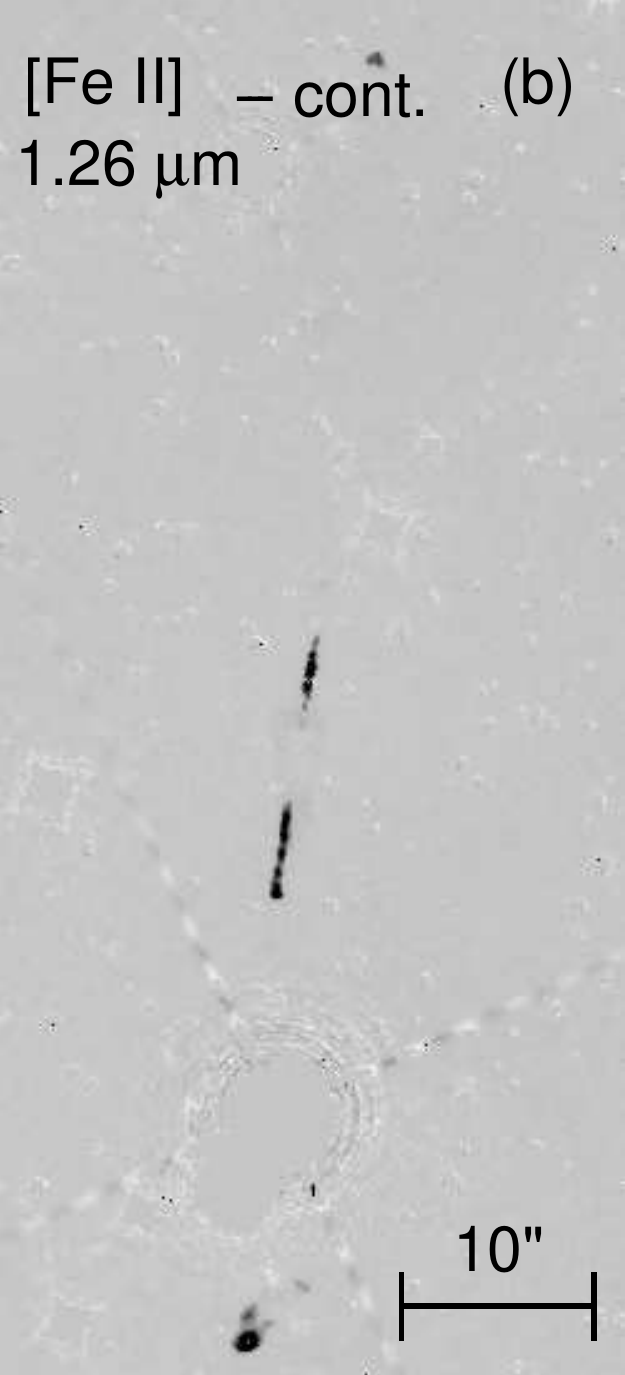} \\
\includegraphics[trim=0mm 0mm 0mm 0mm,angle=0,scale=0.45]{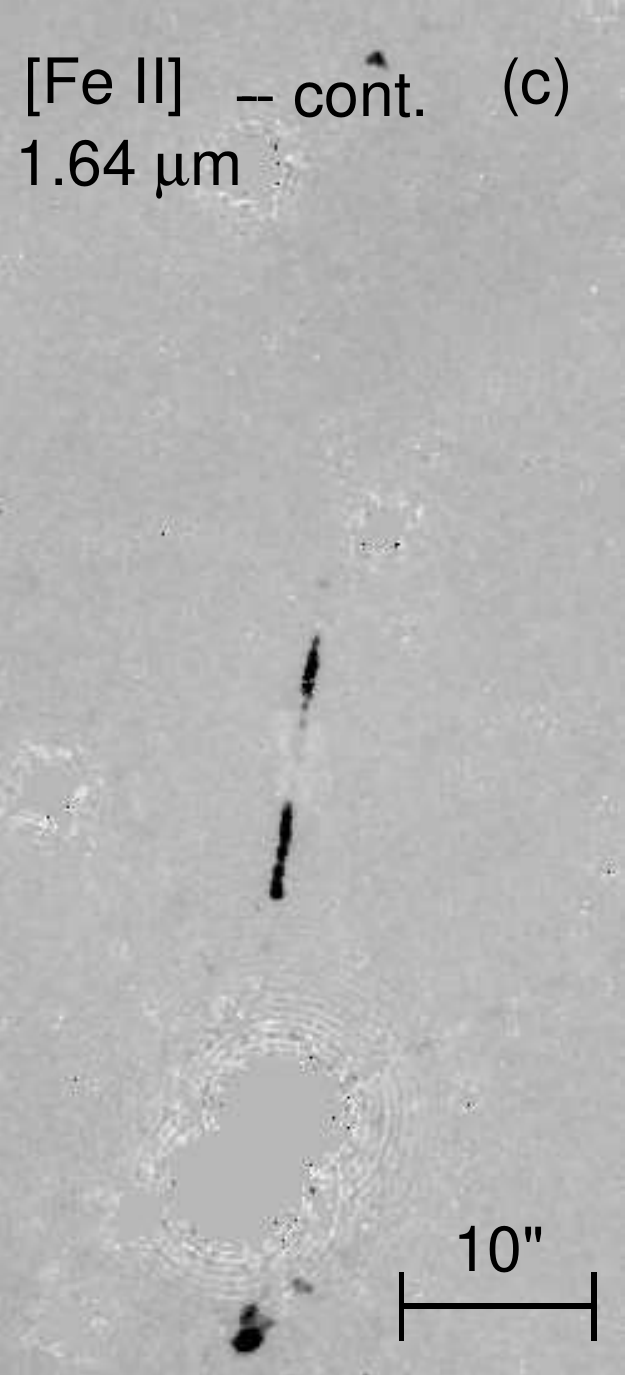} & 
\includegraphics[angle=0,scale=0.45]{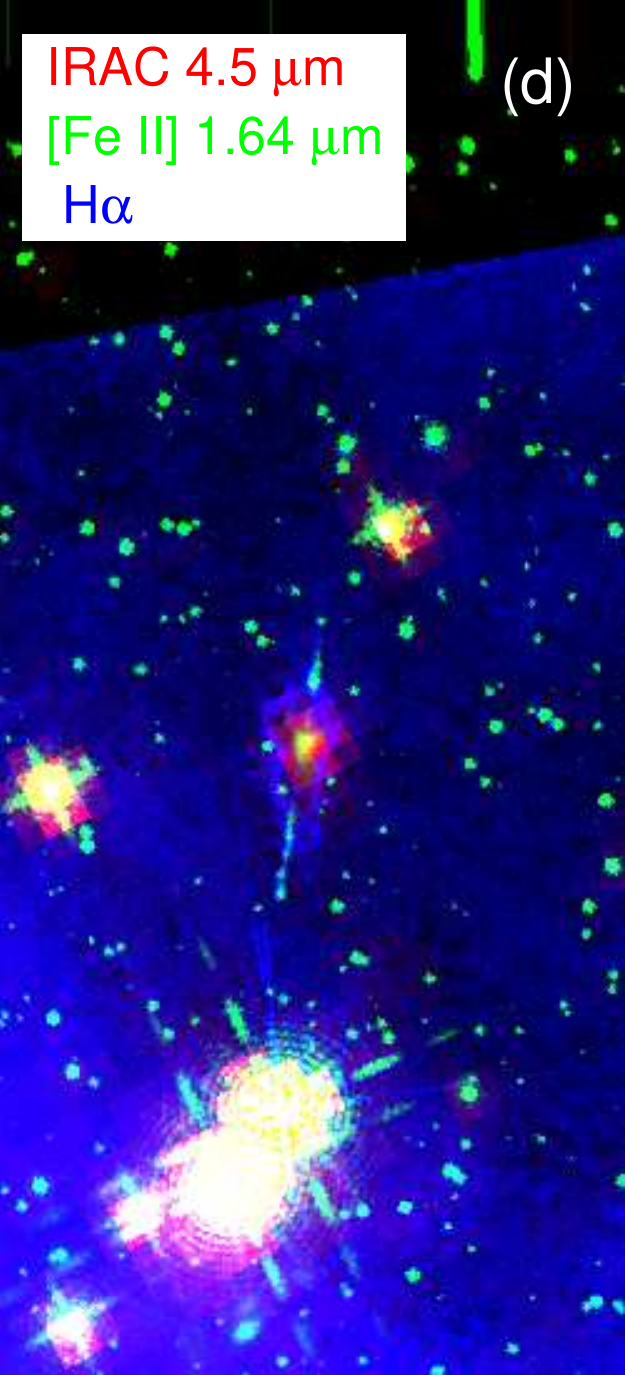} \\ 
\includegraphics[trim=5mm 0mm 0mm 0mm,angle=0,scale=0.45]{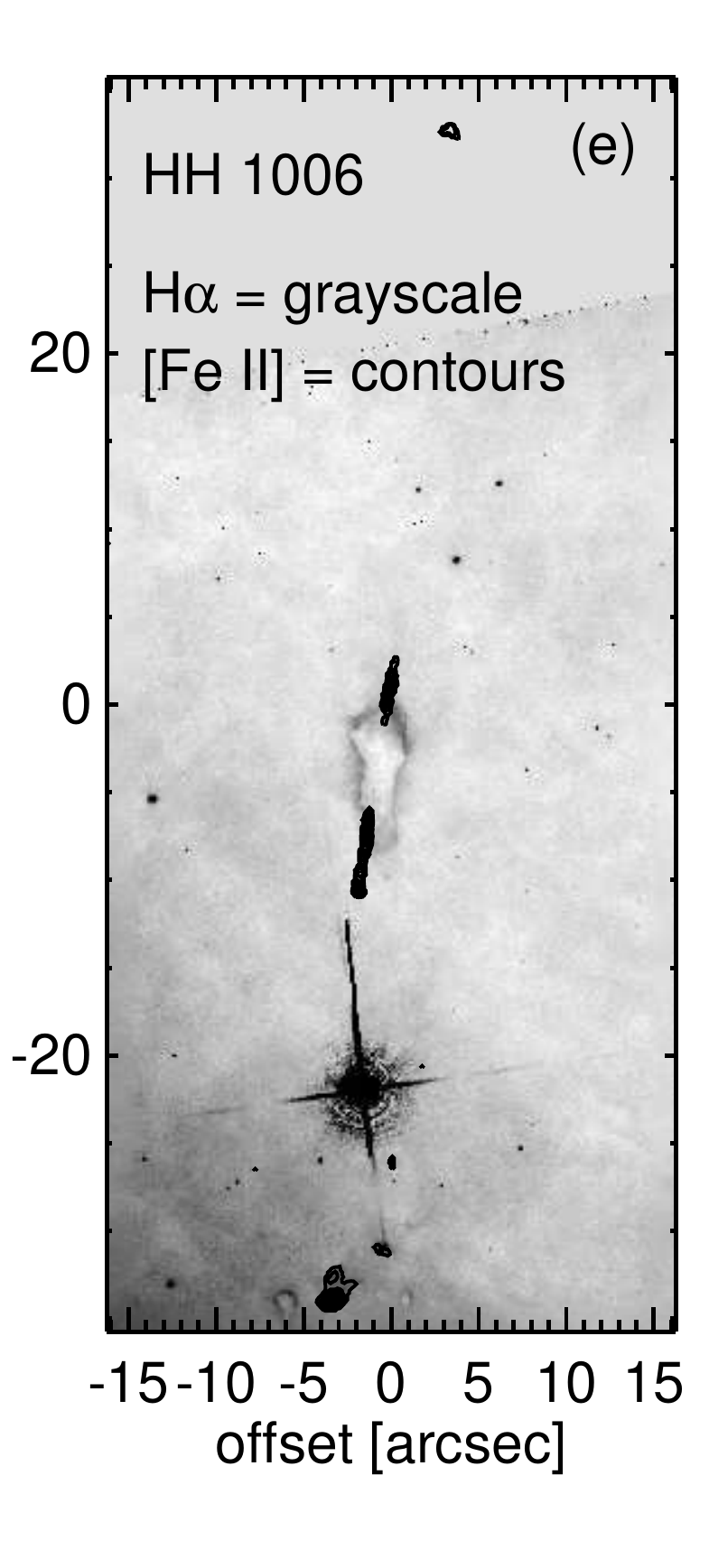} & 
\includegraphics[trim=5mm 0mm 0mm 0mm,angle=0,scale=0.45]{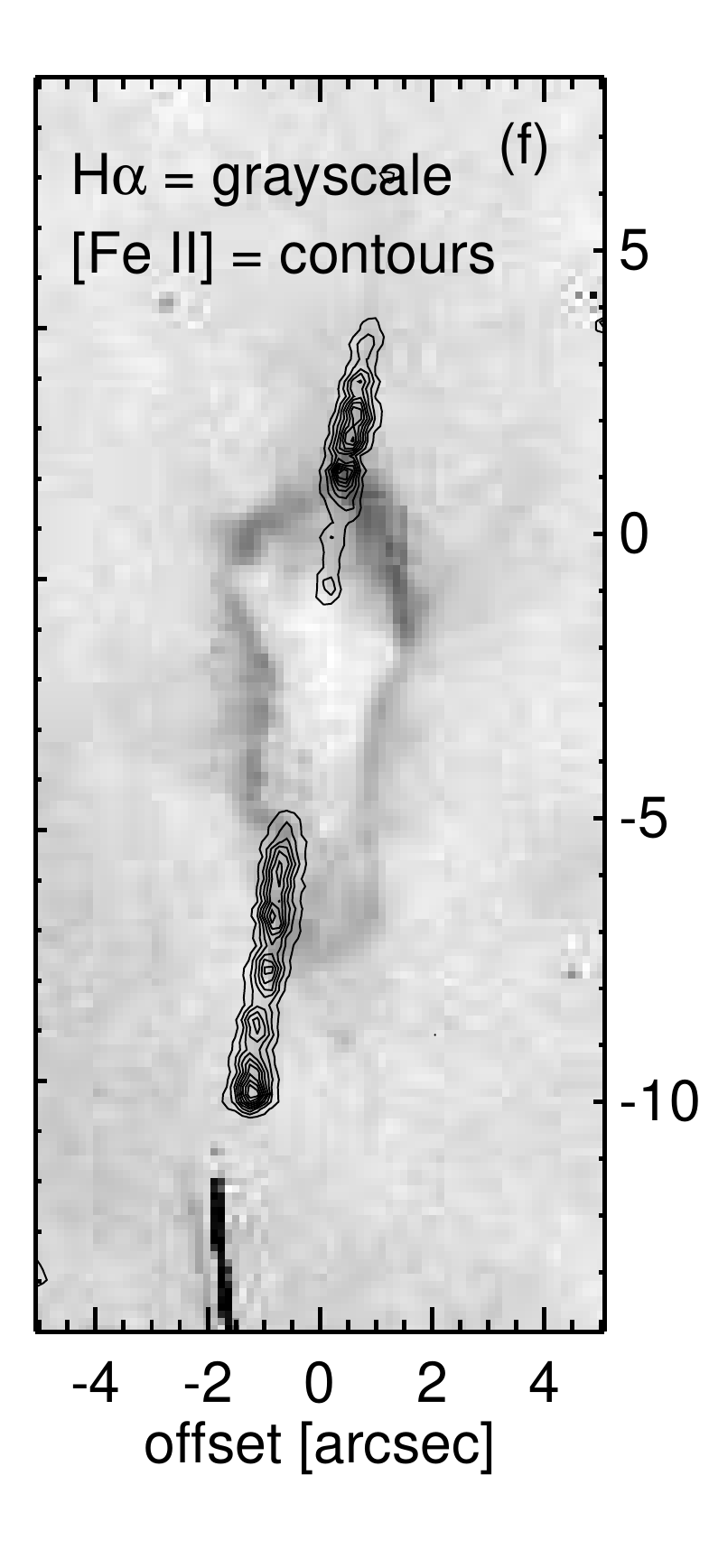} \\
\end{array}$ 
\caption{Like H$\alpha$ \citep[from][shown in panel a]{smi10}, [Fe~{\sc ii}] emission (b,c,d,e) from HH~1006 traces a highly collimated bipolar jet. HH~1006 is one of the few examples of a jet that is aligned nearly parallel to the major axis of its natal globule. Panel (f) shows a zoom of the highly collimated [Fe~{\sc ii}] jet emerging from the small globule. }\label{fig:hh1006_feii} 
\end{figure}

Figure~\ref{fig:hh1006_feii} shows that [Fe~{\sc ii}] and H$\alpha$ trace the same jet morphology outside the cloud. 
[Fe~{\sc ii}] emission from the northern limb of the jet extends inside the globule and connects the jet to IR-bright driving source \citep[PCYC~1172, see Table~\ref{t:jets_ysos} and][]{pov11}. 
This suggests that the northern limb of the jet is tilted slightly toward the observer (blueshifted), providing a clear view of the jet all the way to the driving protostar. 
In contrast, [Fe~{\sc ii}] emission from the southern lobe of the jet begins where there appears to be a notch taken out of the southeast side of the globule, $\sim 3$\arcsec\ south of the putative driving source.

New [Fe~{\sc ii}] images also reveal a bow shock to the north of HH~1006 (the northern bow shock lies beyond the edge of the H$\alpha$ image; see Figure~\ref{fig:hh1006_feii}), complementing the bow shock found to the south of the jet by \citet{smi10}. 
The northern and southern bow shocks are both offset from the driving source by $\sim 30$\arcsec, making the total outflow length $\sim 1$ pc, twice as long as the $\sim 0.45$ pc length estimated by \citet{smi10}. 

\begin{figure}
\centering
$\begin{array}{c}
\includegraphics[trim=0mm 0mm 0mm 0mm,angle=0,scale=0.45]{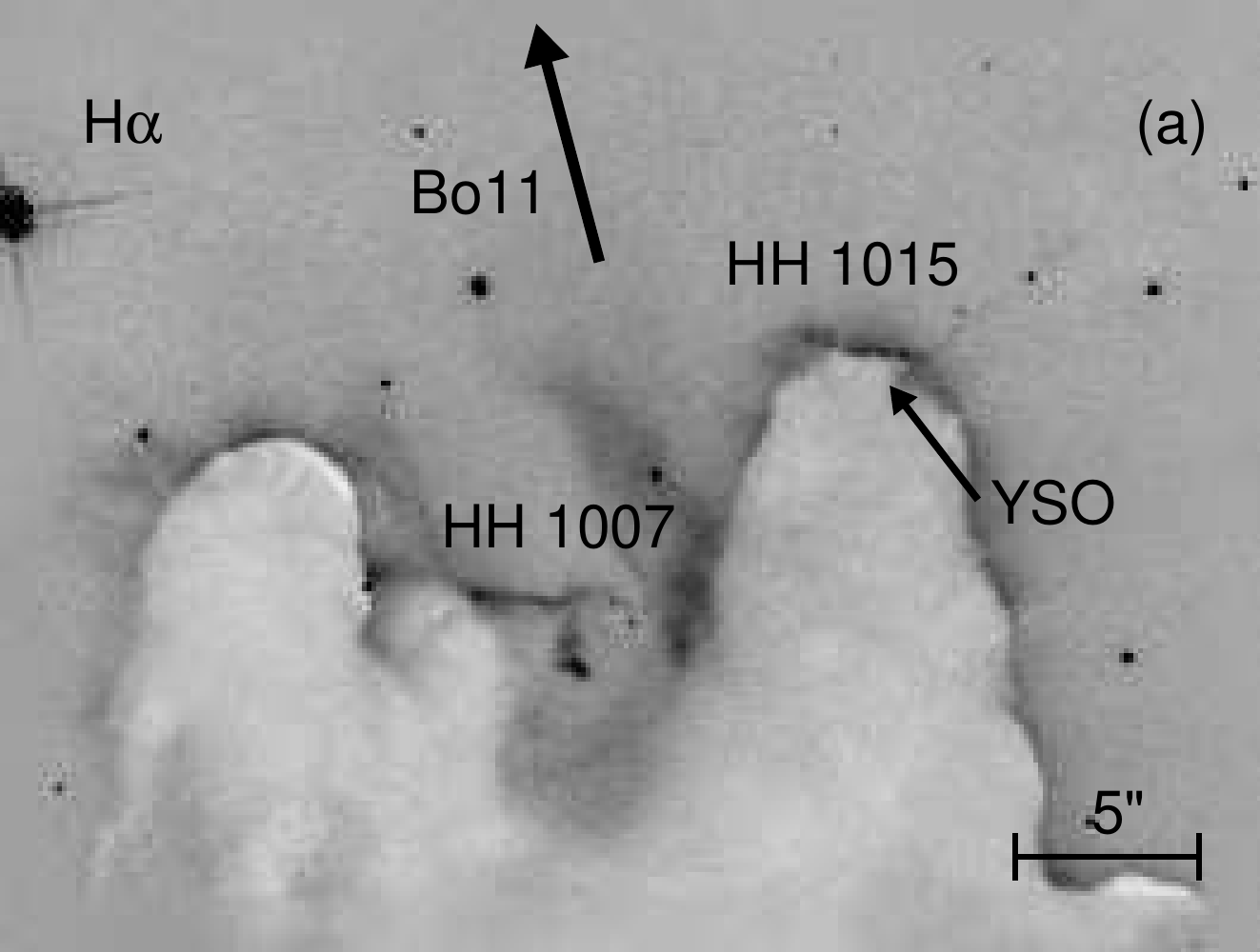} \\
\includegraphics[trim=0mm 0mm 0mm 0mm,angle=0,scale=0.45]{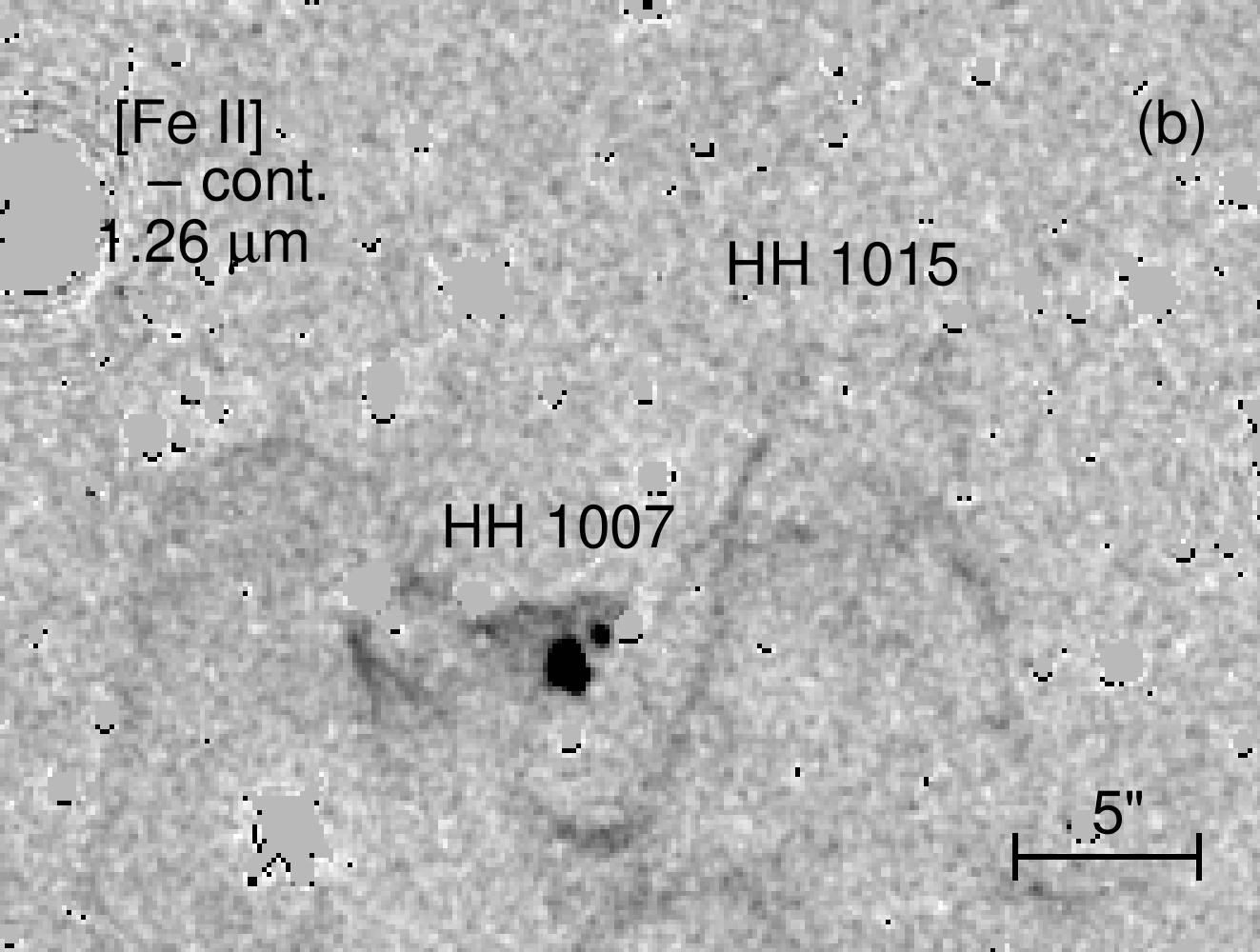} \\ 
\includegraphics[trim=0mm 0mm 0mm 0mm,angle=0,scale=0.45]{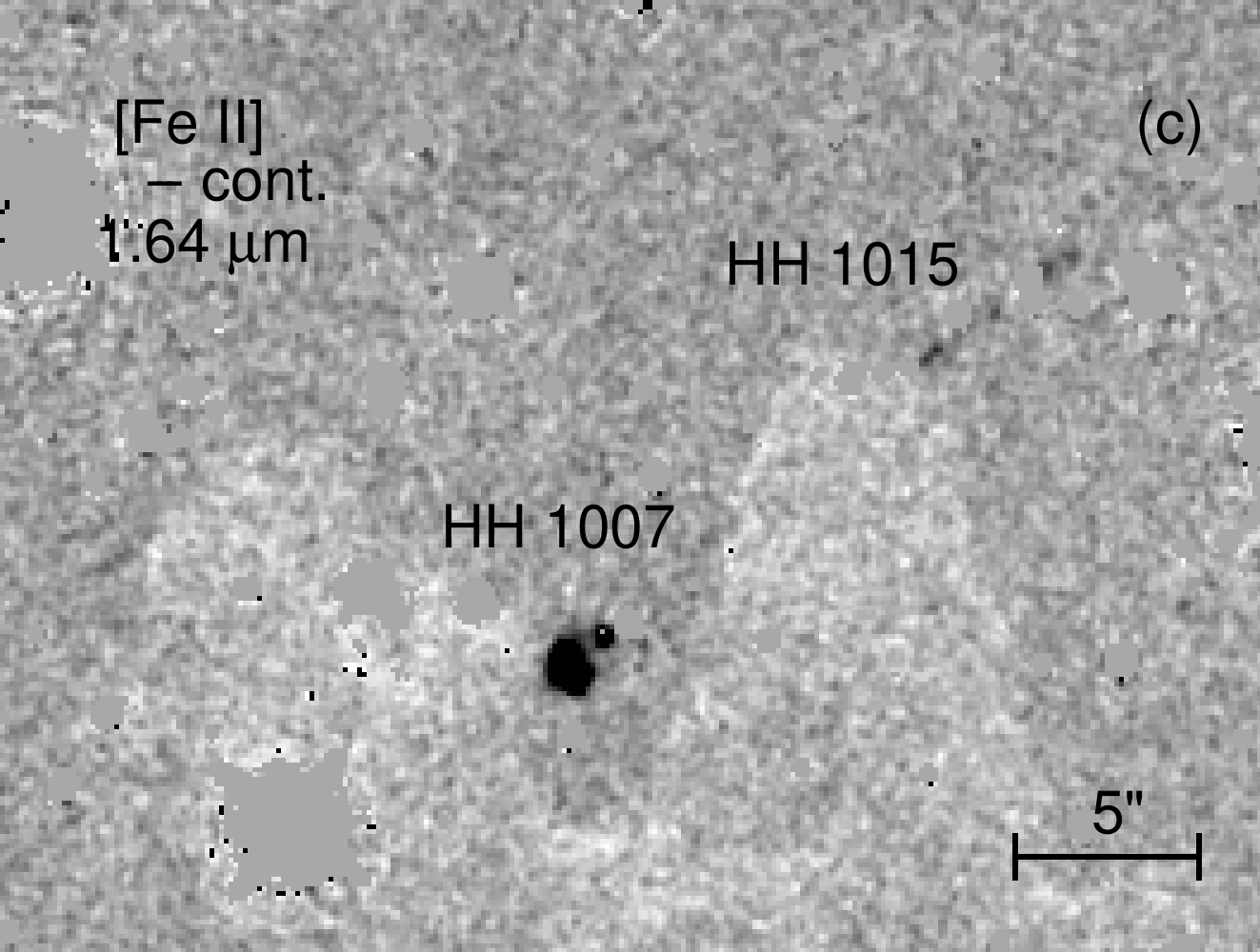} \\ 
\includegraphics[trim=20mm 0mm 0mm 0mm,angle=0,scale=0.45]{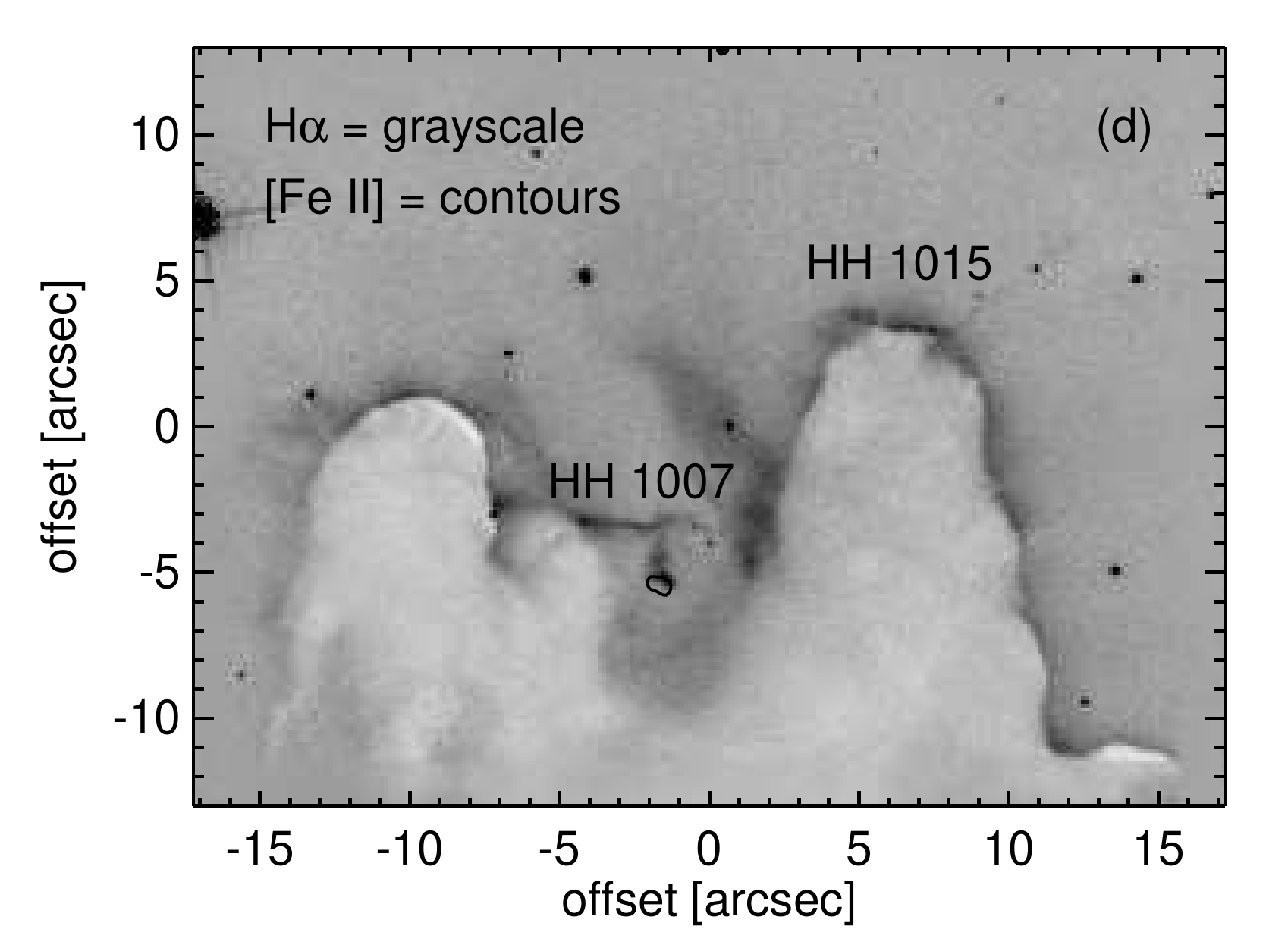} \\ 
\end{array}$ 
\caption{H$\alpha$ (a) and new [Fe~{\sc ii}] (b,c) images of the twin pillars that house HH~1007 (near the center of the image) and HH~1015 (top right). HH~1015 shows the weakest [Fe~{\sc ii}] of all the jets in this sample, unlike the bright knots of [Fe~{\sc ii}] emission seen in HH~1007 (d). 
}\label{fig:hh1007_feii} 
\end{figure}

\clearpage
\textit{HH~1007 and HH~1015:} \citet{smi10} identify HH~1007 as a dense jet body in the saddle between two neighboring dust pillars. 
HH~1007 emerges from the western edge of the eastern pillar and points toward the neighboring cloud. 
Only the brightest H$\alpha$ knot located between the dark dust clouds shows [Fe~{\sc ii}] emission (see Figure~\ref{fig:hh1007_feii}). 
[Fe~{\sc ii}] emission breaks up into a bright, central knot with a smaller knot extending to the northwest. 
These knots lie immediately next to an unobscured star located between the pillars. 
However, this star is not in the PCYC and likely not related to HH~1007. 
There is no evidence for collimated [Fe~{\sc ii}] emission from any other part of HH~1007.

HH~1015 resides in the same cloud complex, emerging from the tip of the western of the two pillars. 
Like HH~1006, HH~1015 does not emerge perpendicular to the major axis of the pillar. 
Instead, the jet axis makes a 47$^{\circ}$ angle with the major axis of the pillar. 
Faint H$\alpha$ emission reveals a collimated jet body that extends $\sim 6$\arcsec\ into the H~{\sc ii} region. 

Like H$\alpha$, [Fe~{\sc ii}] is faint, but clearly traces the collimated jet. 
An IR-bright YSO lies just inside the pillar edge at the base of HH~1015, and is the likely driving source \citep[PCYC~538,][]{pov11}. 
Extrapolating the jet axis defined by H$\alpha$ and [Fe~{\sc ii}] to estimate the position of the counterjet, we find that the southeast limb HH~1015 may intersect HH~1007. 
In fact, the knots in HH~1007 all lie along the HH~1015 jet axis, but are offset from the H$\alpha$ emission that \citet{smi10} identified as HH~1007. 

\textit{HH~1010:} H$\alpha$ emission traces a bipolar jet emerging perpendicular to the major axis of a pillar head \citep{smi10}. 
The two limbs appear symmetric, tracing a jet that extends $\sim 15$\arcsec\ beyond the pillar edge. 
\citet{smi10} identify a putative bow shock, HH~1010~A, to the southwest, suggesting a total jet length $\gtrsim 60$\arcsec. 

\begin{figure}
\centering
$\begin{array}{cc}
\includegraphics[trim=-5mm 2mm 4mm 0mm,angle=0,scale=0.45]{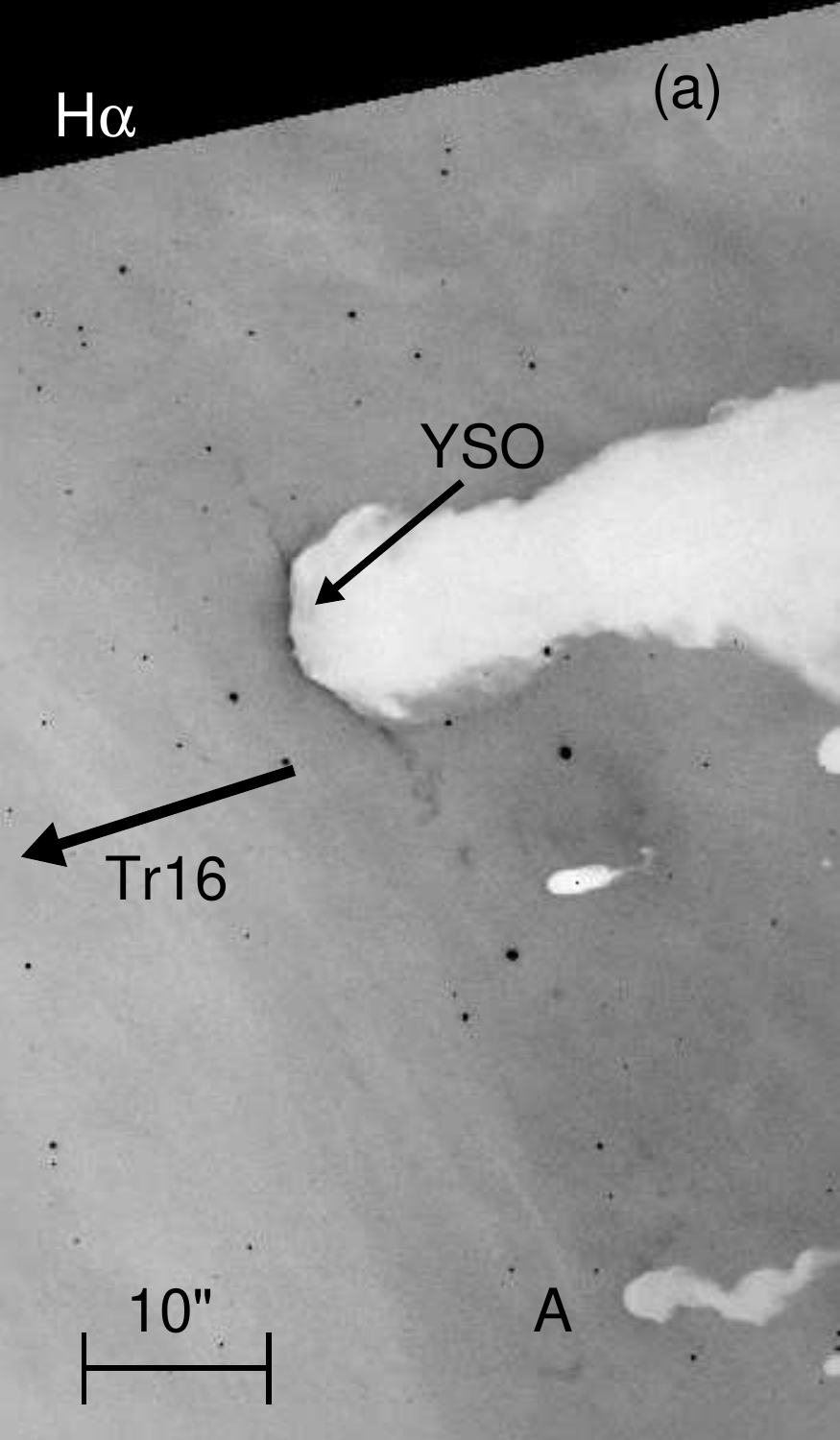} &
\includegraphics[trim=4mm 2mm 0mm 0mm,angle=0,scale=0.45]{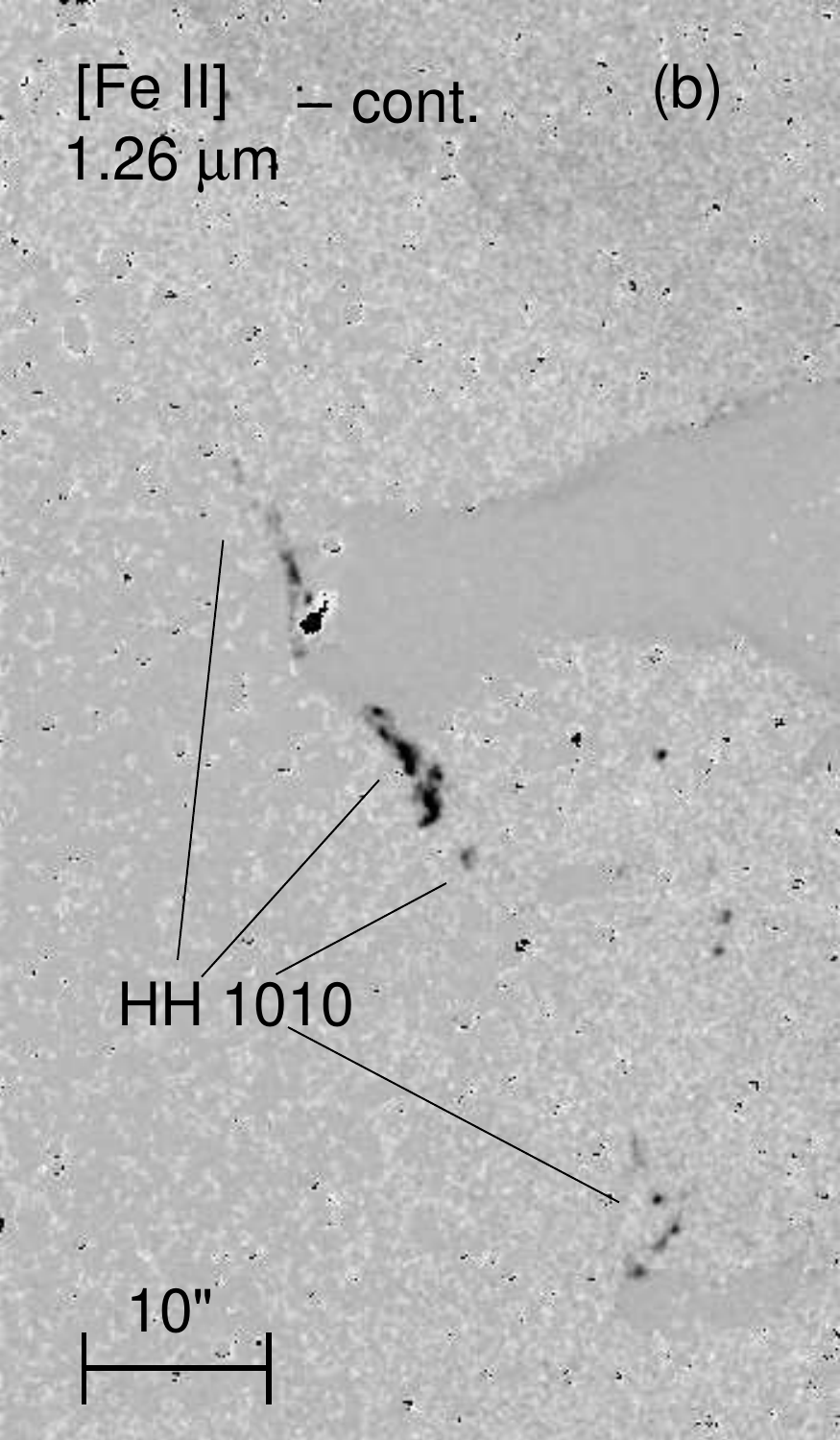} \\ 
\includegraphics[trim=5mm 15mm 4mm 2mm,angle=0,scale=0.45]{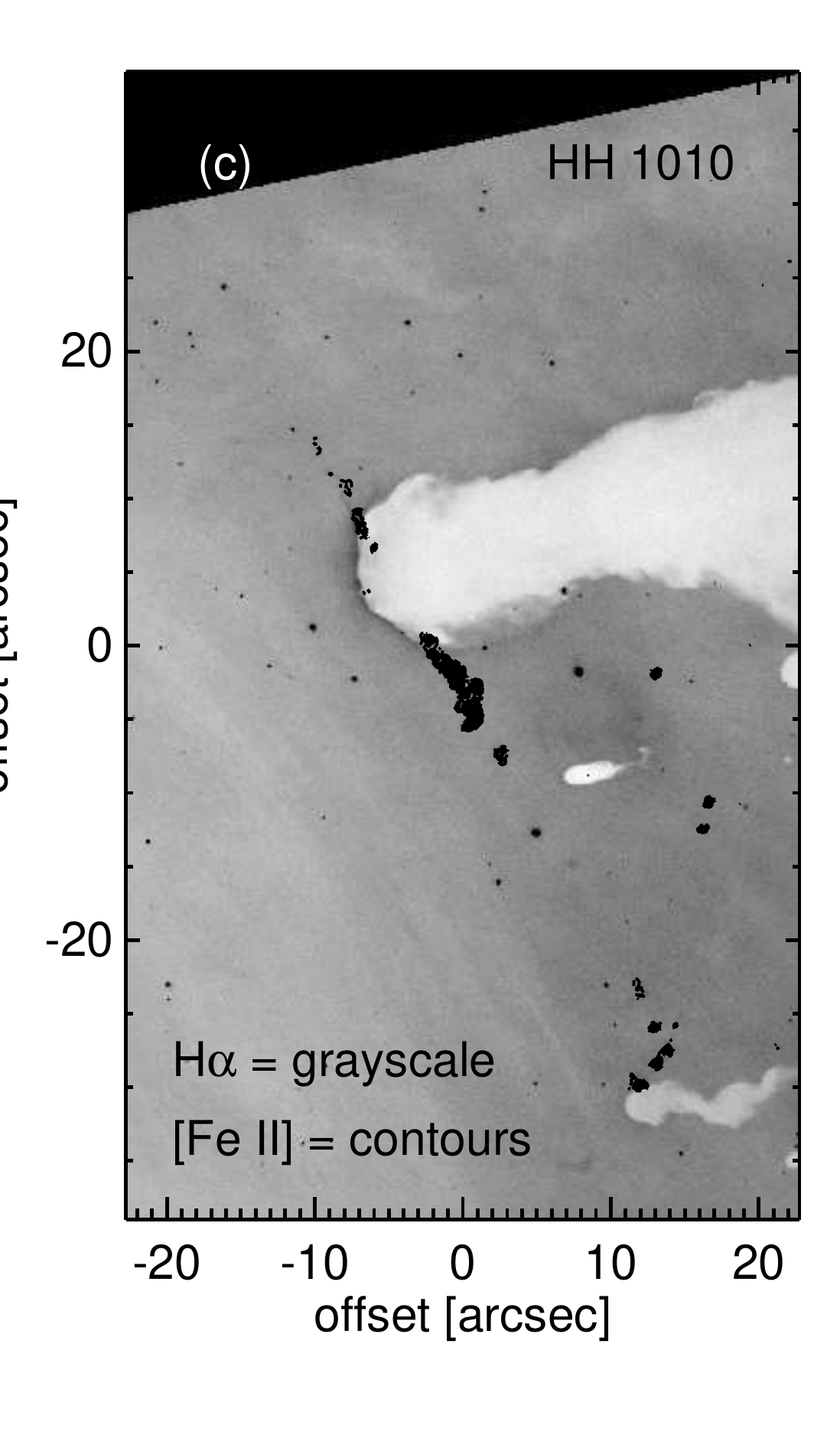} & 
\includegraphics[trim=4mm -14mm 0mm 2mm,angle=0,scale=0.45]{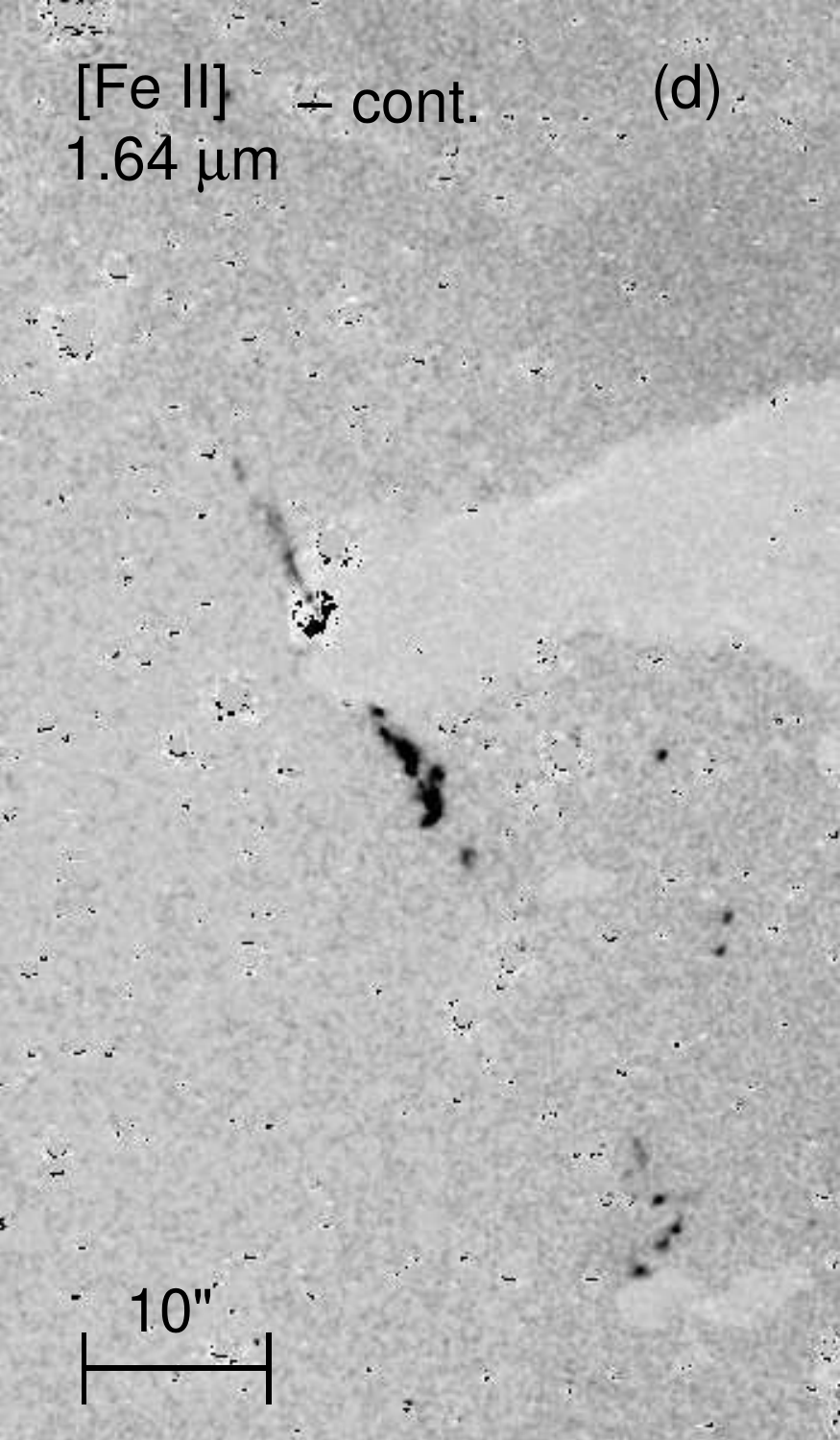} \\ 
\end{array}$ 
\caption{HH~1010 is a bipolar jet that emerges from the head of a dust pillar, as seen in H$\alpha$ (a). [Fe~{\sc ii}] (b,d) emission from the jet extends inside the pillar (c), connecting the larger-scale jet to the \emph{Spitzer}-identified driving source. }\label{fig:hh1010_feii} 
\end{figure}

Our new WFC3-IR images show that bright [Fe~{\sc ii}] emission in HH~1010 delineates the same collimated jet morphology seen in H$\alpha$ images (see Figure~\ref{fig:hh1010_feii}). 
As with HH~1006, [Fe~{\sc ii}] emission extends inside the dust pillar, connecting the extended H$\alpha$ outflow with the embedded IR driving source \citep[PCYC~55, see Table~\ref{t:jets_ysos} and][]{pov11}. 
The driving source does not lie halfway between the two H$\alpha$ jets limbs, but closer to the north, roughly $1/3$ of the way into the pillar. 
The YSO and the beginning of the northern jet limb coincide with the position of the faint optical source noted within the boundaries of the pillar by \citet{smi10}. 
[Fe~{\sc ii}] emission from the southern limb of HH~1010 is offset from the driving source, beginning $\sim 6$\arcsec\ away. 
This geometry suggests that the southern limb of HH~1010 is redshifted and the northern limb blueshifted. 

A few additional [Fe~{\sc ii}] knots lie to the south of the pillar. 
Several knots trace a backward ``L'' shape and can also be seen in H$\alpha$ images. 
These knots lie $\sim 9$\arcsec\ to the northwest of HH~1010~A, 
but closer to the jet axis defined by the [Fe~{\sc ii}]. 
The stem of the backward ``L'' lies along the jet axis, while the foot may trace the culminating bow shock. 
We do not detect [Fe~{\sc ii}] emission from HH~1010~A. 

\textit{HH~1014:} HH~1014 emerges from the head of a dust pillar in the dark molecular ridge that borders the bright inner part of the Carina Nebula. 
Like HH~1006, the axis of the jet is aligned to within a few degrees of the pillar semi-major axis. 
No counterjet can be seen in H$\alpha$ images. 
Both HH~1014 and HH~900 lie close to $\eta$ Carinae. 
However the two jets have very different morphologies. 
While H$\alpha$ emission from HH~900 traces the wide-angle stream of an entrained outflow, H$\alpha$ emission from HH~1014 clearly traces a narrow, collimated jet.  

\begin{figure*}
\centering
$\begin{array}{cc}
\includegraphics[trim=10mm 0mm 0mm 0mm,angle=0,scale=0.325]{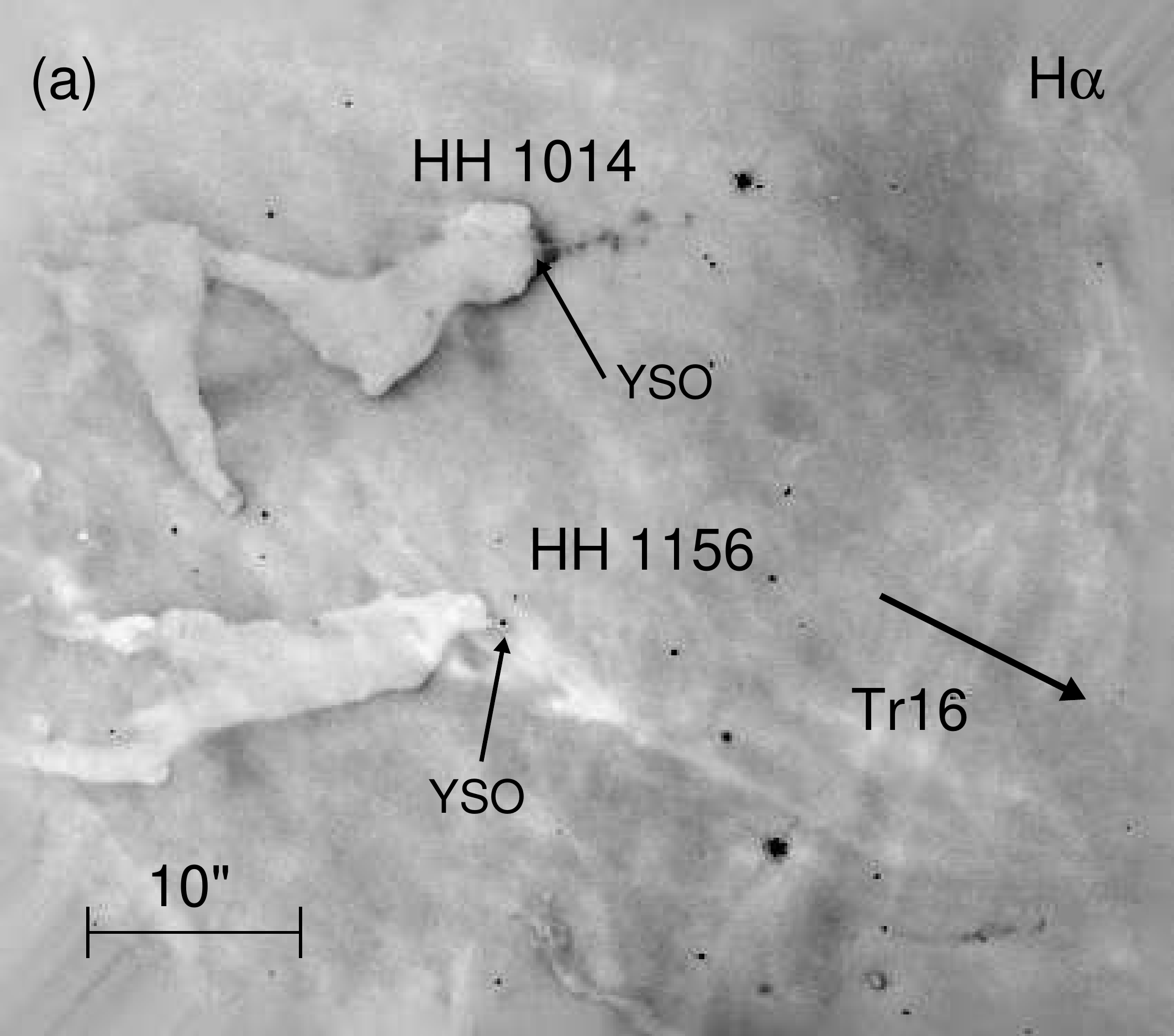} & 
\includegraphics[angle=0,scale=0.325]{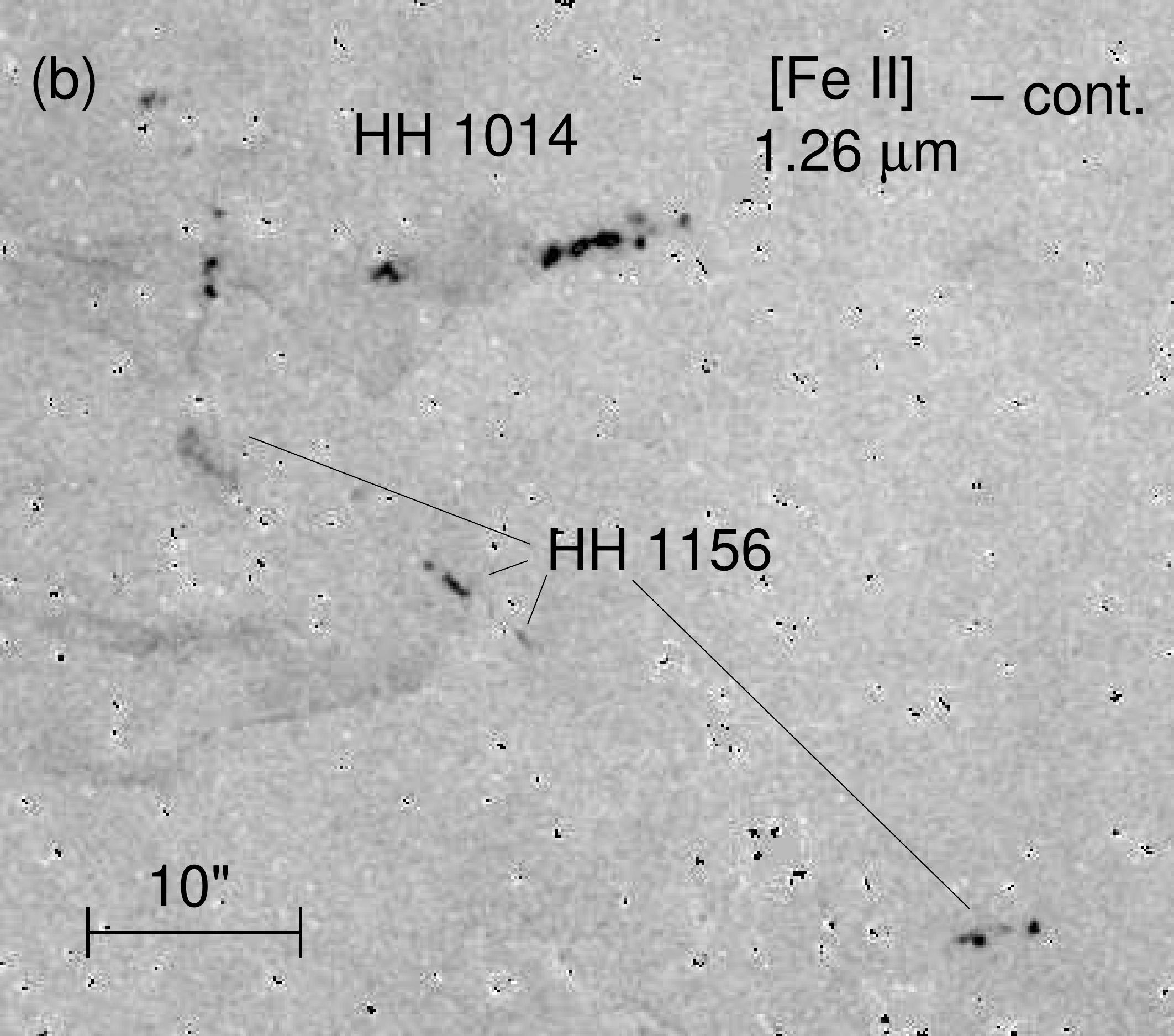} \\ 
\includegraphics[trim=35mm 20mm 0mm 0mm,angle=0,scale=0.305]{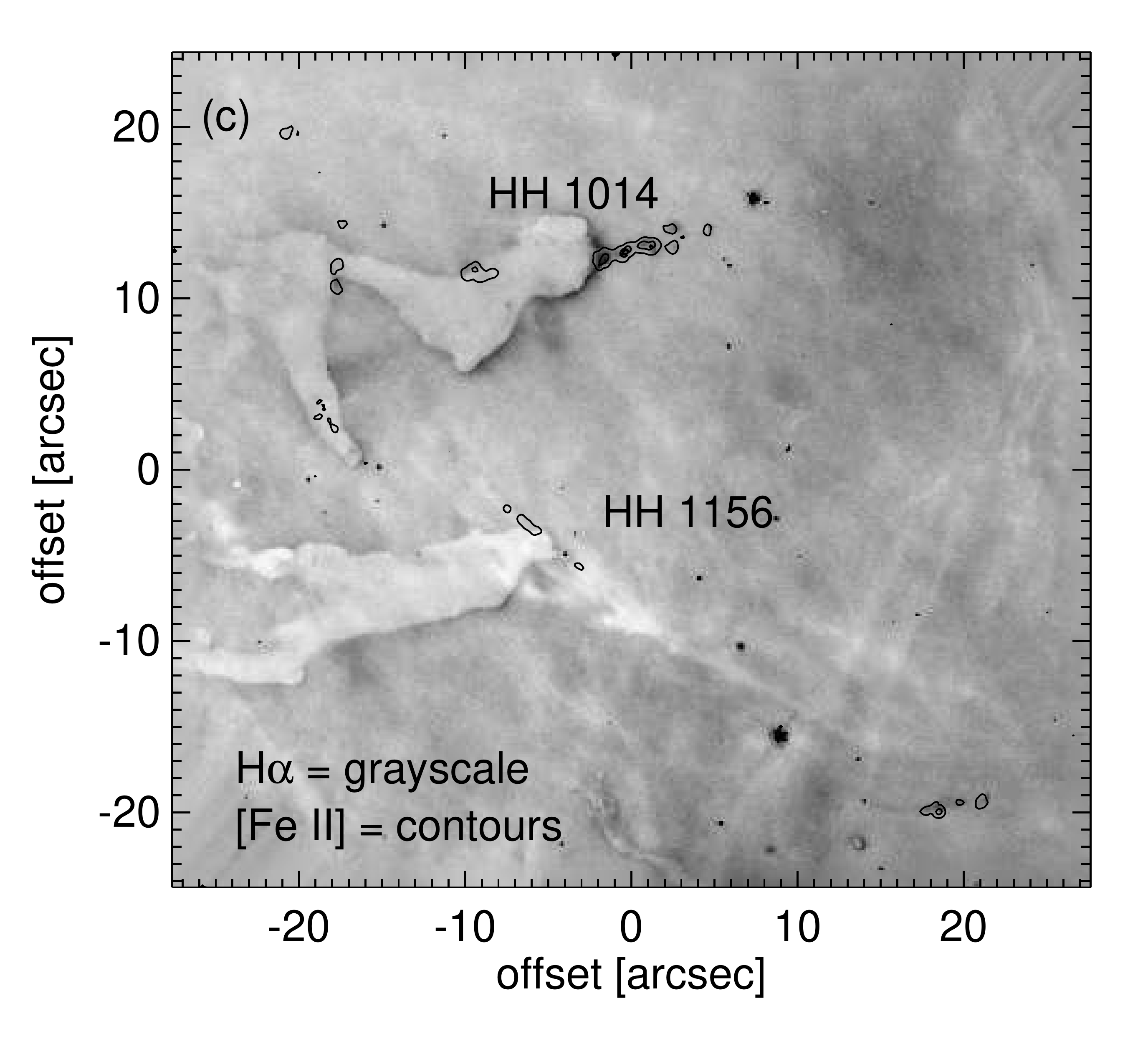} & 
\includegraphics[trim=0mm -20mm 0mm 0mm,angle=0,scale=0.325]{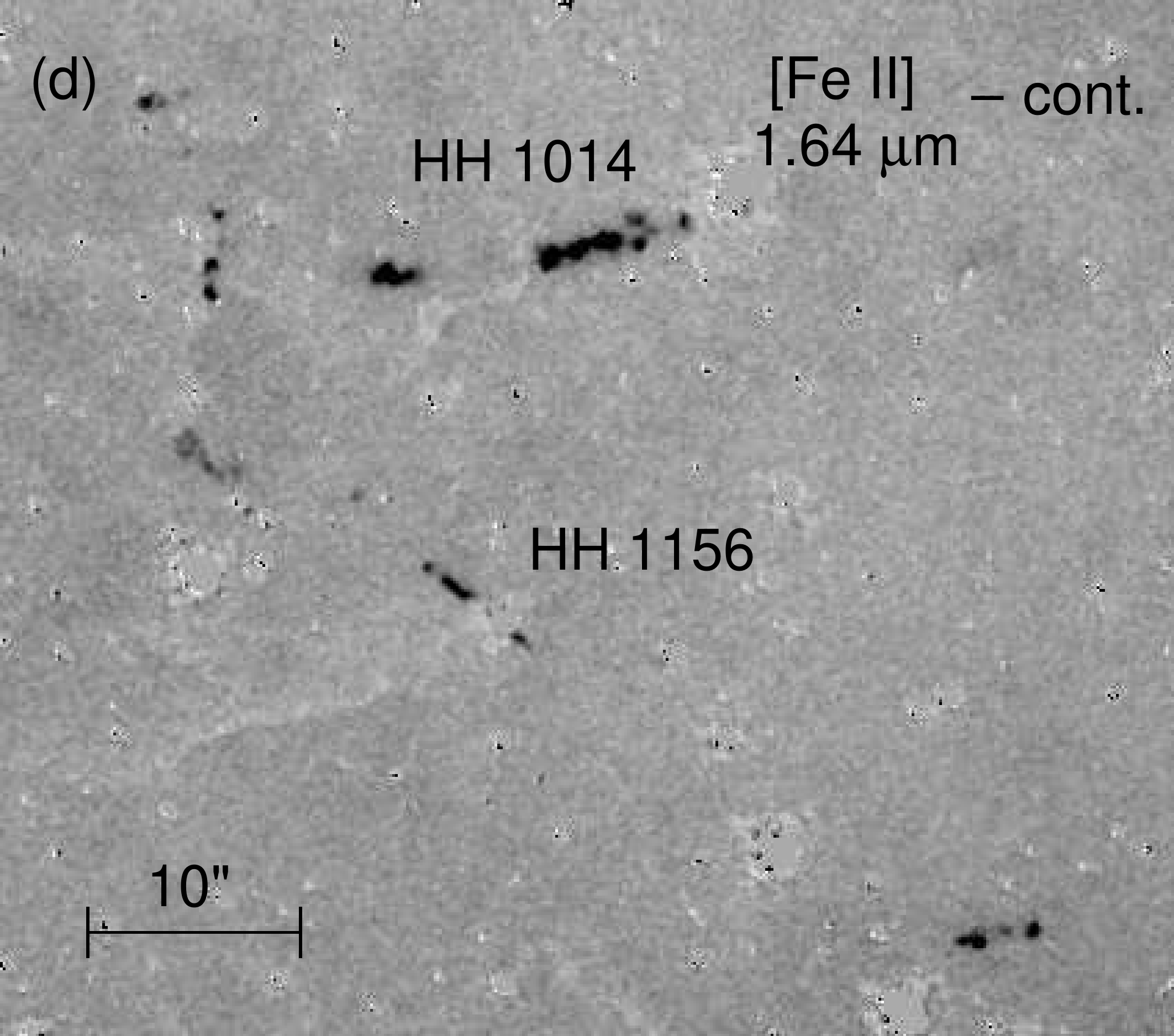} \\ 
\end{array}$ 
\caption{An H$\alpha$ image (a) from \citet{smi10} shows HH~1014 emerging from the northern of two neighboring dust pillars. 
New, continuum-subtracted [Fe~{\sc ii}] images (b,d) reveal the HH~1014 counterjet. 
More surprising is the collimated, bipolar outflow that emerges from the head of the southern pillar, HH~1156, that appears to have no H$\alpha$ counterpart (c). }\label{fig:hh1014_feii} 
\end{figure*}

New [Fe~{\sc ii}] images shown in Figure~\ref{fig:hh1014_feii} show bright emission from the collimated jet and confirm that it is driven by a protostar at the tip of the pillar \citep[PCYC~984,][]{pov11}. 
Inside the pillar, another [Fe~{\sc ii}] knot lies along the same axis, $\sim 5$\arcsec\ to the east of the driving source, likely tracing the counterjet. 
Two additional [Fe~{\sc ii}] knots located further east may also be part of HH~1014. 
However, these two knots are offset slightly north of the jet axis. 
Other jets in Carina have been observed to bend toward (e.g. HH~900) or away (HH~901) from nearby O-type stars \citep[see also][]{smi10,rei13,rei15a}. 

\textit{HH~1156:} \citet{smi10} noted an H$\alpha$ feature with shock-like morphology that lies to the southwest of the pillar housing HH~1014. 
The shock does not lie along the HH~1014 jet axis and is instead identified as candidate jet HH~c-14. 
\citet{smi10} suggest that it may be driven by an IR source embedded in another pillar in the region. 

New WFC3-IR images reveal [Fe~{\sc ii}] emission from HH~c-14. 
More striking is the bright well-collimated [Fe~{\sc ii}] jet that emerges from the dust pillar $\sim 27$\arcsec\ to the northeast of the HH~c-14 bow shock. 
This bipolar jet lies at the tip of a pillar immediately south of HH~1014 and points toward the putative bow shock.  
H$\alpha$ images offer no hint of a jet at the head of this pillar. 
Nevertheless, the jet appears to emerge from an unobscured and IR-bright protostar \citep[PCYC~986,][]{pov11}. 
Additional [Fe~{\sc ii}] knots lie along the jet axis to the northeast and may be the complementary bow shock. 
Both putative bow shocks lie slightly offset from the axis defined by the inner jet, but altogether may trace an S-shaped jet. 
Given the well-defined jet morphology seen in new [Fe~{\sc ii}] images, we assign this jet an HH number of 1156.

\begin{figure}
\centering
$\begin{array}{cc}
\includegraphics[trim=-15mm 0mm 0mm 0mm,angle=0,scale=0.35]{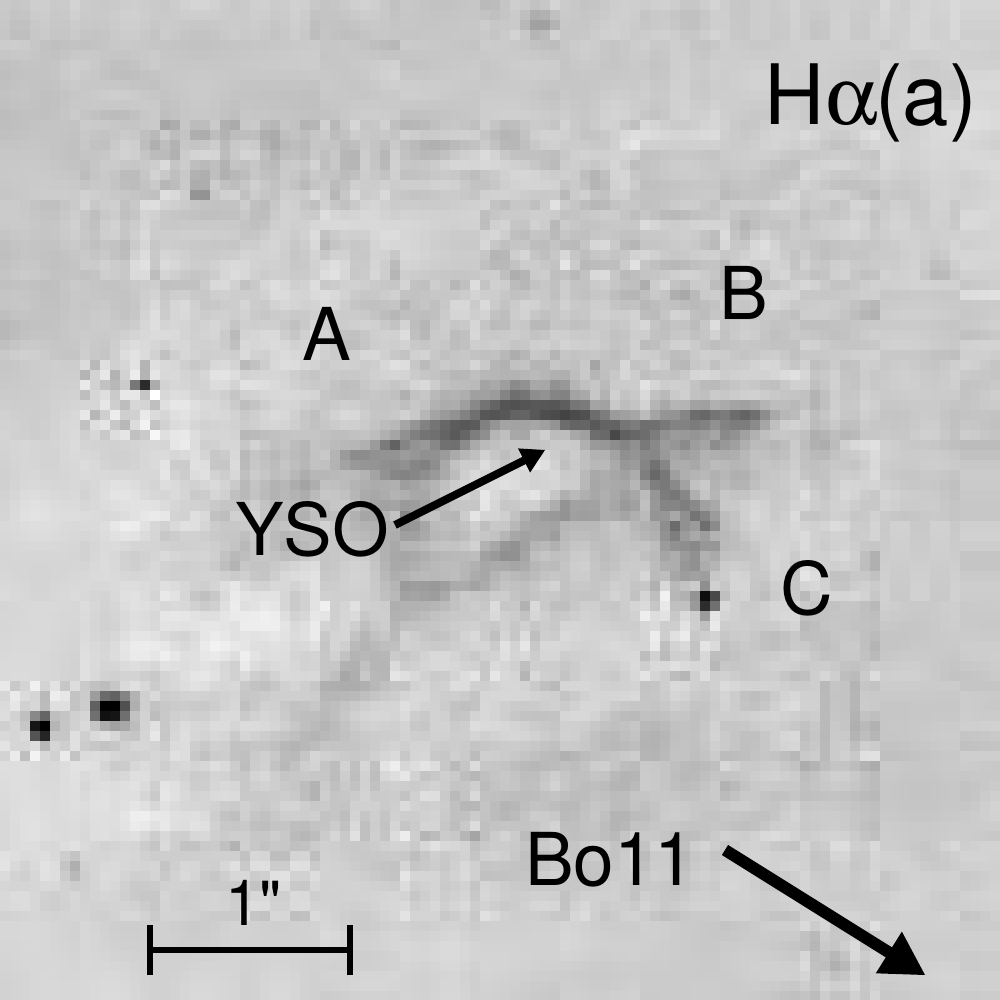} &  
\includegraphics[angle=0,scale=0.35]{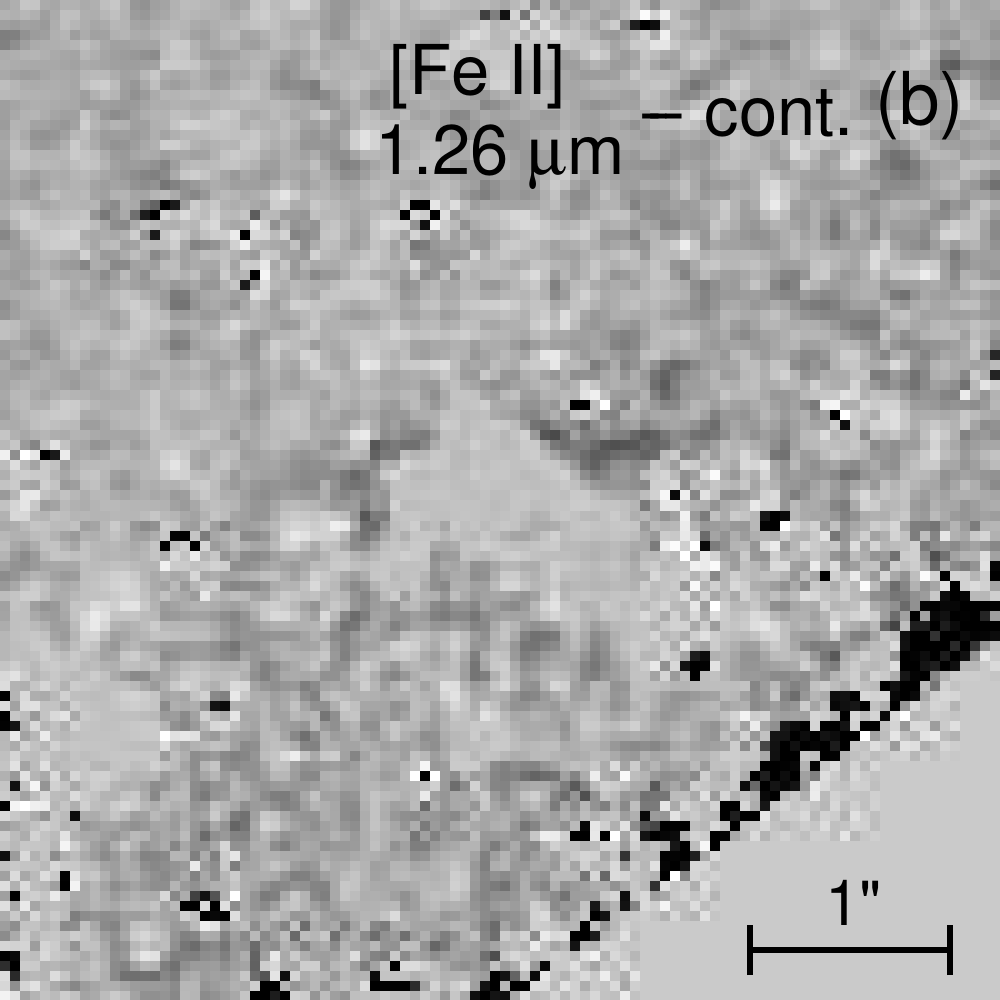} \\  
\includegraphics[trim=0mm 10mm 0mm 0mm,angle=0,scale=0.35]{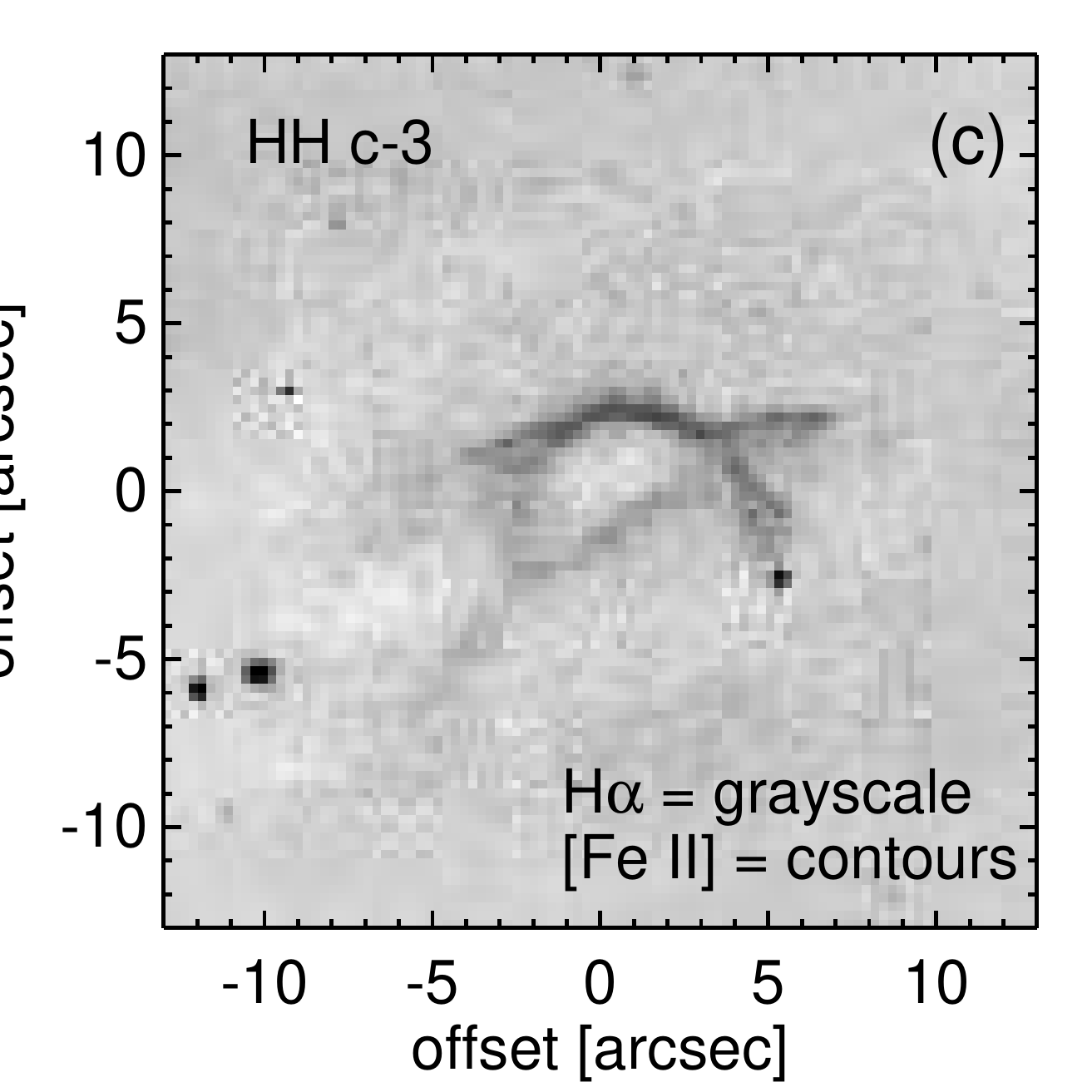} & 
\includegraphics[trim=0mm -15mm 0mm 0mm,angle=0,scale=0.35]{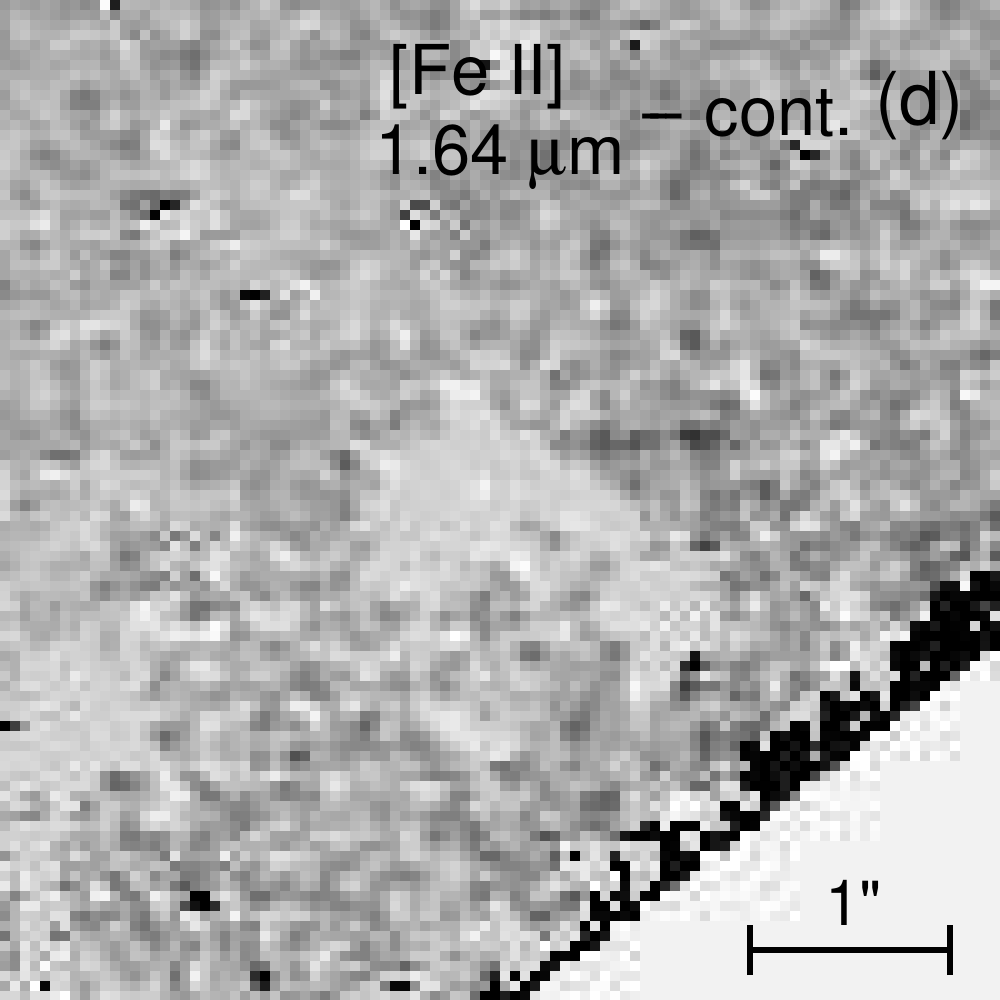} \\ 
\end{array}$ 
\caption{HH~c-3 is a candidate jet, with extended H$\alpha$ emission (a) possibly tracing a bipolar outflow. New, continuum-subtracted [Fe~{\sc ii}] images (b,c,d) trace collimated emission in feature B, although the poor signal-to-noise near the edge of the frame prohibits confident identification of the nature of this source. }\label{fig:hhc3_feii} 
\end{figure}

\textit{HH~c-3:} Multiple protrusions extend off the head of a cometary cloud tracing candidate jet (or jets) HH~c-3. 
No bow shocks are found nearby to indicate the outflowing nature of the gas. 
Extremely faint [Fe~{\sc ii}] emission from HH~c-3~B traces the same morphology seen in H$\alpha$ (see Figure~\ref{fig:hhc3_feii}). 
An IR-bright point source is detected near the apex of the cloud, consistent with the position of the base of the putative jet. 
With no velocity information and the poor signal-to-noise achieved near the edge of the frame where HH~c-3 is located, it remains unclear whether HH~c-3 is, in fact, a jet. 

\begin{figure*}
\centering
$\begin{array}{cc}
\includegraphics[trim=10mm 5mm 0mm 5mm,angle=0,scale=0.525]{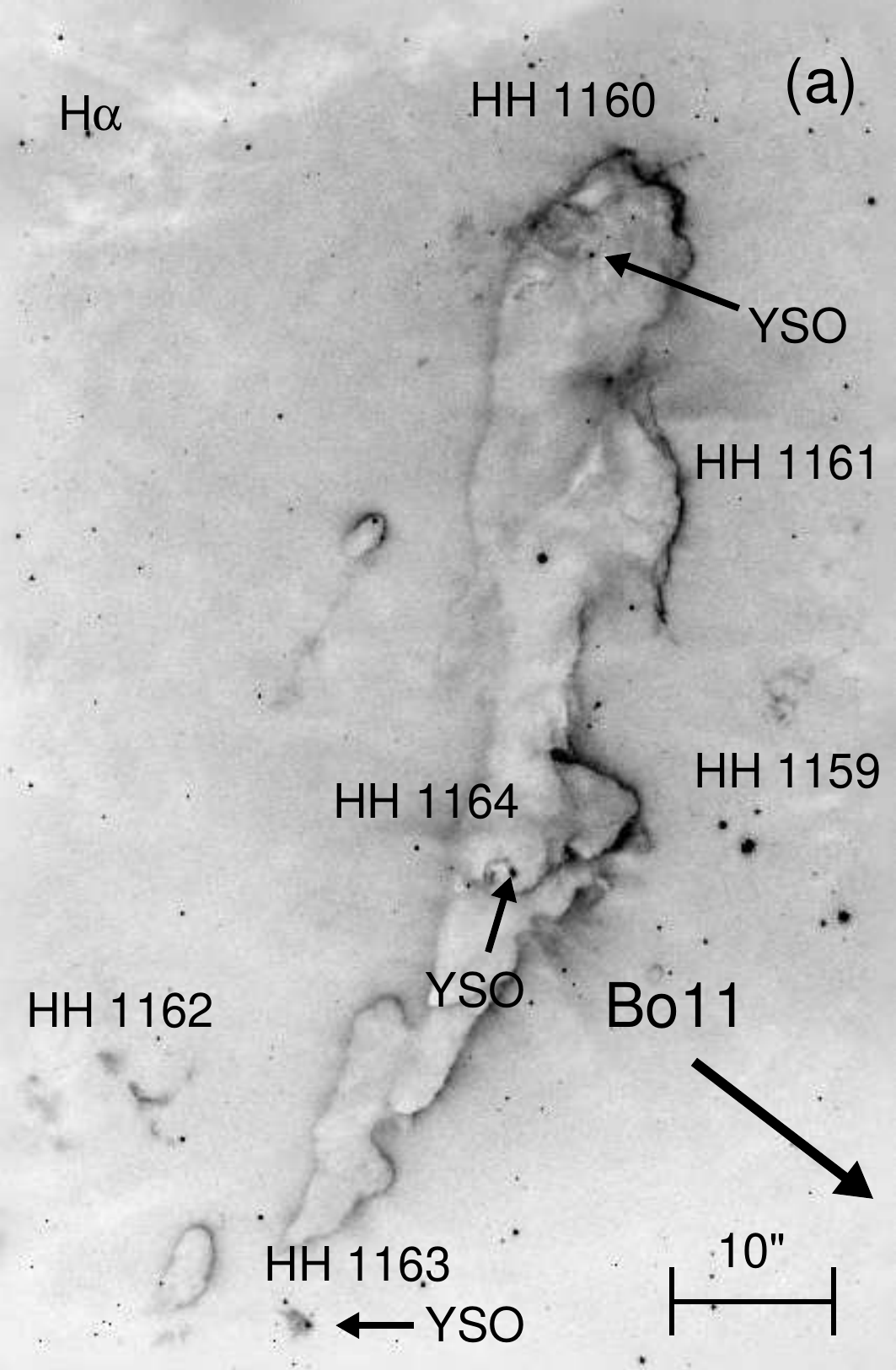} &  
\includegraphics[trim=0mm 5mm 0mm 5mm,angle=0,scale=0.525]{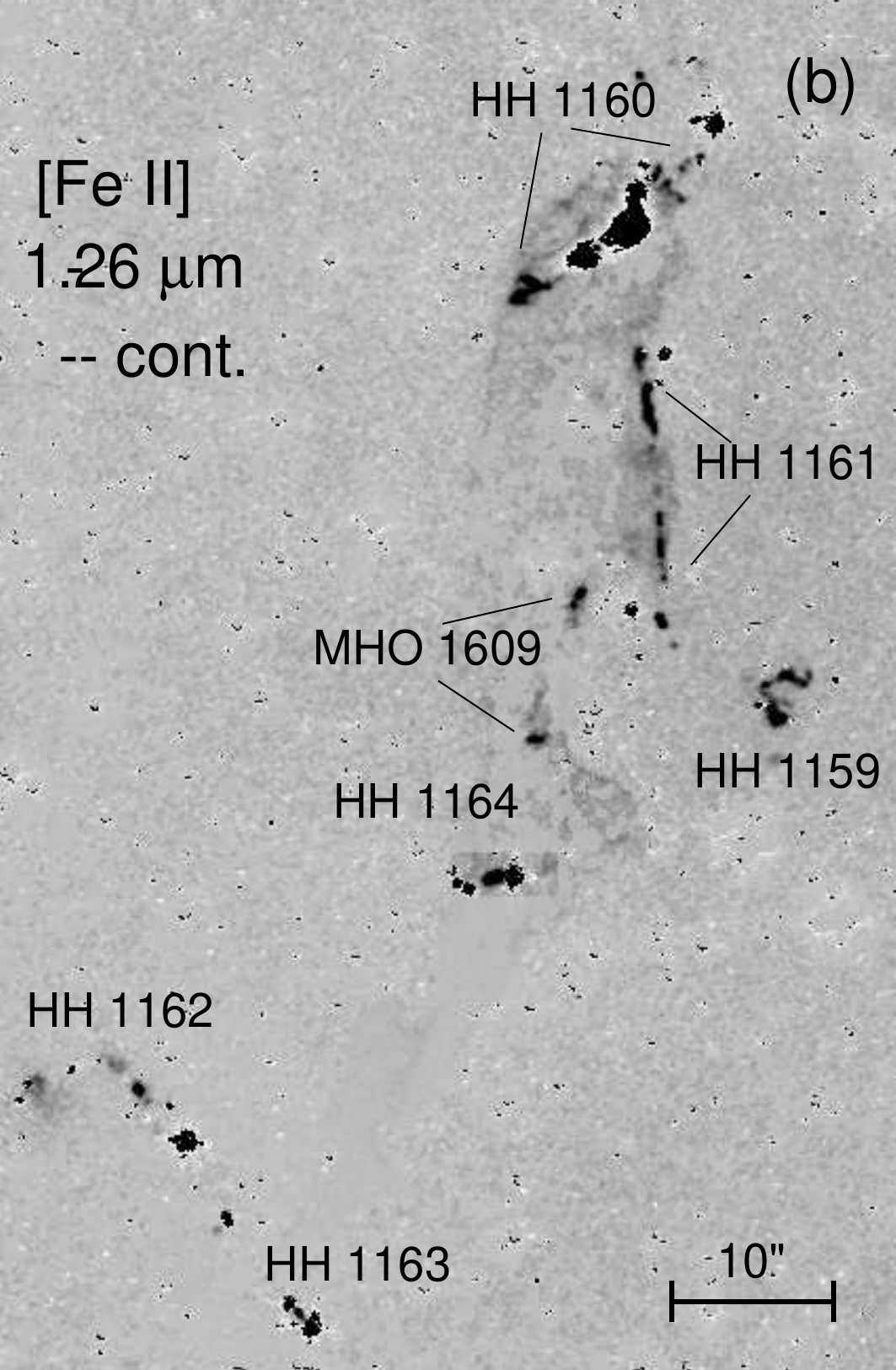} \\  
\includegraphics[trim=22mm 15mm 0mm 0mm,angle=0,scale=0.525]{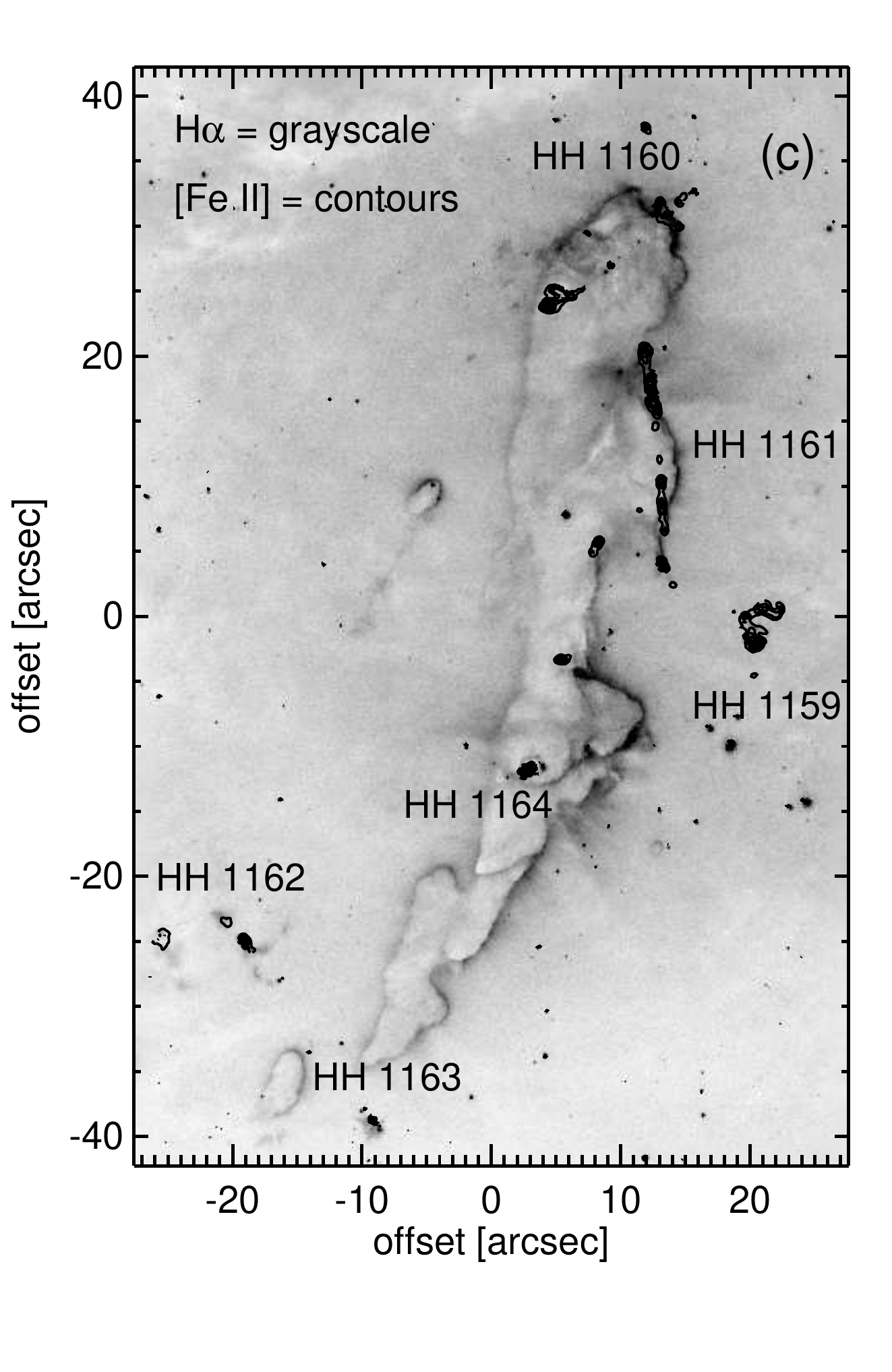} & 
\includegraphics[trim=0mm -17.5mm 0mm 0mm,angle=0,scale=0.525]{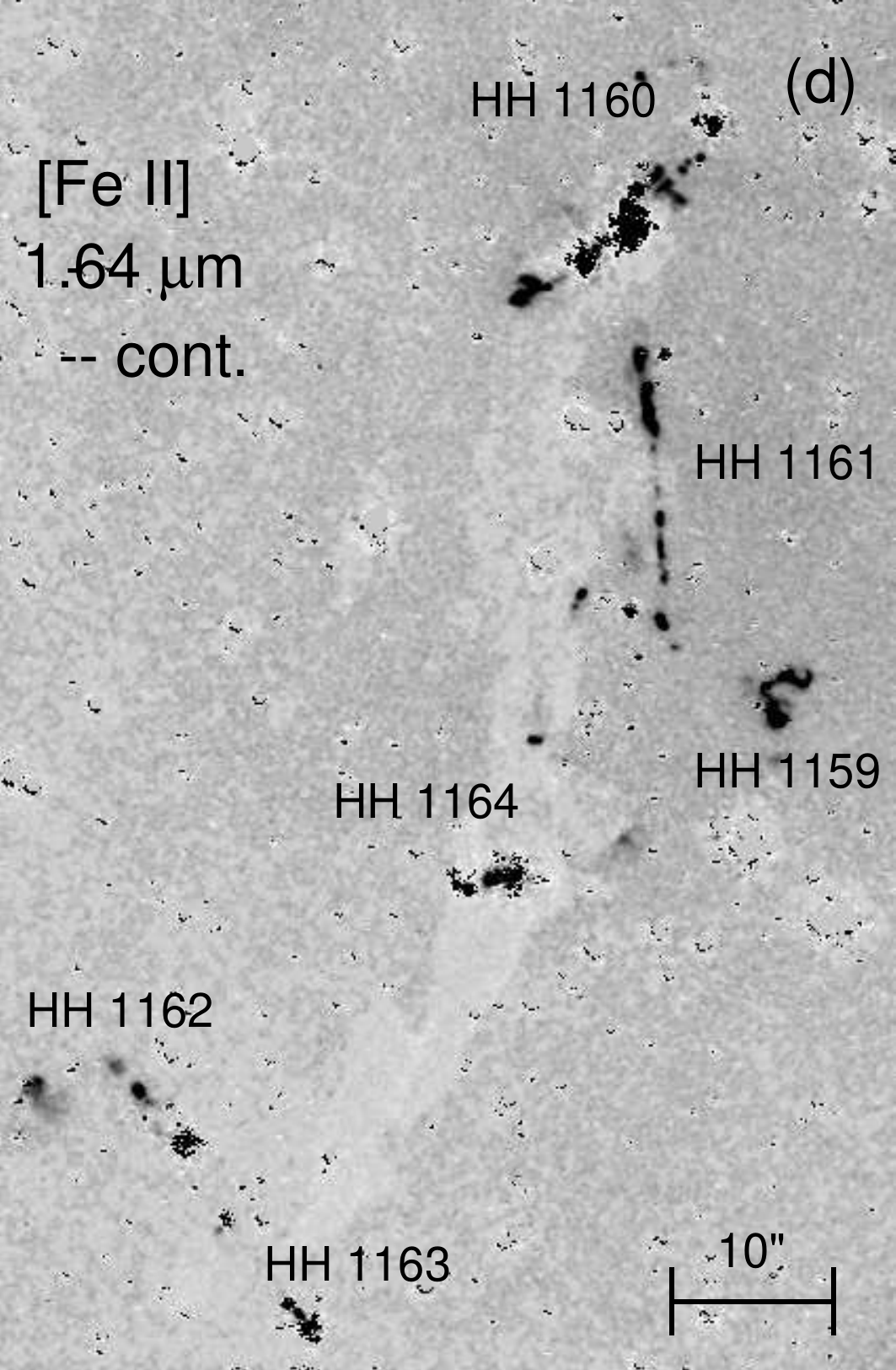} \\ 
\end{array}$ 
\caption{H$\alpha$ images (a) from \citet{smi10} and new near-IR [Fe~{\sc ii}] images (b,c,d) of a globule in the south pillars with several candidate jets identified by \citet{smi10}. 
[Fe~{\sc ii}] observations and our forthcoming proper motion study confirm that these are bonafide HH objects. 
Features are labeled with their new HH numbers (see Section~\ref{ss:new_w_cont}). 
}\label{fig:hhc4_feii} 
\end{figure*}

\textit{HH~1159 through 1164:} 
\citet{smi10} identified several H$\alpha$ knots clustered around an extended globule in the south pillars. 
Five separate HH objects and jet-like features likely trace multiple outflows, although the relationship between each feature is unclear from H$\alpha$ images alone. 
New [Fe~{\sc ii}] images shown in Figure~\ref{fig:hhc4_feii} reveal a network of collimated jets that reach inside the pillar, defining the individual outflows that create the complicated shock structure surrounding the globule. 
Together with proper motion measurements of each feature (Reiter et al.\ in preparation), we can connect each feature with outflow phenomena. Therefore, these 5 candidate jets are assigned HH numbers, as described for each object individually below. 

\textit{HH~1159} is a bow-shock-like knot that lies to the west of the extended globule (see Figure~\ref{fig:hhc4_feii}) that was identified as HH~c-4 by \citet{smi10}. 
Bright [Fe~{\sc ii}] emission traces the same C-shaped morphology as H$\alpha$ and confirms that the cup of the C faces away from the globule (to the west). 
HH~1159 therefore resembles a bow shock moving \textit{toward} the globule, as if launched by a source outside the globule. 
No extended [Fe~{\sc ii}] emission reveals an accompanying jet body or otherwise indicates the origin of the feature. 

\textit{HH~1160} \citep[HH~c-5 in][]{smi10} emerges from the apex of the extended globule and appears to be a one-sided jet in H$\alpha$ images (see Figure~\ref{fig:hhc4_feii}). 
Like HH~1014, [Fe~{\sc ii}] images reveal the counterjet inside the dusty globule. 
Together, the two limbs trace a $\sim 15$\arcsec\ long jet oriented at a $\sim 50^{\circ}$ angle compared to the major axis of the globule. 
The northwest and southeast sides of the jet appear to propagate away from a protostar that lies along the jet axis defined by [Fe~{\sc ii}] \citep[PCYC~787,][]{pov11}. 

\textit{HH~1161} bursts from a curved surface protruding off the side of the extended globule. 
H$\alpha$ images reveal a clear bipolar morphology tracing a $\sim 18$\arcsec\ long jet. 
However, H$\alpha$ emission from the putative jet blends smoothly with the ionization front along the curved surface of the pillar, leading \citet{smi10} to identify it as candidate jet HH~c-6. 

Continuum-subtracted [Fe~{\sc ii}] images clearly separate jet emission from the PDR, revealing a straight bipolar jet. 
The jet axis delineated  by [Fe~{\sc ii}] emission does not appear to bend and is aligned nearly parallel to the major axis of the extended globule. 
In more distant portions of the jet, H$\alpha$ and [Fe~{\sc ii}] emission from the collimated jet coincide (see Figure~\ref{fig:hhc4_feii}). 
Inside the globule, however, the [Fe~{\sc ii}] jet remains straight, unlike the H$\alpha$ emission which curves away from the jet axis, instead tracing the arc of the ionization front along the globule edge. 
No PCYC YSOs lie near HH~1161, although emission at $\lambda \geq 4.5 \mu$m suggests a deeply embedded protostar at the base of the southern limb of the jet. 

\textit{HH~1162} \citep[HH~c-7 in][]{smi10} lies on the opposite side of the globule from HH~1159, to the southeast of HH~1160 and HH~1161. 
Like HH~1159, HH~1162 appears to be a bow shock, located just off the edge of the globule. 
The curved morphology points to the south in this case, as though driven by a source located outside and possibly above the globule. 
\citet{smi10} posit that a star $\sim 30-40$\arcsec\ to the northeast may be the driving source. 

[Fe~{\sc ii}] emission traces the brightest H$\alpha$ emission from HH~1162.  
[Fe~{\sc ii}] emission from HH~1162 breaks the feature into two layers of [Fe~{\sc ii}] knots with a single knot offset $\sim 4.5$\arcsec\ east of the chain of two knots. 
The three [Fe~{\sc ii}] knots (see Figure~\ref{fig:hhc4_feii}) do not trace the curve of the putative bow shock. 
No [Fe~{\sc ii}] emission is seen to the northeast of HH~1162 nor near the candidate driving source identified by \citet{smi10}.

\textit{HH~1163} is an LL~Ori-type object that lies immediately south of the extended globule that \citet{smi10} identified as HH~c-8. 
A cone of H$\alpha$ emission fans open to the east, with no obvious collimated jet body. 
Like the H$\alpha$, [Fe~{\sc ii}] emission from HH~1163 extends $\sim 0.75$\arcsec\ to the east (see Figure~\ref{fig:hhc4_feii}). 
Unlike H$\alpha$, the stream of [Fe~{\sc ii}] emission is collimated, tracing the top edge of the cone of H$\alpha$ emission. 
An additional [Fe~{\sc ii}] knot lies $\sim 7$\arcsec\ to the northeast (see Figure~\ref{fig:hhc4_feii}), along the jet axis defined by extended [Fe~{\sc ii}] emission.

\textit{MHO~1609:} 
A few bright knots stand out inside the globule in continuum-subtracted [Fe~{\sc ii}] images. 
Two bright knots inside the narrow waist of the globule, just south of HH~1161, do not appear to be related to any of the HH objects identified by \citet{smi10}. 
However, \citet{har15} identified a molecular hydrogen outflow, MHO~1609, at this position in ground-based near-IR H$_2$ images. 
The two [Fe~{\sc ii}] knots seen in our WFC3-IR images appear to be at the same position as MHO 1609 knots ``a'' and ``b.'' 
Both [Fe~{\sc ii}] knots clearly lie inside the ionization front tracing the edge of the globule, suggesting that these knots may be shock-excited knots in the embedded molecular jet.

\textit{HH~1164:} 
Further south in the globule, beneath MHO~1609, lies a loop of H$\alpha$ emission emerging from a star in the globule. 
H$\alpha$ emission on the right (west) side of the loop coincides with the position of a \emph{Spitzer}-identified protostar (PCYC~790, see Table~\ref{t:jets_ysos}). 
Our new continuum-subtracted [Fe~{\sc ii}] images reveal a $\sim 0.5$\arcsec\ long jet that extends to the east of the protostar. 
The [Fe~{\sc ii}] jet extends through the middle of the H$\alpha$ loop, tracing the familiar jet-cocoon morphology seen in, e.g. HH~666~O. 
We therefore assign this jet an HH number of 1164.

\begin{figure}
\centering
$\begin{array}{cc}
\includegraphics[trim=0mm 0mm 5mm 0mm,angle=0,scale=0.3625]{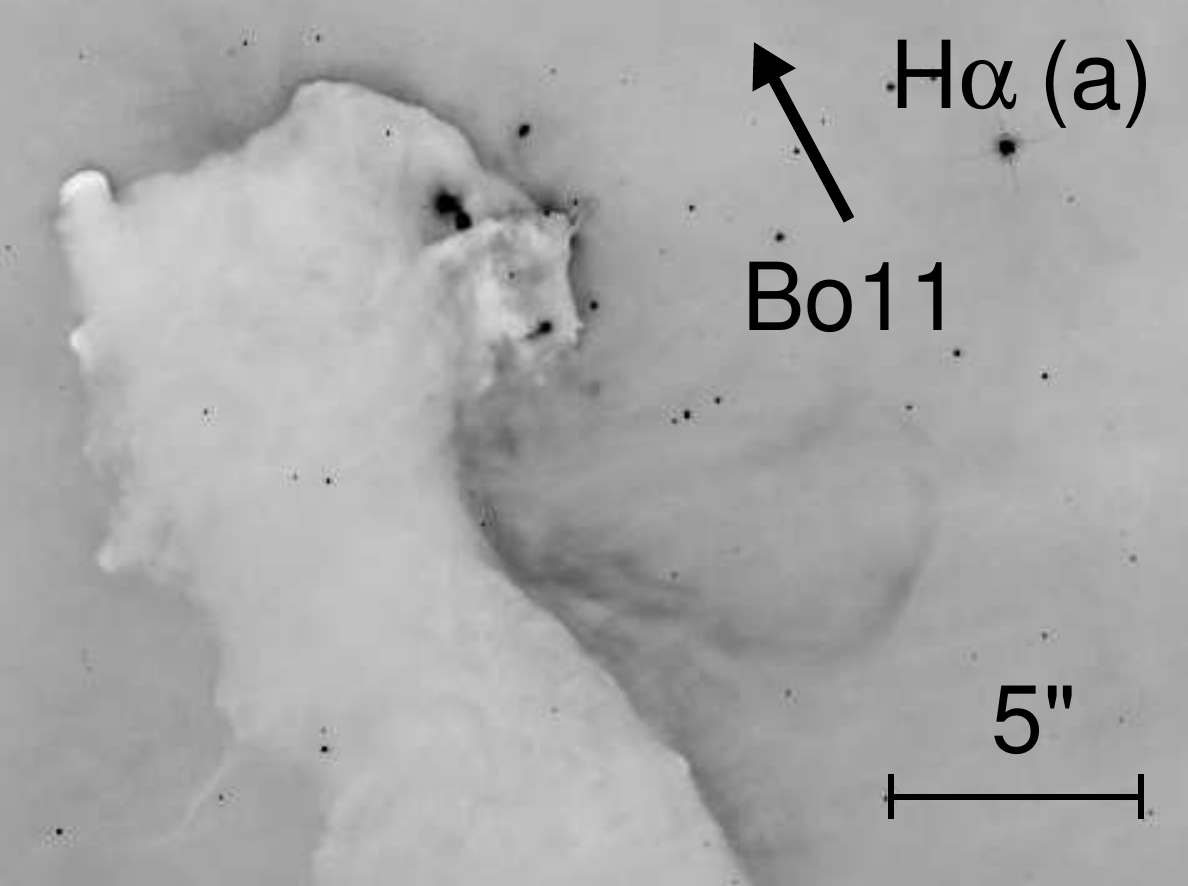} &
\includegraphics[trim=5mm 0mm 0mm 0mm,angle=0,scale=0.3625]{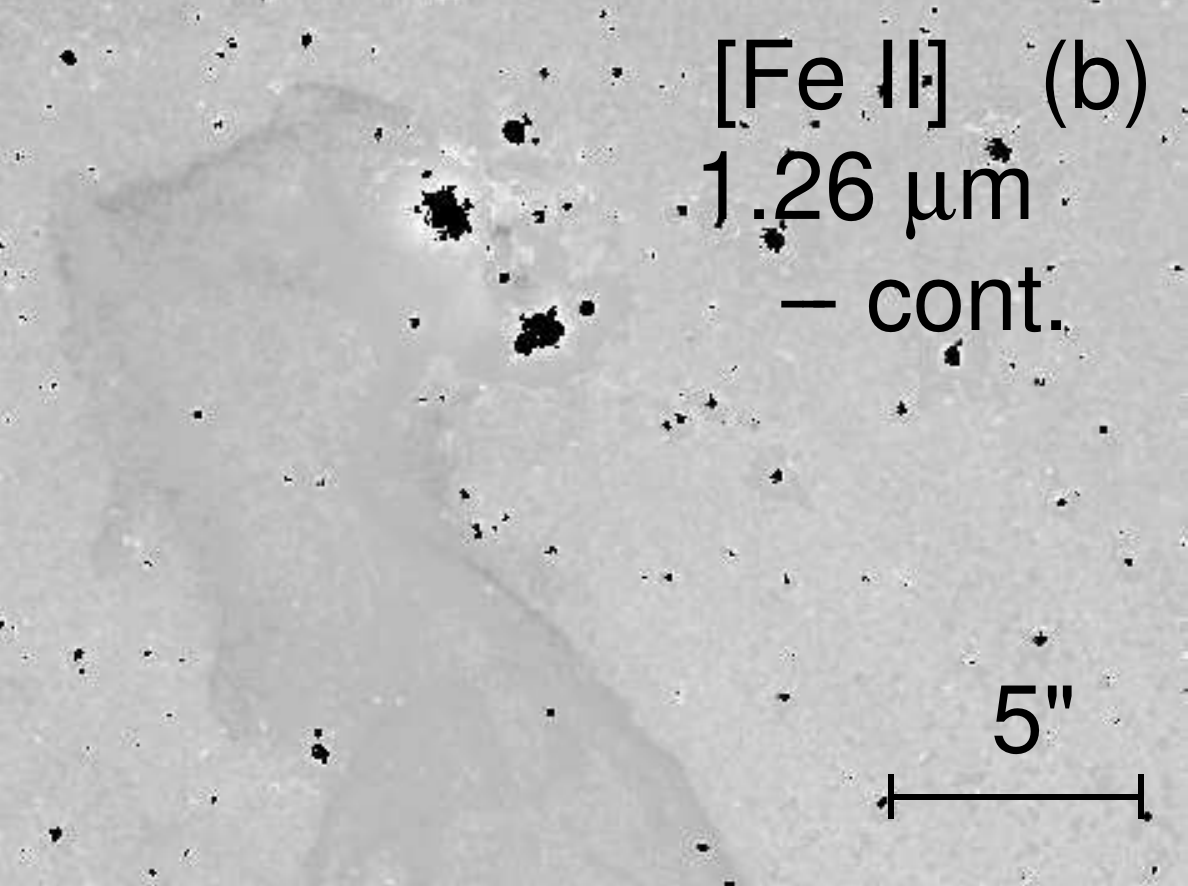} \\
\includegraphics[trim=15mm 0mm 5mm 0mm,angle=0,scale=0.405]{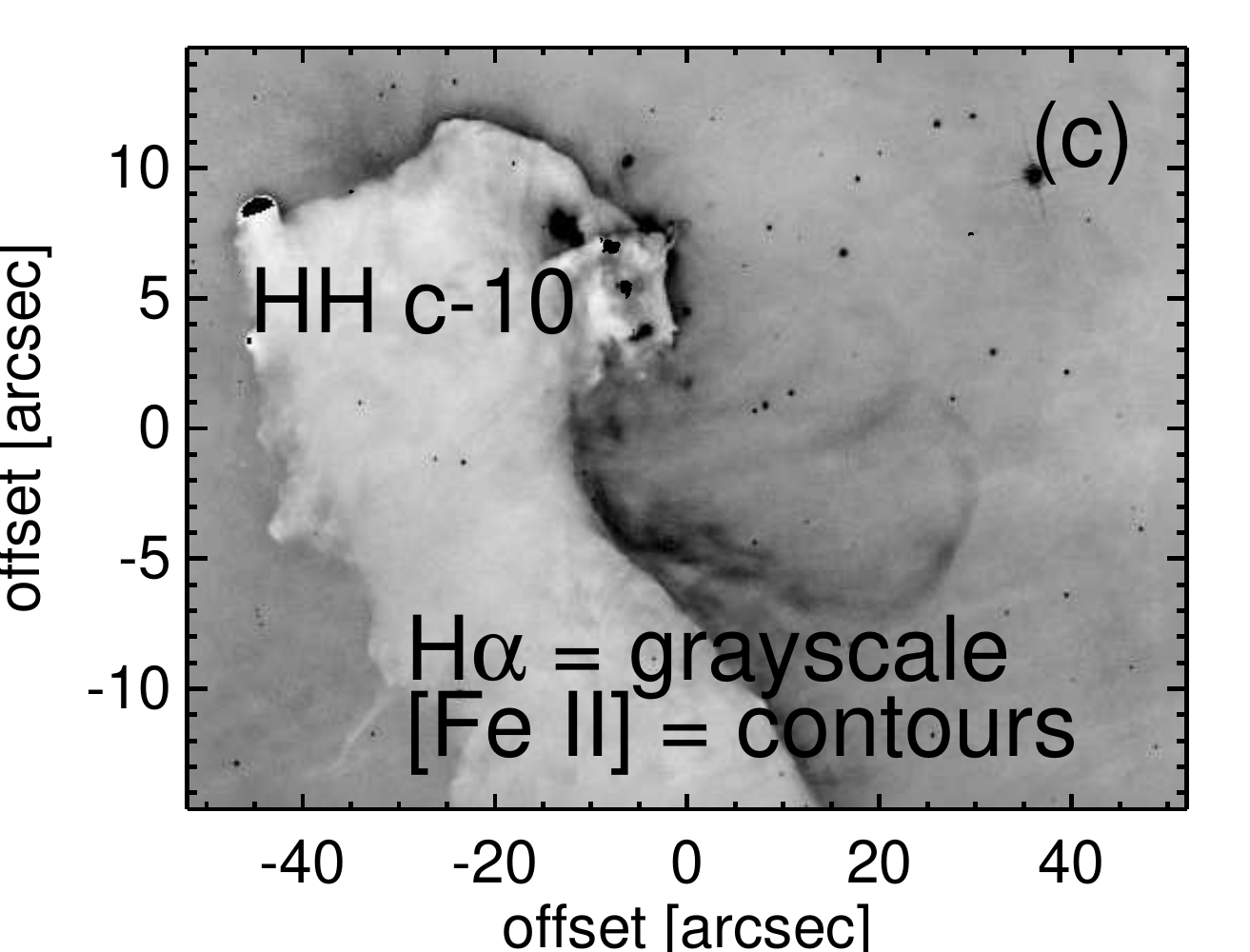} & 
\includegraphics[trim=5mm -18mm 0mm 0mm,angle=0,scale=0.3625]{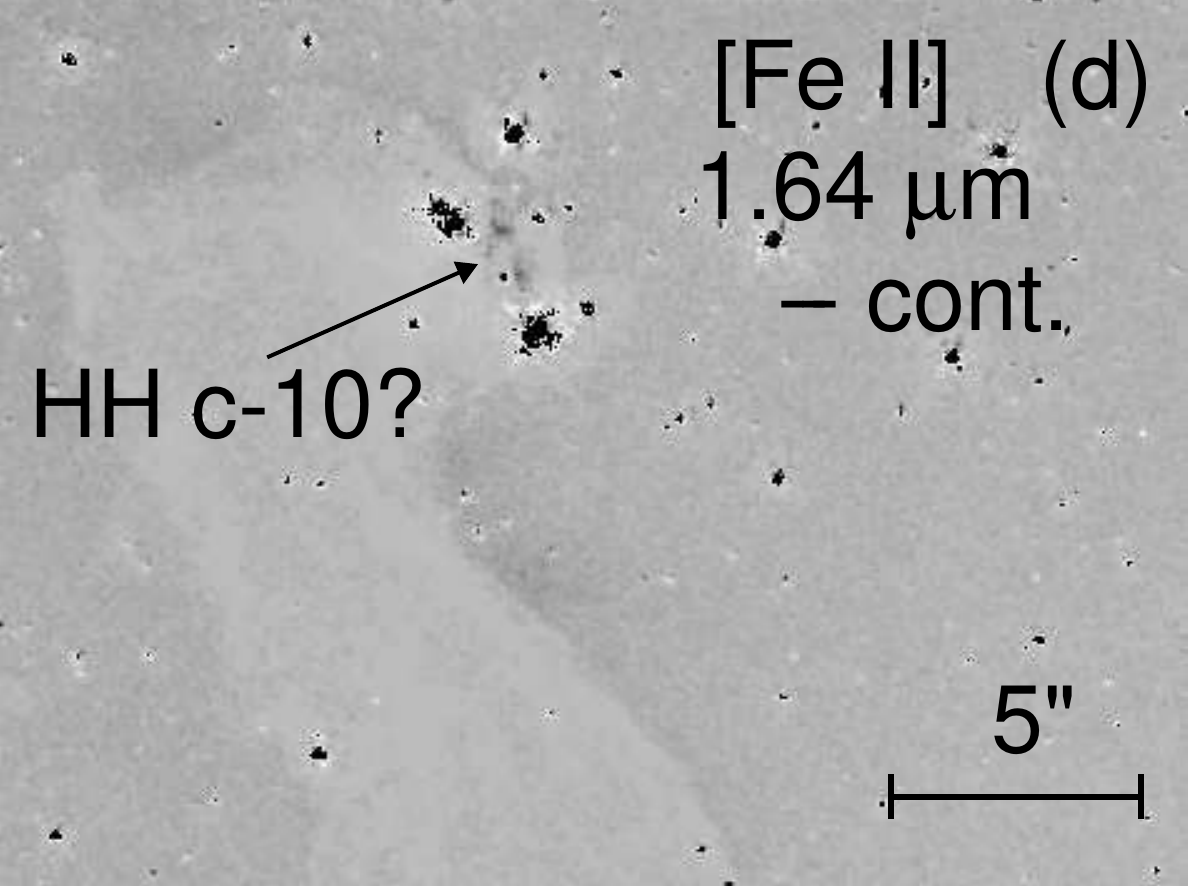} \\ 
\end{array}$ 
\caption{HH~c-10 is a candidate jet identified as a bow shock in H$\alpha$ images (a) by \citet{smi10}. 
Faint [Fe~{\sc ii}] 1.64 $\mu$m emission (b,c,d) may trace the body of the corresponding jet that emerges from the pillar head. }\label{fig:hhc10_feii} 
\end{figure}

\textit{HH~c-10:} A series of curved shock structures that look like bow shocks in H$\alpha$ are seen near the head of the same pillar that houses HH~903 \citep{smi10}. 
The implied outflow axis points to the southwest, toward HH~903 which emerges from the middle of the pillar $\sim 90$\arcsec\ south of HH~c-10. 
No emission from the putative HH~c-10 bow shock is seen in [Fe~{\sc ii}] emission (see Figure~\ref{fig:hhc10_feii}). 
However, two IR bright point sources are detected at the head of the pillar. 
Two knots of [Fe~{\sc ii}] emission emerge in the continuum-subtracted image and fall along the outflow axis implied by the H$\alpha$ morphology. 
[Fe~{\sc ii}] emission is brighter at 1.64~\micron\ than 1.26~\micron, suggesting that this may be an embedded jet body. 
However, the bright point sources lead to imperfect continuum subtraction, so the true nature of HH~c-10 remains unclear from images alone. 

\subsection{Comparison of the H$\alpha$ and [Fe~{\sc ii}] intensity}\label{ss:emission_comp}

\begin{figure}
\centering
$\begin{array}{c}
\includegraphics[trim=5mm 10mm 5mm 10mm,angle=0,scale=0.275]{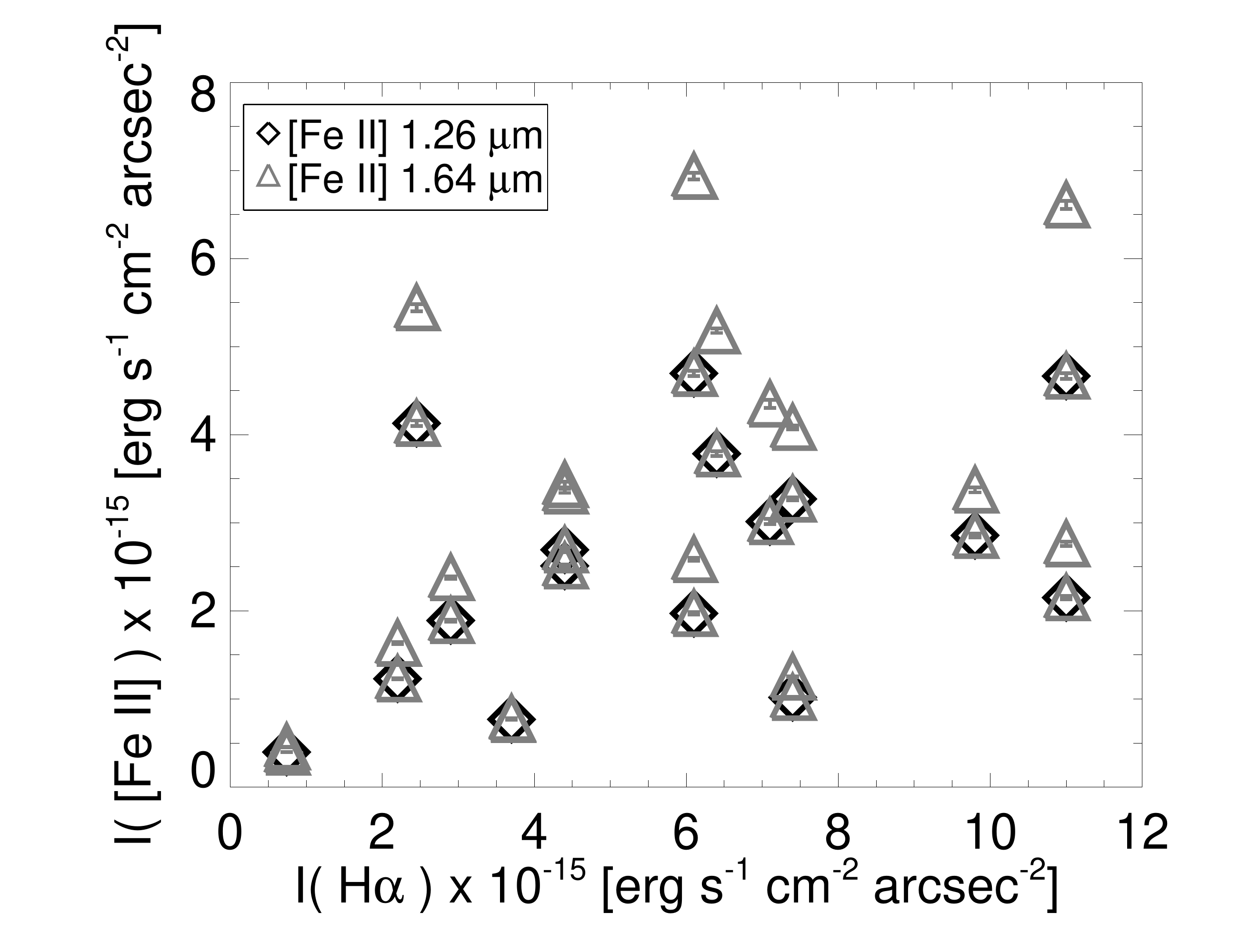} \\
\includegraphics[trim=5mm 10mm 5mm 10mm,angle=0,scale=0.275]{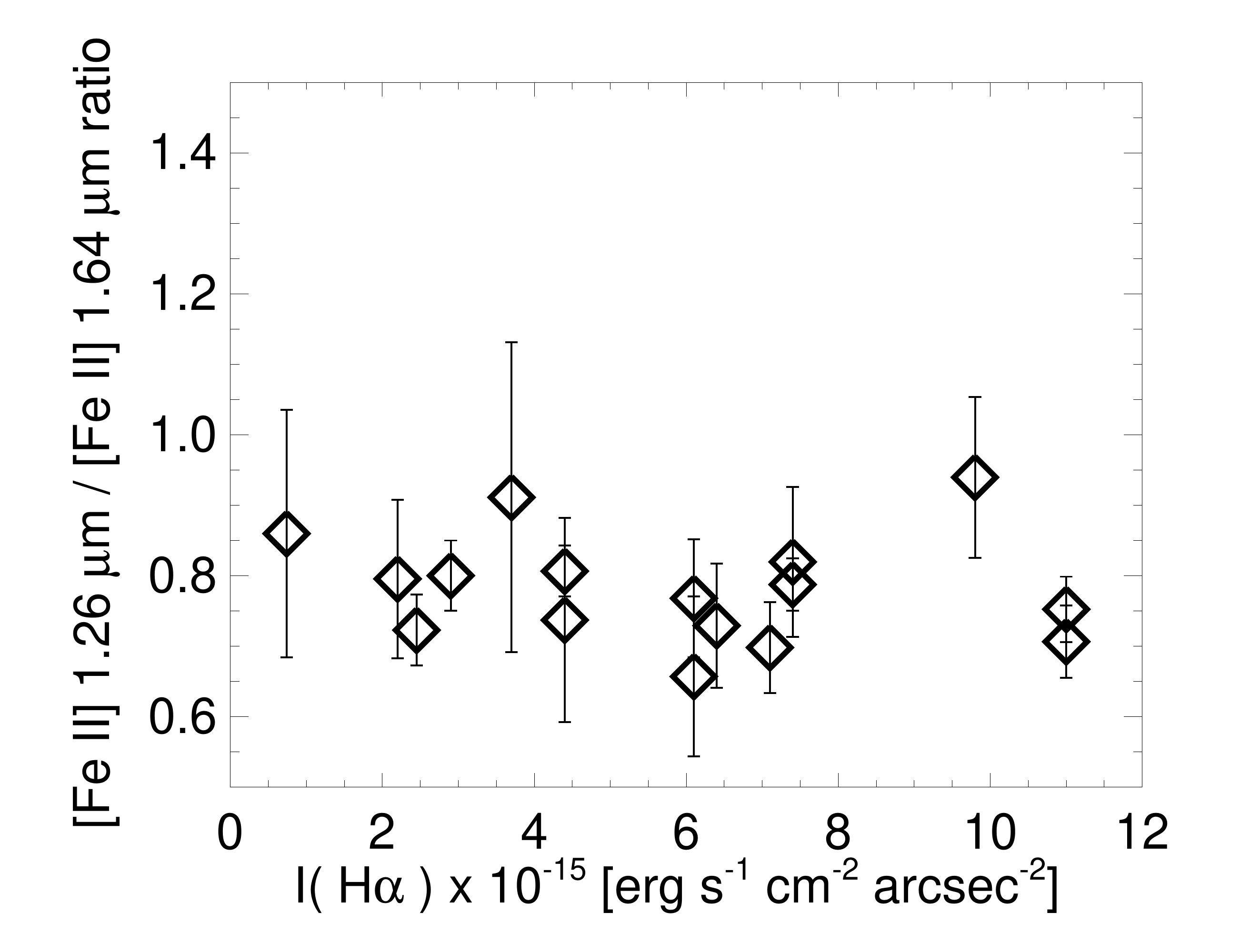} \\
\includegraphics[trim=5mm 10mm 5mm 10mm,angle=0,scale=0.275]{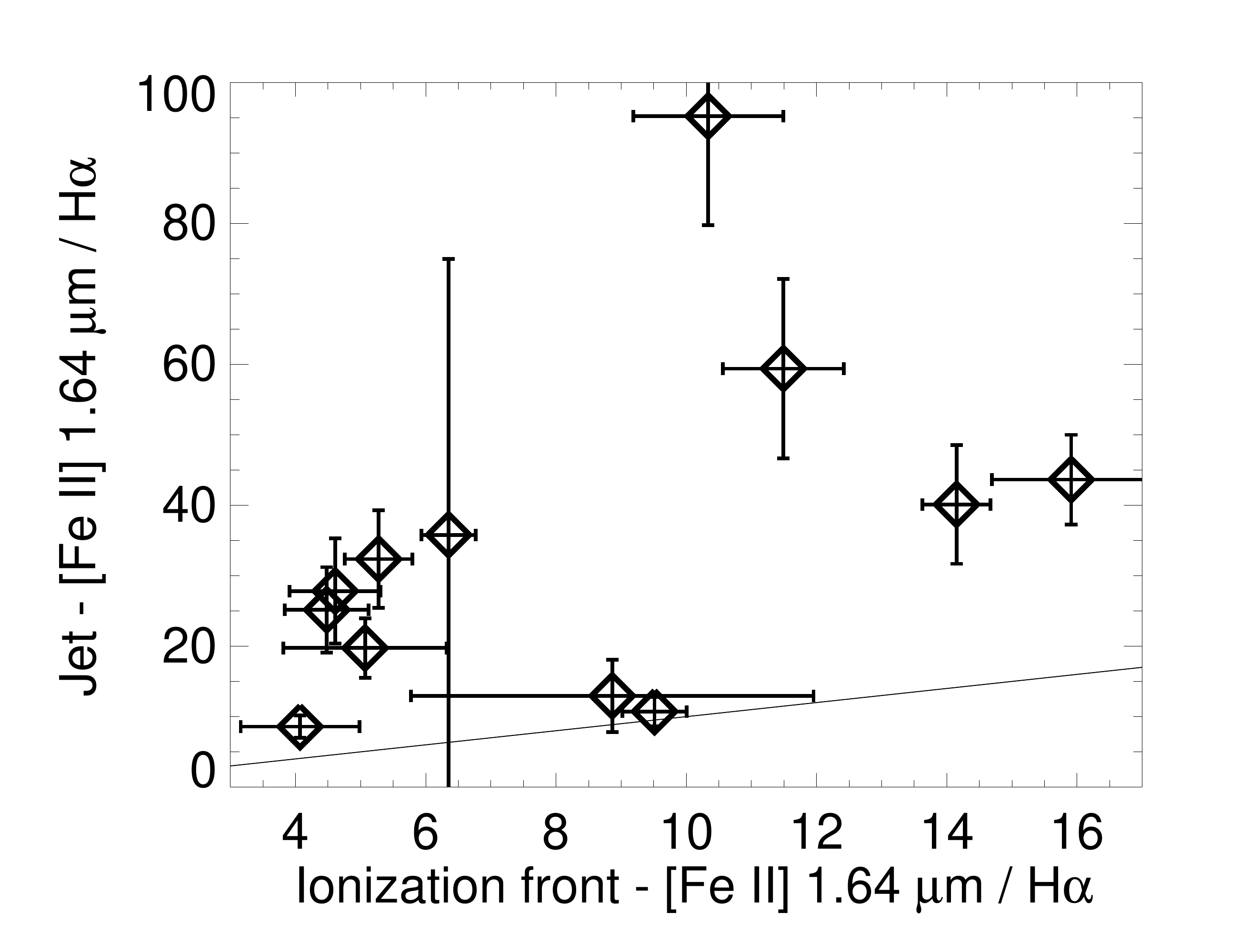} \\
\end{array}$ 
\caption{ 
\textit{Top:} [Fe~{\sc ii}] 1.26 \micron\ (black diamonds) and 1.64 \micron\ (gray triangles) jet intensity plotted against the H$\alpha$ intensity measured for a similar portion of that jet by \citet[][see Table~4]{smi10}. Error bars are smaller than the plot symbols.  
\textit{Middle:} [Fe~{\sc ii}] ratio $\mathcal{R}$ measured in a bright portion of the jet plotted as a function of the H$\alpha$. 
\textit{Bottom:} Comparison of the flux ratio [Fe~{\sc ii}] 1.64 \micron/H$\alpha$ measured in the jet compared to the adjacent ionization front for jets embedded in a globule. The solid line shows where the two are equal. 
}\label{fig:int_comp} 
\end{figure}

Figure~\ref{fig:int_comp} shows comparisons between the H$\alpha$ and [Fe~{\sc ii}] emission in the HH jets with WFC3-IR observations. 
Both the jets and the ionization fronts along the pillar edges tend to be bright in H$\alpha$. 
This is not the case in [Fe~{\sc ii}], where emission is enhanced in the jets and may not be detected at all in the ionization front. 
To quantify this, we measure the [Fe~{\sc ii}] 1.64 \micron/H$\alpha$ ratio and find that it is $\gtrsim 5$ times higher in the jet than the adjacent ionization front (see Figure~\ref{fig:int_comp}).  

\begin{figure}
\centering
\includegraphics[angle=0,scale=0.275]{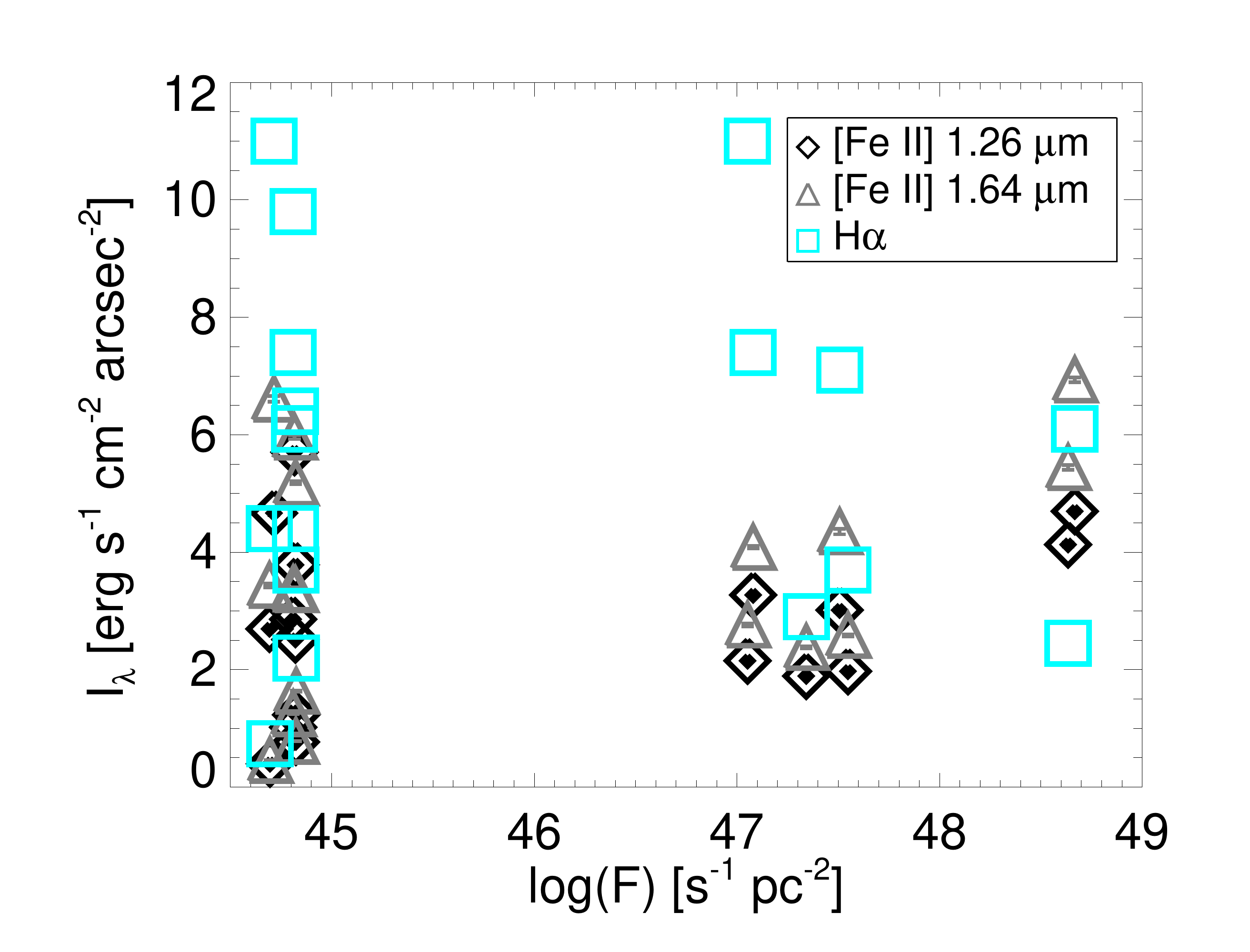} 
\caption{ 
Representative jet intensity in [Fe~{\sc ii}] 1.26 \micron\ (black diamonds), 1.64 \micron (gray triangles), and H$\alpha$ \citep[from][cyan squares]{smi10} plotted versus the incident ionizing photon flux (listed in Table~\ref{t:jet_prop}). 
}\label{fig:int_ostar_comp} 
\end{figure}

We also compare the H$\alpha$ intensities measured by \citet{smi10} to the intensity of both [Fe~{\sc ii}] lines and their ratio over roughly the same portion of the jet. 
Neither quantity is correlated with the intensity of the H$\alpha$ emission.   
However, as shown by \citet{rei13}, H$\alpha$ and [Fe~{\sc ii}] trace different material in the jet and are often bright in different places along the length of the jet. 
H$\alpha$ traces the density of material in the ionized outer radii of the outflow while [Fe~{\sc ii}] traces the density of the jet core. 
These densities are not necessarily correlated with each other or the strength of the incident ionizing flux (see Figure~\ref{fig:int_ostar_comp}). 
In addition, the H$\alpha$ intensity scales as $n_e^2$, so relatively small offsets in density will enhance the spread in jet intensities.

\subsection{[Fe~{\sc ii}] ratio tracings}\label{ss:feii_ratio}


\begin{figure}
\centering
$\begin{array}{c}
\includegraphics[trim=0mm 10mm 0mm 0mm,angle=0,scale=0.315]{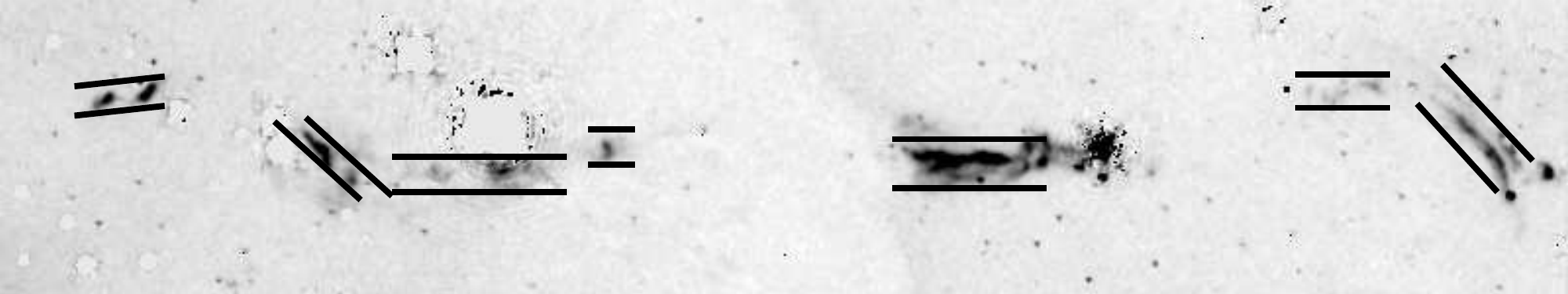} \\ 
\includegraphics[angle=0,scale=0.275]{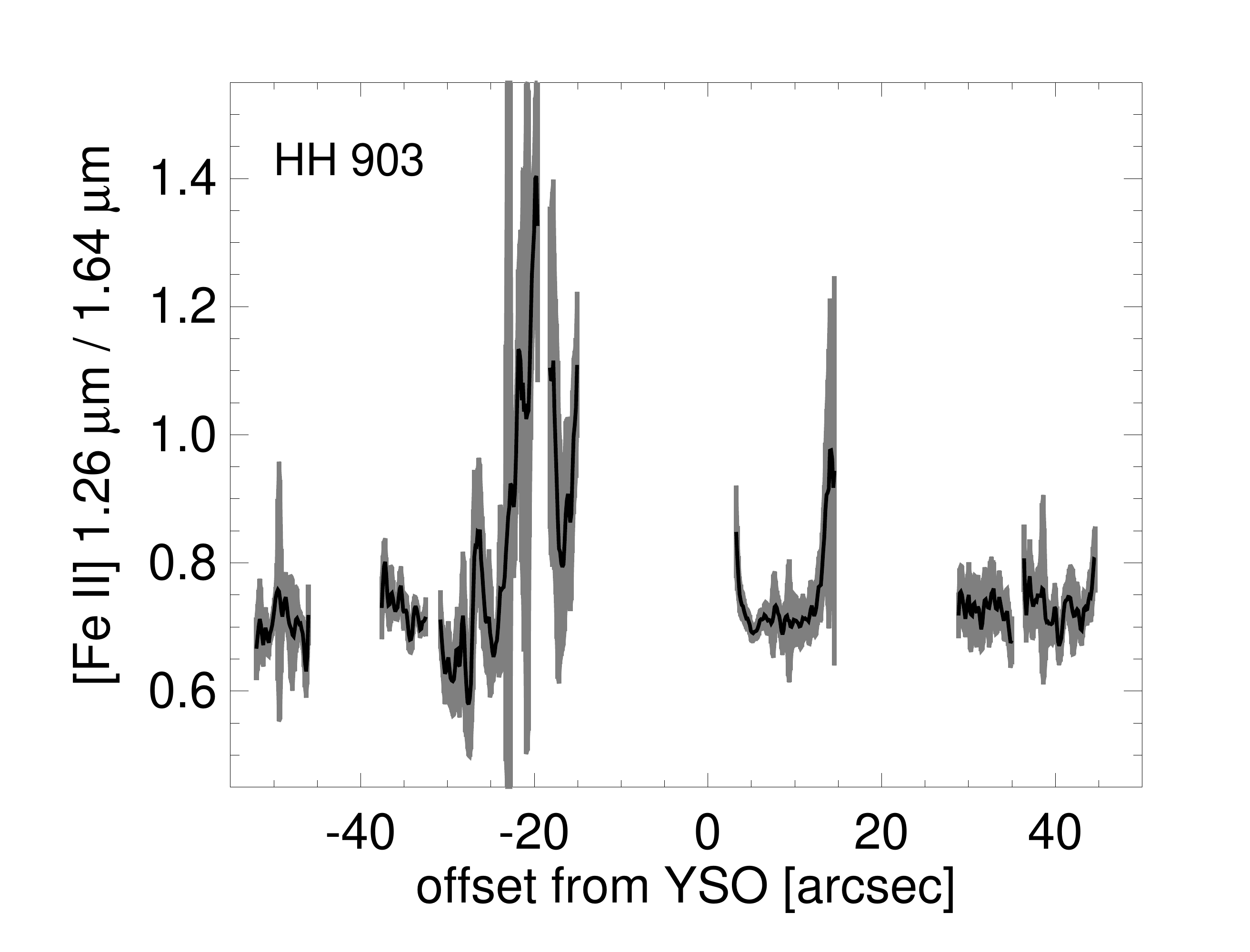} \\ 
\end{array}$ 
\caption[{[Fe~{\sc ii}] ratio tracing of HH~903.}]{
\textit{Top:} [Fe~{\sc ii}] 1.64 \micron\ image of HH~903 with lines that show the area extracted to measure the [Fe~{\sc ii}] ratio along each portion of the jet. Bright stars in the field have been masked. 
\textit{Bottom:} Tracing of the [Fe~{\sc ii}] ratio $\mathcal{R} = \lambda 12657 / \lambda 16437$ along HH~903. 
The gray shaded area surrounding the line shows the 1$\sigma$ uncertainty in the ratio at that point along the jet.  
Ratio tracings for the other jets presented in this work are in the supplementary material available online. 
}\label{fig:hh903_feii_ratio} 
\end{figure}

The [Fe~{\sc ii}] transitions we observed with \emph{HST} are the two brightest lines that originate from the a$^4$D level. 
The intrinsic flux ratio, $\mathcal{R} = \lambda 12657 / \lambda 16437$, will therefore be determined by atomic physics. 
In the absence of any reddening, the 1.26 \micron\ line will be brighter than the 1.64 \micron\ line. 
However, the intrinsic flux ratio $\mathcal{R}$ is only known as well as the transition probabilities, which remain uncertain \citep[e.g.][]{nus88,qui96,bau98,rod04,sh06,deb10,bau13,bau15}. 
Observational estimates of the intrinsic flux ratio $\mathcal{R}$ have yielded values from $\mathcal{R} = 1.11$ in HH~1 \citep{gia15} to $\mathcal{R} = 1.49$ in P~Cygni \citep{sh06}. 
Lower values for the intrinsic ratio $\mathcal{R}$ correspond to less extinction for the same measured ratio \citep[see discussion in][]{rei15b}.

We measure the ratio $\mathcal{R}$ along the length of the jet in new, continuum-subtracted [Fe~{\sc ii}] images (see, e.g.\ Figure~\ref{fig:hh903_feii_ratio}; ratio tracings for the other jets in the sample are available in the online supplement). 
For each jet, we indicate the regions of the jet considered on the [Fe~{\sc ii}] 1.64~\micron\ image. 
To obtain a representative value for each jet, we take the ratio of a bright knot in roughly the same portion of the jet that \citet{smi10} extracted in H$\alpha$, although this is not possible in all cases (e.g.\ HH~1005 and HH~1007). 
We have not corrected for line-of-sight reddening toward Carina \citep{smi87,smi02}. 

Measuring the [Fe~{\sc ii}] ratio in images is most illustrative of how the reddening changes along the length of the jet. 
Unlike spectra, narrowband images may be contaminated with other emission lines that fall within the filter bandwidth.
Unequal widths of the narrowband filters may result in different amounts of contamination from other emission lines in the images (i.e. 15.2~nm for F126N compared to 20.9~nm for F164N).
For example, Br12 and Br11 fall in the F164N and F167N filters, respectively.
He~I 1.253 $\mu$m may contaminate images obtained with the F126N filter; indeed, this line is observed to be brighter than theoretical expectation in the Orion bar and Orion S regions \citep[see, e.g.][]{luh98}.

Regardless of potential contamination in the absolute [Fe~{\sc ii}] line ratio, images allow us to trace the \textit{relative} reddening as a function of position along the jet.
All of the jets have ratios that remain below the intrinsic value along the length of the jet, often below $\mathcal{R} \approx 1$ corresponding to A$_V \sim 5$~mag for $\mathcal{R} = 1.49$ \citep[see Figure~9 in][]{rei13}, or A$_V \sim 1-2$~mag for $\mathcal{R} = 1.11 - 1.20$ \citep{gia15}.  
In the H~{\sc ii} region, the ratio stays near a constant value, similar to the steady $\mathcal{R}$ values found by \citet{rei15b} in HH~666. 
\citet{rei13} used the change in $\mathcal{R}$ in HH~1066 to argue for an optically-thick flared circumstellar disk increasing the reddening near the jet-driving source. 
A similar circumstellar geometry may explain the dip in $\mathcal{R}$ near the position of the driving source in HH~1006 and HH~1156.

\begin{figure}
\centering
\includegraphics[angle=0,scale=0.275]{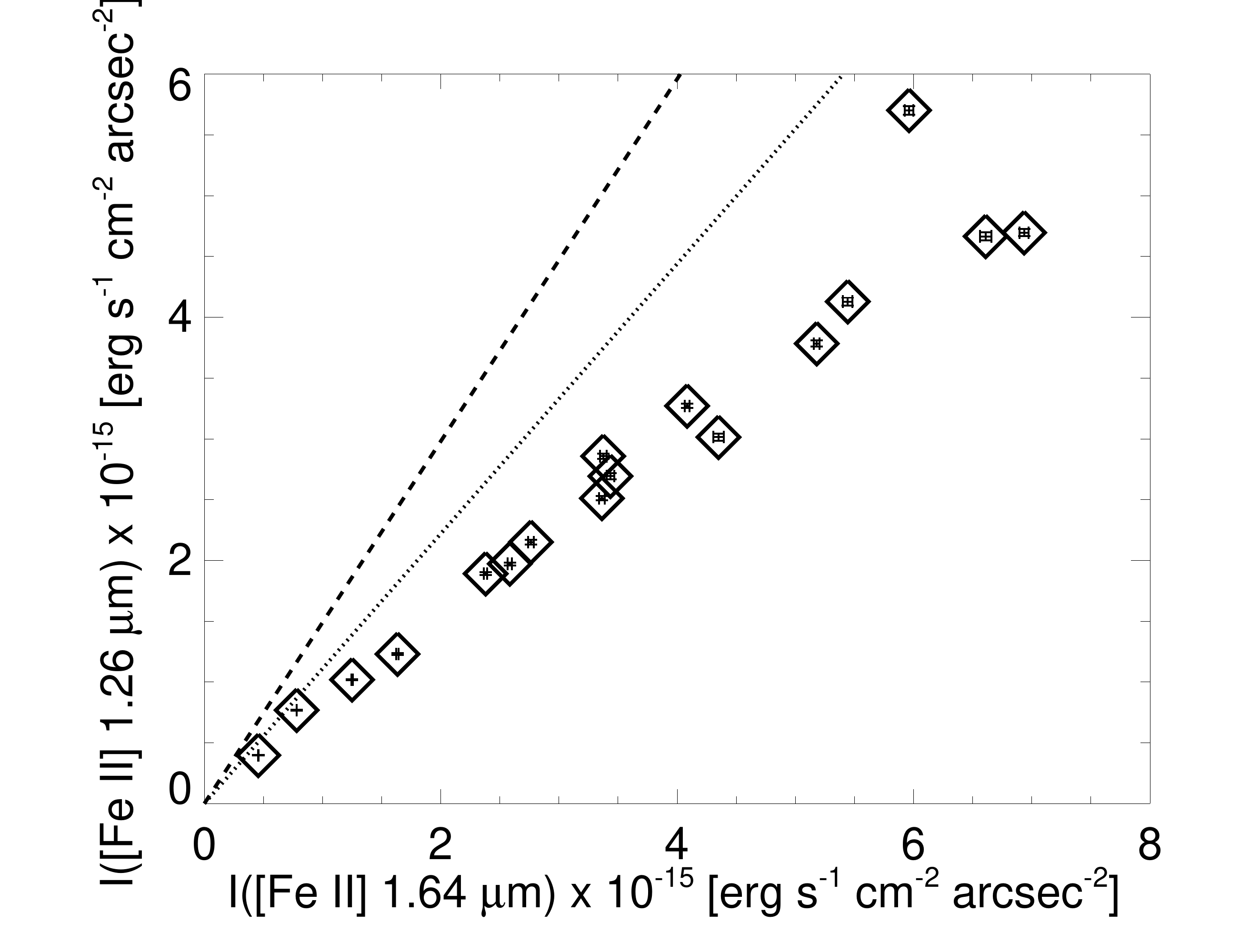} 
\caption{Intensity of [Fe~{\sc ii}] 1.26 \micron\ plotted as a function of the [Fe~{\sc ii}] 1.64 \micron\ intensity. 
Lines corresponding to the expected line fluxes for zero reddening for an intrinsic flux ratio $\mathcal{R} = 1.49$ \citep[][dashed line]{sh06} and $\mathcal{R} = 1.11$ \citep[][dotted line]{gia15} are overplotted. }\label{fig:feii_int_comp} 
\end{figure}

\subsection{Candidate driving sources}\label{ss:ysos}
For this program, we targeted HH jets in the Carina Nebula with a candidate protostar identified along the jet axis. 
[Fe~{\sc ii}] emission from most jets extends inside the dust pillar, clearly connecting the larger outflow seen in H$\alpha$ to the embedded IR YSO. 
We list the PCYC number of the matched driving sources and the physical properties estimated from model fits to the IR SED from \citet{pov11} in Table~\ref{t:jets_ysos}. 
\citet{ohl12} also identified candidate driving sources for some of the HH jets in Carina discovered by \citet{smi10}. 
Their YSO identifications and model fits are also included in Table~\ref{t:jets_ysos}. 
For the 4 sources where the identified driving source has been modeled by both \citet{pov11} and \citet{ohl12}, the derived source properties agree to within a factor of two, and often to within a few percent. 
For the remainder of this analysis, we will focus on the model results obtained by \citet{pov11}.

\begin{figure}
\centering
\includegraphics[angle=0,scale=0.275]{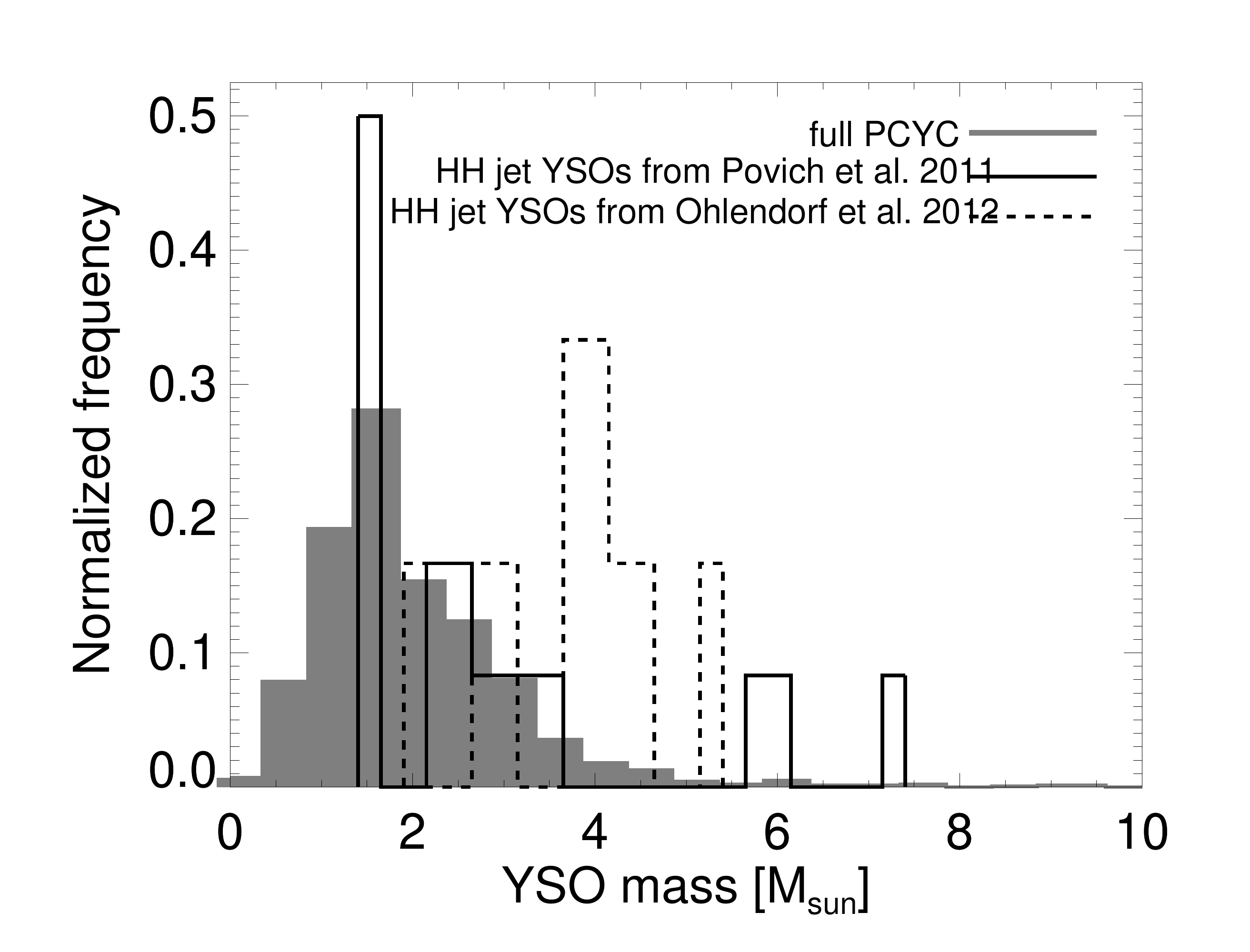} 
\caption[Histogram of protostar masses (with and without jets)]{A normalized histogram of the masses of all PCYC protostars with model fits to the IR SED from \citet{pov11} is shown in gray. 
The overplotted solid line histogram shows the distribution of YSO masses of the jet-driving sources we identified from the PCYC. 
The dashed-line histogram shows the jet-driving protostar masses determined by \citet{ohl12}. 
}\label{fig:yso_masses} 
\end{figure}

Figure~\ref{fig:yso_masses} compares the masses of the jet-driving protostars in Carina to all the YSOs in Carina modeled by \citet{pov11}. 
One of the goals of the \emph{HST} survey for [Fe~{\sc ii}] emission from the HH jets in Carina is to test the physical properties of the jets driven by protostars sampling the intermediate-mass ($\sim 2-8$ M$_{\odot}$) range. 
Estimated driving source masses range from $1.4-7.5$ M$_{\odot}$. 
The SEDs of the jet-driving sources are consistent with young YSOs (although 4/12 have an ambiguous evolutionary classification). 

For 5 jets in this sample, the driving source remains somewhat ambiguous or unknown. 
In the case of HH~1005, point-like IR emission consistent with being a YSO has been identified near the outflow axis \citep[see][]{ohl12}. 
However, [Fe~{\sc ii}] emission from the jet seen inside the natal dust pillar does not clearly connect the larger scale H$\alpha$ outflow to any of the nearby candidate protostellar objects. 
Most of the remaining outflows -- HH~900, HH~901, HH~902 -- lie close to the brightest emission from the H~{\sc ii} region, where the bright and variable background will limit sensitivity to point source emission \citep[see discussion in][]{pov11}. 
No [Fe~{\sc ii}] emission is detected inside the globule from either HH~900 or HH~901, rather it begins $\sim 1$\arcsec\ away from the edge of the globule. 
\citet{rei13} and \citet{rei15a} have explored the possibility that feedback from nearby massive stars may have compressed the globules, leading to high densities that obscure the jet-driving protostar and shield the inner jet. 
However, some jets without a detected driving source lie far from Tr14 and Tr16. 
HH~1161 emerges from a globule deeper in the south pillars. 
While [Fe~{\sc ii}] emission from HH~1161 does reach inside the globule (unlike HH~900 and HH~901), no protostar has been detected near the jet axis.

Jet dynamical ages may be estimated by assuming that a jet knot reached its current position by traveling at a constant velocity. 
This provides an independent estimate of the age and therefore evolutionary stage of the driving source. 
Individual jet velocities have only been measured for 5/19 jets in this work.
Adopting the median knot velocity found by \citet{rei14} for 4 HH jets in Carina -- $v_{jet} = 140$ km s$^{-1}$ -- we find that the median dynamical age of the HH jets presented in this study is $\sim 3500$ yr. 
This age range is very uncertain, but in any case is only a few percent of the Class 0 and Class I lifetimes estimated for low-mass sources \citep[0.16 Myr and 0.54 Myr, respectively;][]{eva09}.


\section{Discussion}\label{s:ir_synth_discussion}

We detect [Fe~{\sc ii}] emission in all of the HH jets in the Carina Nebula targeted for follow-up observation with WFC3-IR. 
In addition, three candidate jets also fell within the area imaged with WFC3-IR. 
Of the three, only HH~1156 (formerly HH~c-14) has a clear bipolar jet morphology. 
HH~c-3 and HH~c-10 also show faint [Fe~{\sc ii}] emission, but their true nature remains unclear. 
Altogether new, continuum-subtracted [Fe~{\sc ii}] images combined with archival images of 4 other sources yield a sample of 21 HH objects corresponding to at least 18 individual jets (see Table~\ref{t:jet_prop}).

\begin{figure*}
\centering
$\begin{array}{cc} 
\includegraphics[trim=5mm 0mm 0mm 0mm,angle=0,scale=0.60]{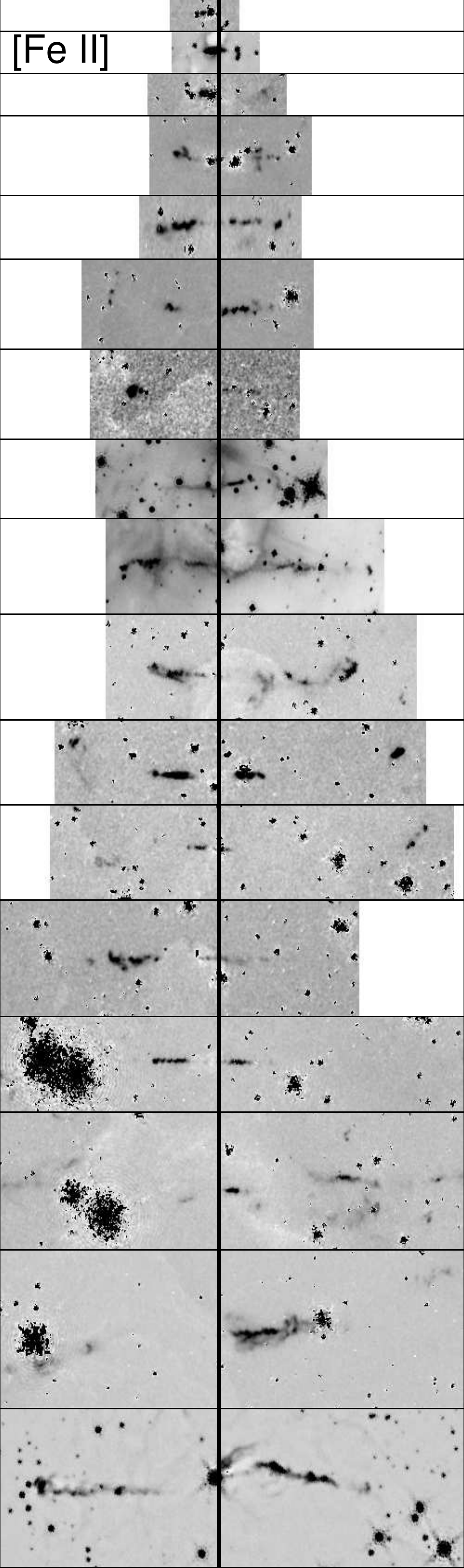} &
\includegraphics[trim=5mm 0mm 0mm 0mm,angle=0,scale=0.60]{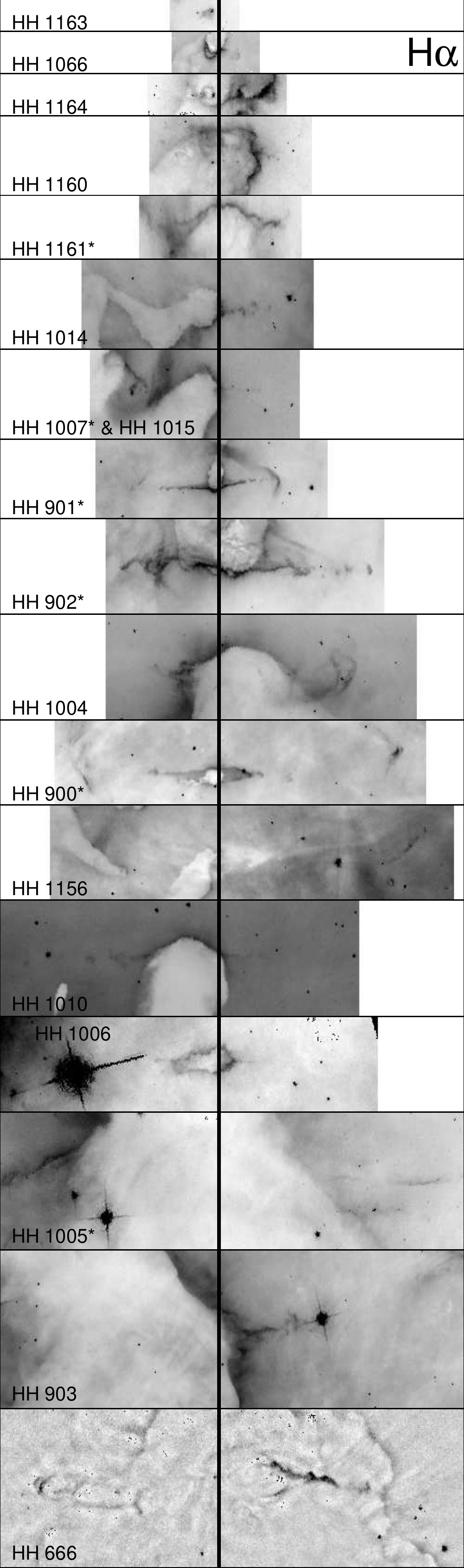} \\
\end{array}$
\caption[Comparison of the HH jets in Carina]{A comparison of 17 HH jets in the Carina Nebula that have been observed in [Fe~{\sc ii}] with WFC3-IR (HH~c-3 is not shown). Jets are shown on the same spatial scale, centered at the driving source position. For jets without a detection of the driving source (indicated with an asterisk next to the name) the YSO location is estimated based on the morphology of the jet emission.}\label{fig:jet_comp} 
\end{figure*}
\clearpage 

\subsection{Strong [Fe~{\sc ii}] emission from externally irradiated HH jets}\label{ss:bright_feii} 
Bright [Fe~{\sc ii}] emission traces a collimated jet in almost every source presented in this paper. 
Unlike H$\alpha$ images, where jet emission is often confused with the pillar edge or surrounding cocoon, continuum-subtracted [Fe~{\sc ii}] images make it easy to distinguish jet emission from the PDR (see Section~\ref{ss:emission_comp} and Figure~\ref{fig:int_comp}). 
For sources with a clear jet-like morphology, the detection of [Fe~{\sc ii}] emission demonstrates that the jet is not completely ionized, but instead maintains a neutral core. 
Therefore, the mass-loss rates estimated by \citet{smi10} are lower limits to the true mass-loss rate in the jet. We revise jet mass-loss rates to account for this in Section~\ref{ss:mdot_jet}.

As pointed out by \citet{bal06}, to first order, the H$\alpha$ intensity of a jet with a neutral core will be proportional to the local Lyman continuum flux. 
We might expect weaker H$\alpha$ emission from jets like HH~1015 that lie far from their ionizing sources. 
Both H$\alpha$ and [Fe~{\sc ii}] emission are weak in HH~1015, suggesting less external excitation of the jet. 
In general, however, the observed strength of the [Fe~{\sc ii}] emission does not correlate with the H$\alpha$ intensity of the jet (see Section~\ref{ss:emission_comp} and Figure~\ref{fig:int_comp}) and does not depend on environment. 
[Fe~{\sc ii}] is often a better tracer of the fast, collimated jet than H$\alpha$ \citep[see Section~\ref{ss:two_comp} and][]{rei15a,rei15b}. 
This is true inside the dust pillars where only [Fe~{\sc ii}] traces the jet and outside the pillar where both H$\alpha$ and [Fe~{\sc ii}] are visible. 
Bright H$\alpha$ emission from ionized gas in the H~{\sc ii} region may obscure fainter jet emission, especially in the complicated environment near $\eta$ Carinae. 
In contrast, bright [Fe~{\sc ii}] emission is confined to the core of protostellar jets where densities are sufficient to shield the Fe$^+$ from further ionization. 
Confusion with the environment may explain why the body of HH~1156 cannot be identified in H$\alpha$ even though bright [Fe~II] emission clearly traces the bipolar jet.

Strong [Fe~{\sc ii}] jet emission is observed from deeply embedded, and therefore undetected, protostars (e.g.\ HH~900 and HH~1161, see Figures~\ref{fig:hh900_feii} and \ref{fig:hhc4_feii}, respectively) as well as those readily apparent in H$\alpha$ images (e.g.\ HH~1163, see Figure~\ref{fig:hhc4_feii}). 
If material entrained from the environment was solely responsible for shielding Fe$^+$ from further ionization, strong [Fe~{\sc ii}] emission in jets would persist only inside the dusty pillars. 
This cannot explain bright [Fe~{\sc ii}] emission from jet limbs that emerge from protostars located near pillar edges (e.g. HH~903, HH~1004, and HH~1006, see Figures~\ref{fig:hh903_feii},~\ref{fig:hh1004_feii}, and~\ref{fig:hh1006_feii}, respectively), nor can it explain strong [Fe~{\sc ii}] emission in jets driven by unobscured protostars (e.g. HH~1163, see Figure~\ref{fig:hhc4_feii}). 
Instead, the ubiquity of strong [Fe~{\sc ii}] emission from the HH jets in Carina suggests that dust in the jet must be introduced locally, either launched into the jet from the disk or created in the outflow. 

\subsection{Two-component jets}\label{ss:two_comp} 

\citet{rei13} showed that [Fe~{\sc ii}] emission traces neutral material in dense jets. 
In some cases (e.g.\ HH~901), the ionization front in the jet can be spatially resolved. 
Bright [Fe~{\sc ii}] emission in the jet peaks behind the H$\alpha$, tracing neutral material located behind the ionized skin of the jet (and further away from the ionizing source). 
For other jets in Carina, H$\alpha$ and near-IR [Fe~{\sc ii}] emission appear to trace two distinct outflow components. 
In both HH~666 and HH~900, a slower, wider-angle cocoon of H$\alpha$ emission surrounds the fast, highly collimated jet seen in [Fe~{\sc ii}] \citep{rei15a,rei15b}.

Four jets presented in this paper also show this two-component, jet-outflow morphology. 
HH~1004 (SW), HH~1161, HH~1164, and HH~1066 all show H$\alpha$ emission that is parallel to, but offset from, the [Fe~{\sc ii}] jet (see Figures~\ref{fig:hh1004_feii}, \ref{fig:hhc4_feii}, and \ref{fig:hh1066_feii}, respectively). 
Both HH~1161 and HH~1164 were originally identified as candidate jets by \citet{smi10} because H$\alpha$ emission from the jet is confused with the ionization front along the surface of the globule. 
New [Fe~{\sc ii}] images reveal steady, collimated jets with morphologies that converge with H$\alpha$ only near the terminus of the continuous inner jet, similar to HH~666~M and HH~900 \citep[see Figure~\ref{fig:hh900_feii} and][]{rei15a}.

HH~903, HH~1004 (NE), and HH~1014 disrupt the smooth morphology of the ionization front as they break out of the pillar. 
Broad H$\alpha$ emission that extends into the H~{\sc ii} region along the body of the jet may trace entrained material that is being dragged out of the pillar. 
Unlike the H$\alpha$ sheaths seen in HH~1004 (SW) and HH~666, these broad H$\alpha$ components survive for only a short length. 
H$\alpha$ emission from more distant portions of the jet are more collimated, and well-matched to the jet morphology seen in [Fe~{\sc ii}]. 
This is similar to HH~666~M where H$\alpha$ traces the wider-angle outflow inside the pillar, and the ionized skin of the bare jet in the H~{\sc ii} region.

Each of these two-component jets emerges from an embedded protostar. 
Deeply embedded sources like HH~900 and HH~1161, where no protostar has been identified, only show two components beyond the edge of the globule. 
When the protostar can be identified in H$\alpha$ images, as in HH~666 and HH~1164, both H$\alpha$ and [Fe~{\sc ii}] trace the outflow from the source. 
HH~1163 illustrates a third situation, where a two-component system is observed from a protostar that is not embedded in a pillar. 
However, \citet{smi10} note that HH~1163 resembles an LL~Ori object, suggesting that wide-angle H$\alpha$ emission may come from interaction with a side wind, rather than being entrained by the jet.

The only other jet seen emerging from an exposed protostar, HH~1156, also has a strong, bright, collimated [Fe~{\sc ii}] jet. 
However, only the bow shock identified by \citet{smi10} shows any H$\alpha$ emission. 
The immediate vicinity of HH~1156 is dark in H$\alpha$ images, hinting at uneven foreground extinction that may obscure visual wavelength emission from the jet. 
However, the line-of-sight reddening, estimated from the [Fe~{\sc ii}] ratio $\mathcal{R}$, does not vary significantly between the inner jet and the shock that can be seen in H$\alpha$. 
It remains puzzling that the protostar can be seen at visual wavelengths, but no part of the inner jet can be identified in H$\alpha$ images.

\subsection{Knot structure}\label{ss:ir_synth_structure}

New [Fe~{\sc ii}] images trace portions of the HH jets in Carina not seen in H$\alpha$. 
The morphology of the [Fe~{\sc ii}] emission in some HH objects suggest that these separate knots may be part of a larger coherent jet. 
For example, HH~1007 lies along the outflow axis defined by HH~1015, and appears to be the counterjet bow shock. 
Velocity measurements are required to determine how these knots relate to individual coherent jet structures. 

Smooth, continuous emission from inner jets (suggesting no strong shocks) hint at a sustained accretion outburst that powers the jet. 
The median inner jet length is $\sim 6$\arcsec, corresponding to an outburst duration of $475$ yr (assuming $v_{jet} = 140$ km s$^{-1}$ for all jets). 
Only the shortest jets are consistent with the observed duration of FU Orionis outbursts \citep{hk96}. 
It is unclear how the typical decay time of an accretion burst from an intermediate-mass star compares to an FU~Orionis outburst, as accretion outbursts have only been detected in a few intermediate-mass sources \citep[e.g.][]{hin13}.

An alternate way to estimate how the duration of an accretion outburst scales with protostellar mass is to consider the viscous evolution. 
Viscosity determines the structure and evolution of an accretion disk. 
Therefore, the outburst duration will be governed by the viscous timescale, 
\begin{equation}
t_{vis} \sim \frac{R_{out}^2}{\nu}, 
\end{equation} 
where $R_{out}$ is the outermost radius in the disk that is unstable to accretion and $\nu$ is the viscosity. 
The viscosity itself is a function of the local sound speed $c_s$, and the Keplerian angular velocity $\Omega_K  = \sqrt{\frac{G M_{\star}}{R_{out}^3}}$ at radius $R_{out}$ for a star of mass $M_{\star}$ giving 
\begin{equation}
\nu = \frac{\alpha c_s^2}{\Omega_K}
\end{equation}
where $\alpha$ is the viscosity parameter \citep{ss73}. 
If we assume that $\alpha$ and $c_s$ are the same in protostellar accretion disks, regardless of the spectral type of the central star, then we can estimate the mass-dependence of the viscous time as 
\begin{equation}
t_{vis} \sim R_{out}^2 \sqrt{ \frac{G M_{\star}}{R^3_{out}} } \propto R_{out}^{1/2} M_{\star}^{1/2}. 
\end{equation}
Thus, the viscous time for a $M_{\star} = 10$ M$_{\odot}$ object will be $\sim 3 \times$ longer than for a $1$ M$_{\odot}$ star at the same $R_{out}$.

Magneto-rotational instability \citep[MRI,][]{bal91} is the most promising source of viscosity in protostellar disks. 
Only in portions of the disk where the gas is coupled to the magnetic field will undergo MRI and be unstable to accretion. 
If the spectral type of the star meaningfully changes the radius in the disk where this occurs, then the outermost radius that is unstable to accretion, $R_{out}$, might be larger for higher-mass stars, leading to longer outbursts. 
This estimate is clearly too simplistic to reflect the complicated physics of real disks \citep[e.g.\ non-zero viscosity in the ``dead zone'' of the disk,][]{bae13} but nevertheless show that longer accretion outbursts are plausible in intermediate-mass protostars.

\subsection{Revised mass-loss rate estimates}\label{ss:mdot_jet} 
\begin{figure}
\centering
$\begin{array}{c}
\includegraphics[trim=15mm 10mm 15mm 10mm,angle=0,scale=0.275]{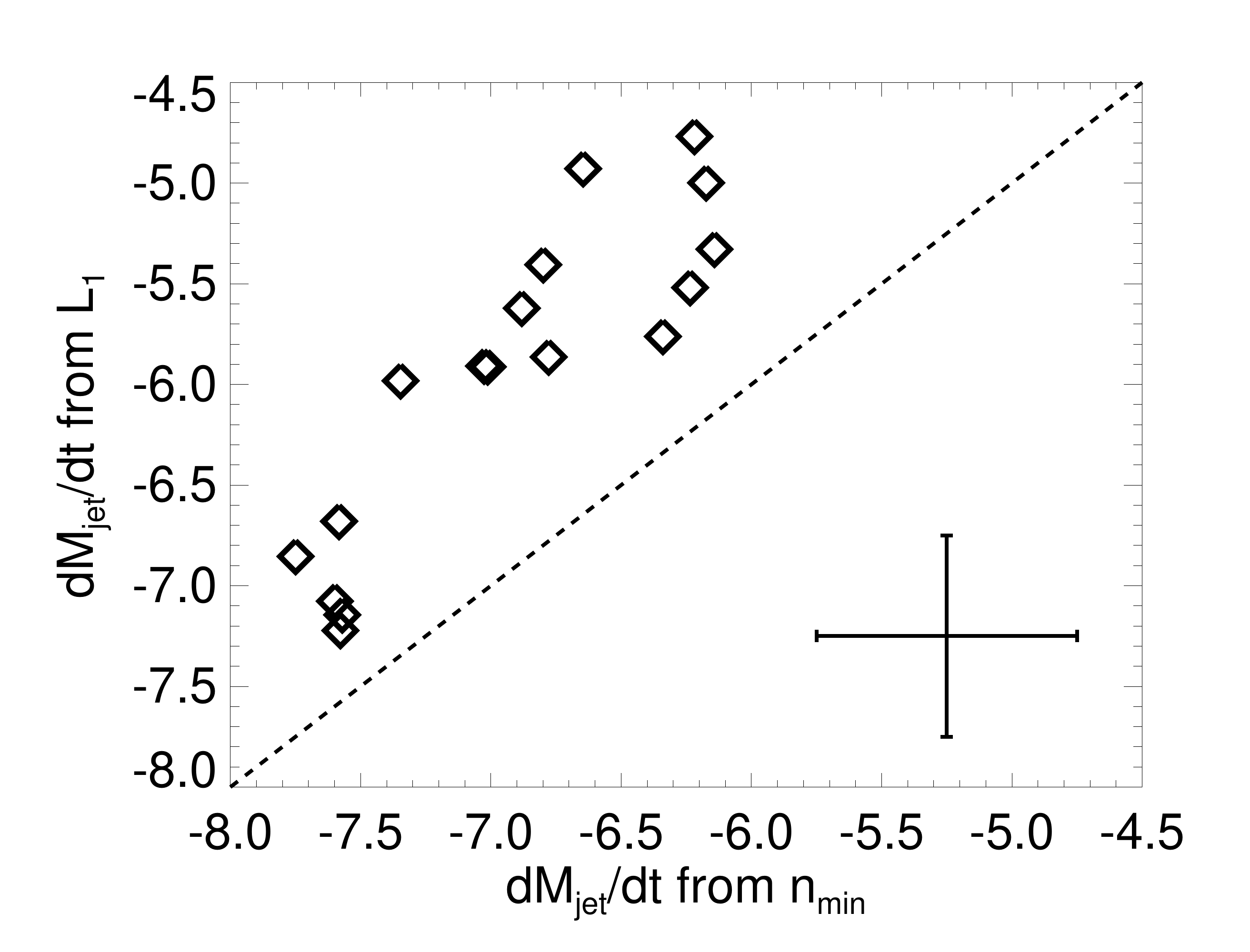} \\
\includegraphics[trim=15mm 10mm 15mm 10mm,angle=0,scale=0.275]{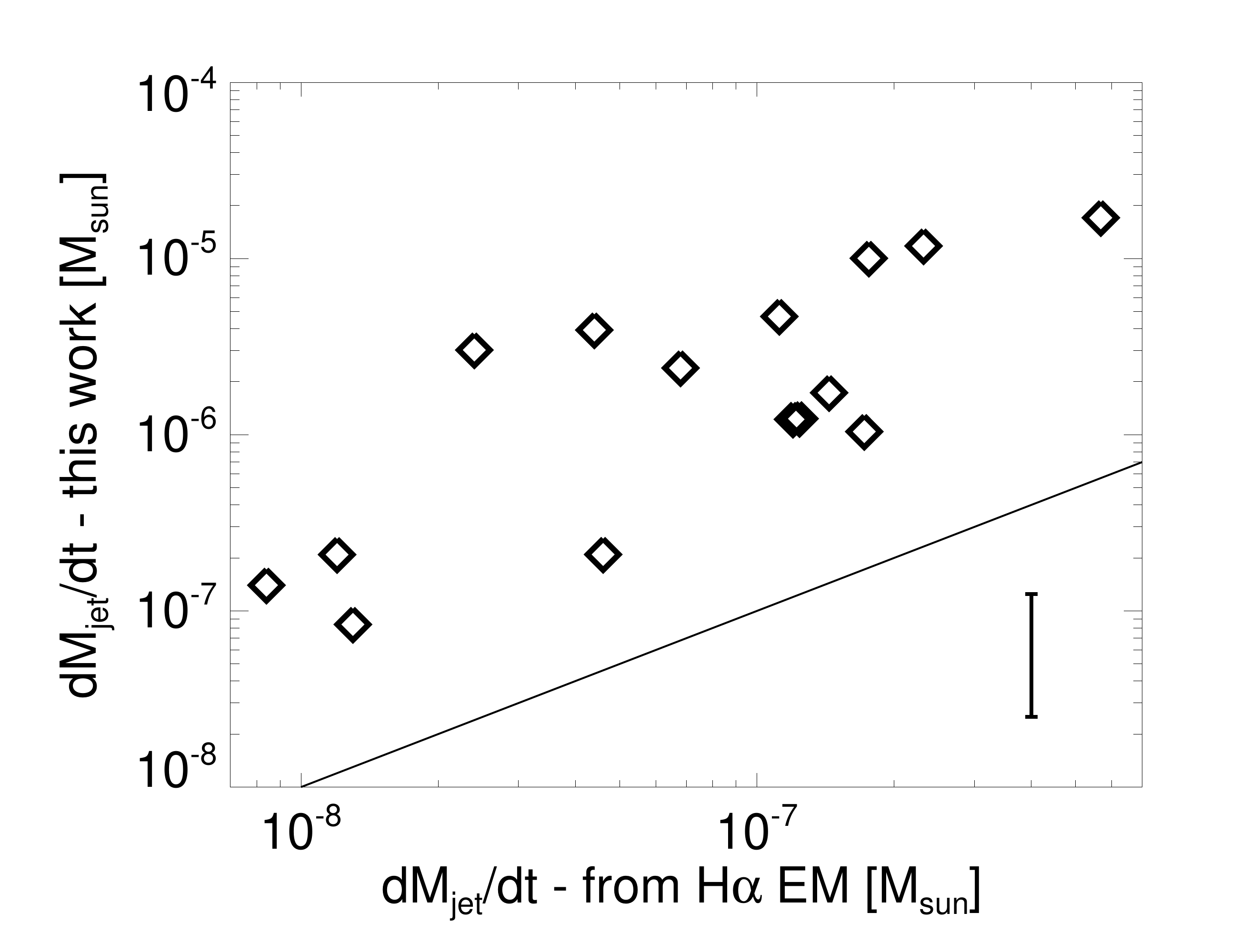} \\
\includegraphics[trim=15mm 10mm 15mm 10mm,angle=0,scale=0.275]{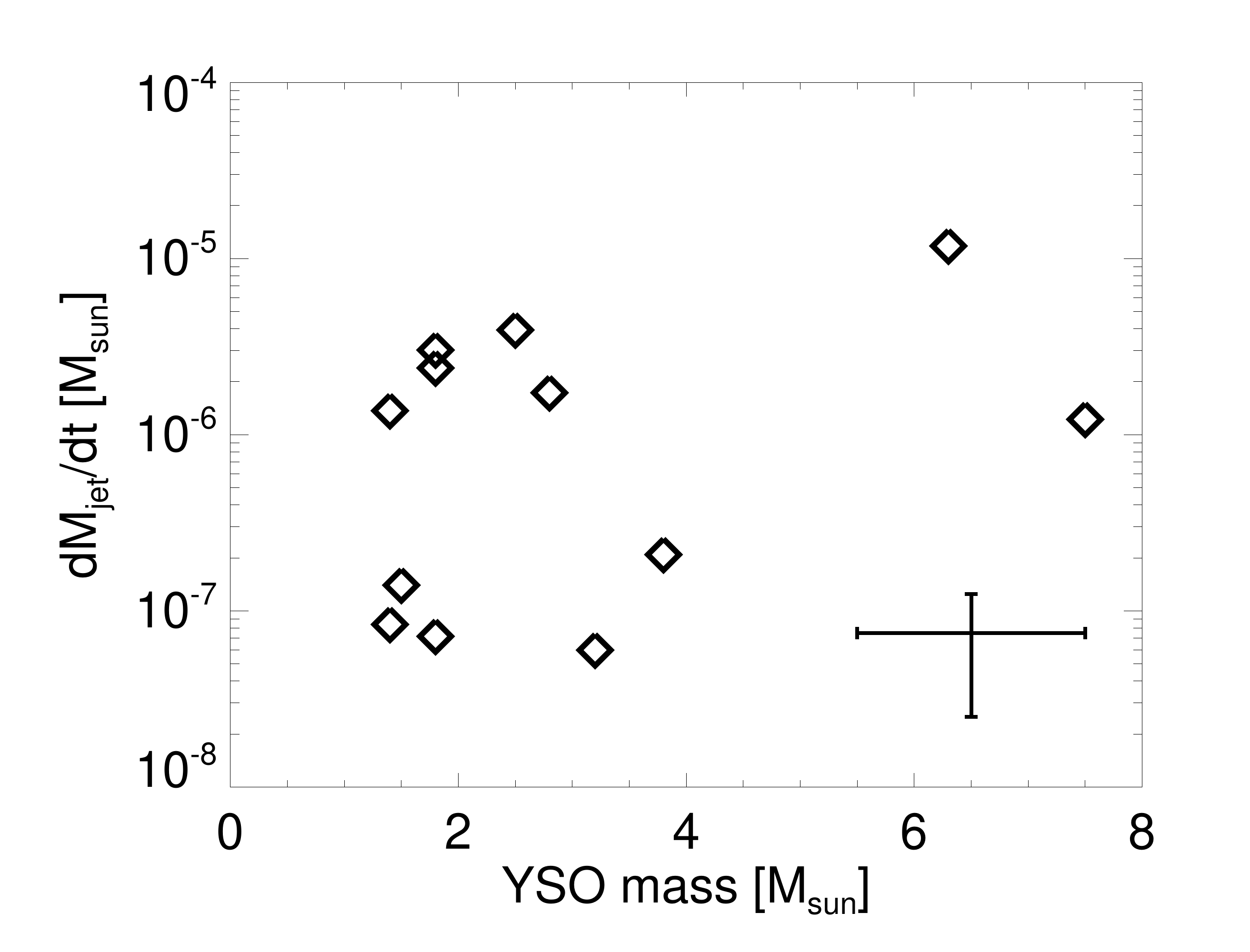} \\
\end{array}$
\caption[Mass-loss rate comparison]{
\textit{Top:} Comparison of $\dot{M}_{jet}$ estimated from the minimum density, $n_{min}$, to maintain a slow-moving ionization front in the jet and $\dot{M}_{jet}$ estimated by requiring that the neutral jet core survives to a distance $L_1$ in the H~{\sc ii} region (see Section~\ref{ss:mdot_jet}). 
The dashed line shows where the two estimates are equal. 
\textit{Middle:} Comparison of $\dot{M}_{jet}$ estimated by requiring that the neutral jet core survives to a distance $L_1$ in the H~{\sc ii} region to $\dot{M}_{jet}$ estimated from the H$\alpha$ emission measure by \citet{smi10}. 
\textit{Bottom:} Jet mass-loss rate versus the driving source mass \citet[][see Table~\ref{t:jets_ysos} and]{pov11}. 
For all three panels, error bars showing the typical uncertainty are shown in the lower right. 
}\label{fig:mdot_comp}
\end{figure}
\citet{smi10} estimated the mass-loss rate of the HH jets in Carina from the H$\alpha$ emission measure. 
\citet{rei13} showed that for a jet with a low-ionization (neutral) core that is not traced by H$\alpha$, this may underestimate the mass-loss rate by as much as an order of magnitude. 
Following \citet{bal06} and \citet{rei13}, we compute new mass-loss rate estimates that include the neutral jet core in two ways. 
First, we estimate the minimum density, $n_{min}$, required to maintain a slow-moving ionization front in the jet. 
Second, we estimate the mass-loss rate in the neutral jet by comparing it to the photoablation rate, using the known ionizing photon luminosity from the O-type stars in the Carina Nebula \citep{smi06a}.

The minimum density required for a jet to remain optically thick to Lyman continuum radiation is 
\begin{equation}
n_{min} = \sqrt{ \frac{Q_H}{4 \pi D^2} \frac{\mathrm{sin}(\beta)}{2 \alpha_B r} }
\end{equation} 
where $Q_H$ is the ionizing photon luminosity incident on the jet 
at an angle $\beta$ 
from a distance $D$, and 
$\alpha_B \approx 2.6 \times 10^{-13}$ cm$^3$ s$^{-1}$ is the case B recombination coefficient for hydrogen. 
We have assumed a cylindrical jet column with radius $r$. 
By requiring that the density of the neutral jet, $n_H$, is at least as large as this minimum density ($n_H \geq n_{min}$), we can derive a lower limit on the mass-loss rate. 
For a cylindrical jet, the mass-loss rate is 
\begin{equation}
\dot{M} = n_{min} | \pi \mu m_H v_{jet} r^2 |
\end{equation}
where 
$\mu$ is the mean molecular weight ($\approx 1.35$), 
$m_H$ is the mass of hydrogen, and
$v_{jet}$ is the jet velocity. 
Velocities have only been measured for a few of the HH jets in Carina, so we assume the median knot velocity measured by \citet{rei14}, $v_{jet}=140$ km s$^{-1}$, for the remaining jets. 
We find a median mass-loss rate of $\sim 9 \times 10^{-8}$ M$_{\odot}$ yr$^{-1}$, similar to the median mass-loss rate estimated from the H$\alpha$ emission measure, $\sim 1 \times 10^{-7}$ M$_{\odot}$ yr$^{-1}$ \citep{smi10}.  
Comparing these two estimates suggests that most of the HH jets in Carina support densities high enough to shield a neutral jet core, and indeed, we detect [Fe~{\sc ii}] emission in every jet targeted with this sample.

Both of these estimates provide a lower limit on the jet mass-loss rate. 
To further improve the estimated mass-loss rate, we can use the length of the inner jet as a measure of how long the cylinder of ejected mass survives photoablation in the H~{\sc ii} region.  
For a neutral jet core to persist, the jet density must be $n_H \geq n_{min}$ throughout the length of the continuous inner jet. 
Therefore, for a jet in an H~{\sc ii} region to remain visible out to a distance $L_1$ from the driving source, it must have a mass-loss rate of at least $\dot{M}_{jet} \geq L_1 \dot{m}$ where $\dot{m}$ is the photoablation rate of the jet in the H~{\sc ii} region \citep{bal06}. 
For a cylindrical jet, this corresponds to a mass-loss rate 
\begin{equation}
\dot{M} \approx \frac{L_1 f \mu m_H c_s}{2D} \left[ \frac{\alpha_B}{\pi r L_{LyC} sin(\beta)} \right]^{-1/2} 
\end{equation}
where 
$f \approx 1$ is the filling factor for a cylinder of radius $r$ losing mass from one side,  
$c_s \approx 11$ km s$^{-1}$ is the sound speed in photoionized plasma, and 
$\beta$ is the angle between the jet axis and the direction of the ionizing radiation from a source with luminosity $L_{LyC}$ located a distance $D$ from the jet. 
We use the inventory of massive stars and their ionizing photon luminosities cataloged by \citet{smi06a} to calculate the photoablation rate of the irradiated jet. 
Given the uncertainty in the three-dimensional structure of the Carina Nebula, we assume that the angle of the incoming radiation is $90^{\circ}$ for all of the jets. 
Estimated this way, the median mass-loss rate is $\sim 1.4 \times 10^{-6}$ M$_{\odot}$ yr$^{-1}$. 
Mass-loss rates estimated from $L_1$ for each jet are listed in Table~\ref{t:jet_prop}.

Comparing mass-loss rates estimated by \citet{smi10} from H$\alpha$ with $\dot{M}_{jet}$ estimated from [Fe~{\sc ii}], we find that the latter is $\sim 5 - 100$ times higher. 
The ratio of the H$\alpha$- and [Fe~{\sc ii}]-derived mass-loss rates for each jet is listed in Table~\ref{t:jet_prop} and plotted in Figure~\ref{fig:mdot_comp}.

Figure~\ref{fig:mdot_comp} shows a comparison of the two methods used to estimate the mass-loss rate based on the detection of near-IR [Fe~{\sc ii}]. 
Mass-loss rates estimated from the requirement that a jet survives to a length $L_1$ are higher than those calculated from $n_{min}$ for all sources. 
Imposing the more stringent constraint that a jet with density $\geq n_{min}$ survives to a distance $L_1$ requires a higher initial density, and therefore yields a higher mass-loss rate. 
Even with this accounting, the estimated mass-loss rates are still lower limits. 
Inferring the jet density from the length of the continuous inner jet assumes that photoablation eventually truncates the jet. 
However, many jets show evidence for time-variable mass-loss rates (see Section~\ref{ss:ir_synth_structure}). 
In this case, the jet length is a measure of how long the jet has been ``on,'' provided the density is above $n_{min}$. 
A high density jet that has been losing mass at a high rate for only a short time may have a small length, $L_1$, on the sky but support a density much higher than the minimum required to shield Fe$^+$ (see Section~\ref{ss:macc_lacc_est}).

If we instead assume that mass funnels into the jet in a continuous stream, we can estimate the observed length that a jet would attain before being truncated by photoablation in the H~{\sc ii} region. 
Taking the lowest mass-loss rate estimate for an HH jet in Carina, $\dot{M}_{jet} \sim 10^{-9}$ M$_{\odot}$ yr$^{-1}$ for HH~1015 from the H$\alpha$ emission measure \citep{smi10}, 
we expect the jet to extend at least $\sim 310$\arcsec. 
This exceeds the length of the longest one-sided jet length in the sample ($308$\arcsec, see Table~\ref{t:jet_prop}). 
However, the survival of Fe$^+$ in all of these jets requires a higher density, and therefore mass-loss rate. 
Jets with higher mass-loss rates than assumed in this simple estimate would survive in the H~{\sc ii} region even longer. 
Thus, the [Fe~{\sc ii}]-bright jets must reflect a recent change in the mass-loss rate.

\subsection{Implied accretion rate and estimated accretion luminosity}\label{ss:macc_lacc_est}

We list the bolometric luminosity ($L_{bol}$) of the candidate jet driving sources measured by \citet{pov11} and \citet{ohl12} in Table~\ref{t:jets_ysos}. 
\citet{pov11} classify half of the jet-driving sources as Stage 0/I, reflecting their youth (7/12). 
HH~666 is the only jet source identified as Class II, although it still has a strong IR excess \citep{smi04}, indicating that a significant amount of circumstellar material remains. 
Furthermore, HH~666 is tilted $\gtrsim 30^{\circ}$ away from the plane of the sky \citep{rei14}, allowing a clearer view down the envelope cavity opened by the jet. 
The remaining jet sources have an ambiguous classification of their evolutionary stage (4/12). 
Two protostars with an ambiguous evolutionary classification can be seen in H$\alpha$ images (HH~1163 and possibly HH~1010), indicating more clearing of the circumstellar environment that suggests more evolved sources (Class~II). 
The remaining jet-driving sources are only seen at IR and longer wavelengths, suggesting earlier evolutionary stages. 
Four jets that emerge from dense globules do not have an IR point source detected on the jet axis -- HH~901, HH~902, HH~900, and HH~1161.

Many of the putative jet driving sources have luminosities $L_{bol} \sim 20-200$ L$_{\odot}$, similar to the luminosities observed from low-mass stars in an elevated accretion state (see, e.g.\ Table~1 in \citealt{hk96} and \citealt{aud14}). 
\citet{rei97} studied 14 HH jet-driving sources and concluded that most of their sources are consistent with low-mass protostars that are in an elevated accretion phase, similar to FU~Orionis outbursts. 
In outburst, the accretion luminosity will dominate the stellar luminosity, increasing the total observed luminosity by $1-2$ orders of magnitude. 
Outbursting low-mass stars may therefore be detected in surveys that are not sensitive to quiesent low-mass (and thus low-luminosity) protostars.

\begin{figure}
\centering
\includegraphics[angle=0,scale=0.275]{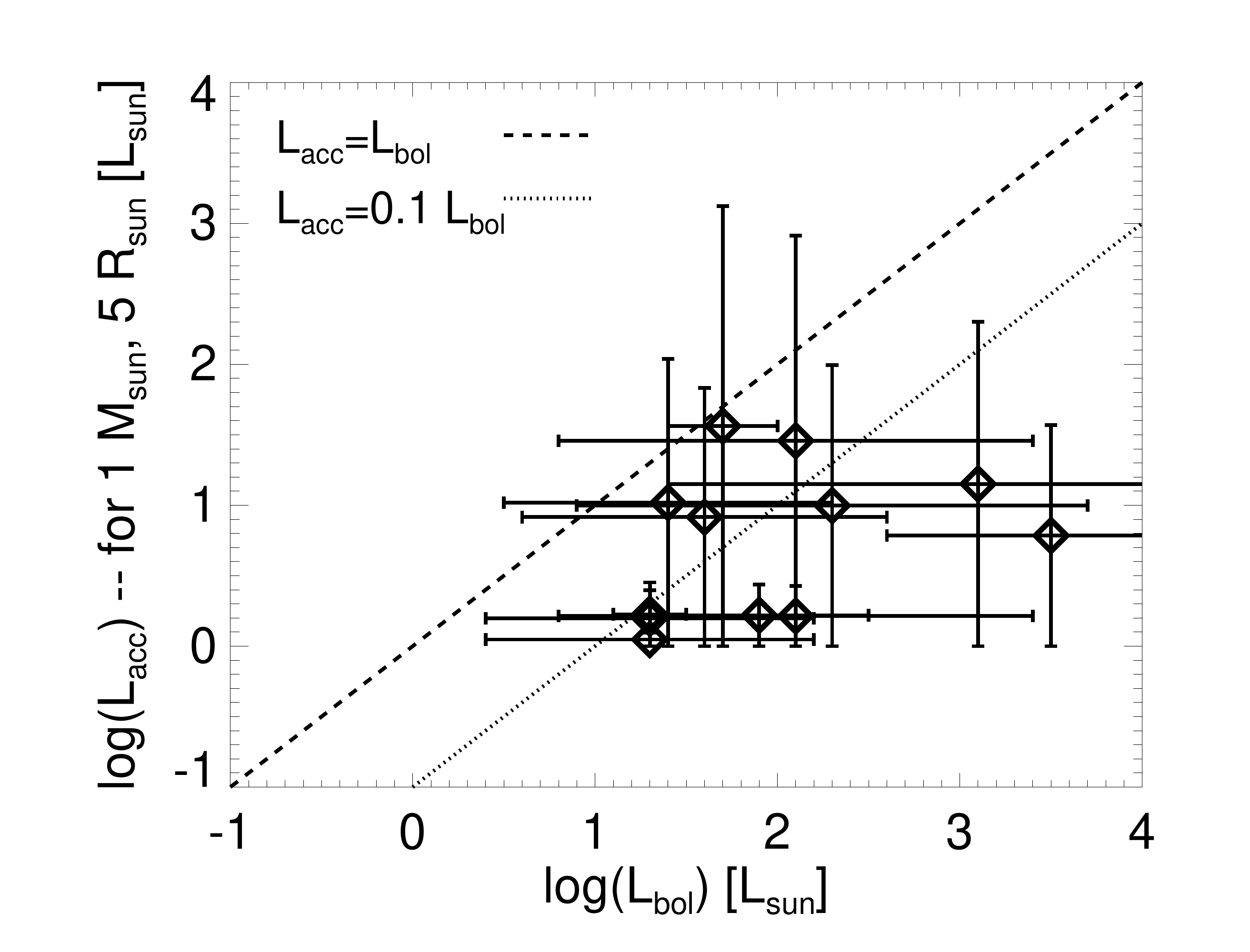} 
\caption[Estimated accretion luminosity]{
Estimated L$_{acc}$ derived by assuming $\dot{M}_{acc} = 10\times \dot{M}_{jet}$ for a 1 M$_{\odot}$ star accreting from $5$ R$_{\odot}$ plotted versus L$_{bol}$ from the PCYC model fit. 
The dashed (dotted) line shows where points would fall if L$_{acc}$ accounted for 100\% (10\%) of L$_{bol}$. 
A higher stellar mass ($>1$ M$_{\odot}$), more compact accretion geometry ($<5$ R$_{\odot}$), or lower outflow efficiency (1\% rather than 10\%) will all increase $L_{acc}$. 
 }\label{fig:lbol_lacc_comp} 
\end{figure}

To test whether the HH jet-driving sources in Carina could be low-mass sources in an outburst, we estimate the fraction of $L_{bol}$ that may be due to accretion. 
Assuming that $\dot{M}_{jet} = 0.1 \times \dot{M}_{acc}$, we can estimate the accretion luminosity, $L_{acc}$. 
For a 1 M$_{\odot}$ star accreting from 5 R$_{\odot}$, the $L_{acc}$ implied by the median $\dot{M}_{jet}$ estimated in Section~\ref{ss:mdot_jet} is $\sim 2$ L$_{\odot}$. 
This $L_{acc}$ is smaller than $L_{bol}$ of all the sources presented in this paper (all points in Figure~\ref{fig:lbol_lacc_comp} fall below the one-to-one line). 
Allowing for a lower outflow efficiency -- i.e.\ if $\dot{M}_{jet}$ is 1\% of $\dot{M}_{acc}$, rather than 10\% -- increases the estimated $L_{acc}$ by an order of magnitude.
This brings the estimated $L_{acc}$ roughly equal to the 4 lowest luminosity jet-driving sources in our sample (see Table~\ref{t:jet_prop}).

Conversely, if we assume that accretion dominates the luminosity of a 100 L$_{\odot}$ protostar, we estimate an accretion rate of $\sim 10^{-5}$ M$_{\odot}$ yr$^{-1}$ (again using the parameters for a low-mass protostar, 1 M$_{\odot}$, 5 R$_{\odot}$). 
This is on the order of the $\dot{M}_{acc}$ implied by the jet if the outflow efficiency $\dot{M}_{jet}$/$\dot{M}_{acc} = 0.01$. 
While this high luminosity and vigorous accretion rate is consistent with observed FU~Ori-like sources \citep{hk96,aud14}, these outbursts are short-lived (typically $\sim 10-100$~yrs). 
Detecting many such sources in a single region is unlikely, even for the large population of low-mass protostars in Carina \citep[$\sim 14000$, see][]{bro11}, especially given that only $\sim 25$ FU~Ori-like sources are known \citep{aud14}.

Assuming an FU~Ori outburst rate of $8 \times 10^{-4}$~yr$^{-1}$~star$^{-1}$ (as \citealt{hil15} infer for Class~I sources from the models of \citealt{bae14}),
we may detect $\sim 100$ FU~Ori-like outbursts among the $\sim 10^4$ sources reported by \citet{bro11} over a period of 10 years. 
However, only $\sim 375$ of those sources fall within the area images with WFC3-IR, making it unlikely that we would detect more than $\sim 3$ outbursts. 
Better estimates that account for the decrease in outburst rate with evolution (i.e.\ $3 \times 10^{-6}$~yr$^{-1}$~star$^{-1}$ for Class~II sources) will reduce the likelihood that we have caught a source in outburst. 
Thus, we consider this possibility to be unlikely.

For intermediate-mass driving sources, the underlying stellar luminosity is higher, leading to larger bolometric luminosities, even in the quiescent state. 
$L_{acc}$ will also increase due to the increased source mass and smaller inner disk radii \citep[see, e.g.][]{muz04}. 
Using the estimated masses from \citet{pov11} and assuming that all intermediate-mass stars accrete from 3 R$_{\odot}$ \citep[the radius used by][]{muz04}, we find a median accretion luminosity of $\sim 24$ L$_{\odot}$. 
This is similar to the smallest luminosities in our sample (see Table~\ref{t:jets_ysos}), although we note that this estimate requires parameters consistent with intermediate-mass stars to produce this agreement.
Altogether, this further supports our argument that the protostars driving the HH jets in Carina are predominately of intermediate-mass.

\subsection{Momentum injection}\label{ss:ir_synth_momentum} 
In this paper, we build on the argument first made in \citet{rei13} that the HH jets in Carina are the unveiled counterparts to the molecular outflows typically observed from intermediate-mass protostars in embedded regions. 
Because much of the obscuring gas and dust has been cleared in the H~{\sc ii} region, the core of the atomic jet is illuminated by nearby O-type stars and can be studied directly, rather than inferred from the properties of the outflow it entrains. 
The existence of two component outflows from some embedded protostars \citep[e.g.][]{rei15a,rei15b} supports this interpretation. 
Another test is to determine whether the HH jets in Carina have enough momentum to power the molecular outflows observed from intermediate-mass protostars in other regions. 
Estimates of whether jets from low-mass stars have sufficient momentum to drive molecular outflows provide conflicting results \citep[e.g.][]{har94,bac99}.

\citet{rei14} showed that the HH jets in Carina have high momentum even though their velocities are similar to jets from low-mass stars because of their high densities. 
All of the HH jets presented in this paper must also have high densities to shield [Fe~{\sc ii}]. 
Combining the median mass-loss rate, $\sim 1.4 \times 10^{-6}$ M$_{\odot}$ yr$^{-1}$, with an assumed velocity, $v_{jet} = 140$ km s$^{-1}$, yields a lower limit on the average momentum rate of $\sim 2 \times 10^{-4}$ M$_{\odot}$ km s$^{-1}$ yr$^{-1}$. 
Assuming the same velocity for all of the jets in Carina, the range of mass-loss rates estimated for the HH jets in Carina correspond to momenta rates ranging from $\sim 10^{-5} - 10^{-3}$ M$_{\odot}$ km s$^{-1}$ yr$^{-1}$. 
This is similar to the range \citet{bel08} find in their study of the CO outflows from intermediate-mass stars, suggesting that the momenta of these jets is sufficient to power the outflows.

\subsection{Statistics}\label{ss:stats} 

In the total area covered by our WFC3-IR images, \citet{pov11} detected 55 IR point sources with spectral energy distributions consistent with being intermediate-mass YSOs ($2$~M$_{\odot} \lesssim $~M~$\lesssim 10$~M$_{\odot}$). 
These model fits include an estimate of the evolutionary stage of the YSO using a system parallel to the empirical classification of T~Tauri stars, although an analogous evolutionary sequence is not yet firmly established for intermediate-mass stars \citep[see discussion in Section~3 of][]{pov11}. 
More than half of the detected sources are classified as Stage 0/I/II from the best-fit SED model -- 23/55 or $42$\% are classified as Stage 0/I, while 12/55 or $22$\% are Stage II; the remaining $36$\% have an ambiguous evolutionary stage. 
Near-IR [Fe~{\sc ii}] clearly traces a collimated HH jet back to 12 of those protostars for a total of 22\% of the protostars observed to be associated with jets (note that the HH~903 driving source was not included in the PCYC, see Section~\ref{ss:new_w_cont}). 
The jet-driving sources appear to be young with 58\% (7/12) classified as Stage 0/I, 8\% (1/12) Stage II sources, and the remaining sources with an ambiguous evolutionary stage (4/12 or 33\%). 
Excluding those sources that can be identified in an H$\alpha$ image, 75\% (9/12) of the jet-driving sources remain embedded.

Assuming that every protostar drives an episodic jet, the fraction of intermediate-mass protostars detected within the area imaged with WFC3-IR that are associated with an [Fe~{\sc ii}] jet suggests that the jets are ``on'' $\sim25$\% of the time. 
This may be an overestimate given the uneven sensitivity to embedded sources across the survey area. 
If the sample is contaminated with a few low-mass sources in an elevated luminosity (FU~Orionis-like) outburst state, then this number will be even lower. 
While we cannot exclude the possibility that some of the HH jets presented here are driven by low-mass sources, it is unlikely to be true for most of them (see Sections~\ref{ss:mdot_jet} and~\ref{ss:macc_lacc_est}).

\section{Conclusions}\label{s:ir_synth_conclusions}
We present narrowband, near-IR [Fe~{\sc ii}] images obtained with \emph{HST}/WFC3-IR of 18 jets and 2 HH objects in the Carina Nebula. 
Bright [Fe~{\sc ii}] emission traces 18 separate collimated bipolar jets. 
This survey targets jets with a candidate driving source identified near the jet axis. 
[Fe~{\sc ii}] emission connects the larger scale H$\alpha$ jet to the intermediate-mass protostar that drives it in 13/18 sources. 
Jets without a detection of their driving source primarily emerge from small, dense globules.

Simultaneous off-line continuum images allow us to remove PDR emission from the irradiated surface of the natal cloud and isolate emission from the jet. 
The dense cores of the jets traced by [Fe~{\sc ii}] appear highly collimated, while this is not always the case for H$\alpha$.
In some cases, the two lines appear to trace different outflow components altogether. 
From these new [Fe~{\sc ii}] images, we report the discovery of two new jets, HH~1156 and HH~1164, that cannot be identified as such from H$\alpha$ images alone.

Bright [Fe~{\sc ii}] emission in externally irradiated protostellar jets requires high densities to shield against further ionization in the H~{\sc ii} region. 
From this minimum density, we estimate high mass-loss rates that point to powerful jets. 
With these new and mass-loss rates and conservative estimates of the jet velocity, we find that the momentum of the jets is similar to the outflow momentum measured in molecular outflows from intermediate-mass stars \citep{bel08}. 
However, both the assumed velocities and estimated mass-loss rates are likely to be underestimates. 
The true jet momentum may be as much as an order of magnitude higher, suggesting that these jets are more than capable of entraining the molecular outflows more typically seen from intermediate-mass protostars.

Altogether, these highly collimated jets look like a scaled-up version of the jets seen from low-mass stars. 
The harsh UV environment in the Carina Nebula offers a rare glimpse of collimated jets from intermediate-mass protostars. 
These jets remain invisible in the absence of external irradiation, but may well be a ubiquitous feature of star formation. 
If so, this offers strong evidence that similar accretion physics governs the formation of stars of all masses.

\section{Online-only material}\label{s:online_material}

In the online supplement, we present [Fe~{\sc ii}] ratio tracings, as in Figure~\ref{fig:hh903_feii_ratio}, of all jets presented in this work.  


\section*{Acknowledgments}
We thank the referee for a thorough and thoughtful report. 
MR would like to thank Joan Najita and Jaehan Bae for helpful discussions. 
Support for this work was provided by NASA through grants AR-12155, GO-13390, and GO-13391 from the Space Telescope Science Institute. 
This work is based on observations made with the NASA/ESA Hubble Space Telescope, obtained from the Data Archive at the Space Telescope Science Institute, which is operated by the Association of Universities for Research in Astronomy, Inc., under NASA contract NAS 5-26555. These \textit{HST} observations are associated with programs GO~10241, 10475, 13390, and 13391.


\bibliographystyle{mnras}
\bibliography{ir_synth_bibliography}

\begin{table*}
\caption{Jet Properties\label{t:jet_prop}}
\begin{tabular}{llllllllll}
\hline\hline
Name & L$_{tot}^*$ & L$_{inner}^*$ & L$_1$ & W$_{pillar}$ & 
ionizing & proj. dist.$^{\dagger}$  & $\dot{M}_{jet}$ & $\Delta \dot{M}$ &  Comment \\
 & [\arcsec] & [\arcsec] & [pc] & [\arcsec] & cluster & [pc] & $M_{\odot}$ yr$^{-1}$ & $^{\dagger\dagger}$ &  \\
\hline
HH~666        & 308  & 16.5 & 0.18 &  34  & Tr16 & 14.2 & 1.2e-5 &  51 &  see also \citet{rei13}  \\
HH~901        &  20  & 3.5  & 0.04 &  2   & Tr14 &  2.0 & 4.7e-6 &  42 &  see also \citet{rei13}  \\
HH~902        &  35  &  8   & 0.09 &  11  & Tr14 &  2.1 & 1.0e-5 &  57 &  see also \citet{rei13}  \\
HH~1066       &   7  &  2   & 0.02 & 1.25 & Tr14 &  2.9 & 1.7e-6 &  12 &  see also \citet{rei13}  \\
\hline
HH~900        &  46  &  10  & 0.11 & 2.35 & Tr16 &  7.9 & 1.7e-5 &  30 &  see also \citet{rei15a}  \\
HH~903        & 167  &  17  & 0.19 &  28  & Bo11 & 13.0 & 1.0e-6 &  6  &  \\
HH~1004       &  27  & 9.6  & 0.11 &  14  & Bo11 &  5.7 & 1.2e-6 &  10 &  \\
HH~1005       &  71  &  10  & 0.11 &  33  & Bo11 &  5.9 & 1.2e-6 &  10 &  \\ 
HH~1006       & 67.5 &  6   & 0.07 &  6   & Bo11 &  7.1 & 3.0e-6 & 126 &  \\
HH~1007$^{**}$ & ...  & ...  &  ... & ...  & Bo11 & 29.9 &  ...   & ... &  no extended jet component \\
HH~1010       &  48  &  7   &  8.5 & 0.08 & Tr16 & 29.1 & 2.4e-6 &  35 &  \\
HH~1014       &  22  &  6   &  7   & 0.07 & Tr16 & 11.4 & 3.9e-6 &  89 &  \\
HH~1015$^{**}$ & 6-18 &  6   &  6   & 0.07 & Bo11 & 30.0 & 1.4e-7 &  17 &  \\
HH~c-3        & ...  & 2.5  &  2.5 & 0.03 & Bo11 & 21.8 & 8.4e-8 &   6 &  only component B from \citet{smi10} \\
HH~1159       & ...  & ...  &  ... & ...  & Bo11 & 21.2 &  ...   & ... &  no extended jet component \\
HH~1160       & 15.5 &  6   &  7   & 0.07 & Bo11 & 21.1 & 2.1e-7 &  17 &  \\
HH~1161       & 18.6 &  6   &  7   & 0.07 & Bo11 & 21.1 & 2.1e-7 &   5 &  \\
HH~1162       & ...  & ...  & ...  & ...  & Bo11 & 20.3 &  ...   & ... &  no extended jet component \\
HH~1163       & ...  & 2    & ...  & 0.02 & Bo11 & 20.5 & 7.2e-8 & ... &  \\
HH~1164       & 1.7  & 1.7  & 9.25 & 0.02 & Bo11 & 20.8 & 6.0e-8 & ... &  not detected by \citet{smi10} \\
HH~c-10       & ...  & ...  & ...  & ...  & Bo11 & 12.9 &  ...   & ... &  no extended jet component \\
HH~1156       & 44.5 &  2   &  1   & 0.02 & Tr16 & 11.5 & 1.4e-6 & ... &  jet body not detected by \citet{smi10} \\ 
\hline 
\multicolumn{10}{l}{$^*$ one-sided length} \\
\multicolumn{10}{l}{$^{**}$ two sides of the same jet, see Section~\ref{ss:new_w_cont}} \\
\multicolumn{10}{l}{$^{\dagger}$ median distance to all stars in the cluster} \\
\multicolumn{10}{l}{$^{\dagger\dagger}$ increase in $\dot{M}_{jet}$ as estimated in this work compared to $\dot{M}_{jet}$ estimated from the H$\alpha$ emission measure by \citet{smi10}} \\ 
\end{tabular}
\end{table*}

\begin{table*}
\caption{Jets and their driving sources}
\footnotesize
\begin{tabular}{lllllllll}
\hline\hline
Jet & PCYC & log(L$_{bol}$) & Mass & Stage & Ohlendorf & log(L$_{bol}$) & Mass & Comment \\
\hline 
HH~666  &  345  &  3.1 (1.7)  &  6.3 (1.3)  &  II  & J104351.5-595521 & 2.6 (2.6-3.1)  &  3.2 (1.7-9.2)  & SBB2004, RS2013 \\
HH~901  & ... & ... & ... & ... & ... & ... & ... &   \\
HH~902  & ... & ... & ... & ... & J104401.8-593030 & ... & ... &  ruled out by RS2014 proper motions \\
HH~1066 &  429  &  2.1 (1.3)  &  2.8 (1.6)  &  A  & J104405.4-592941 & ... & ... & identified in RS2013 \\
\hline
HH~900  & ... & ... & ... & ... & ... & ... & ... & S2013 candidates ruled out by R2015a \\
HH~903  & ... & ... & ... & ... & J104556.4-600608 &  2.4 (2.4-2.4)  &  4.3 (4.3-4.3)  &  \\
HH~1004 &  1198  &  3.5 (0.9)  &  7.5 (3.0)  &  A  & J104644.8-601021 &  2.2 (1.9-3.2)  &  4.6 (2.6-7.0)  &  \\
HH~1005 & ... & ... & ... & ... & J104644.2-601035 &  2.4 (2.4-2.6)  &  5.5 (5.5-7.3)  &  \\
HH~1006 &  1173  &  1.7 (0.3)  &  1.8 (0.9)  &  0/I  & J104632.9-600354 & ... & ... & YSO identified by \citet{sah12} \\ 
HH~1007 & ... & ... & ... & ... & ... & ... &  \\
HH~1010 &  55  &  1.6 (1.0)  &  1.8 (1.4)  &  A  & J104148.7-594338 &  1.8 (1.6-3.1)  &  1.9 (1.3-6.7)  &  \\
HH~1014 &  984  &  2.3 (1.4)  &  2.5 (1.8)  &  0/I  & J104545.9-594106 & ... & ... &  \\ 
HH~1015 &  538  &  1.3 (0.9)  &  1.5 (1.2)  &  0/I  & ... & ... & ... &  \\ 
HH~c-3  &  760  &  1.3 (0.9)  &  1.4 (1.3)  &  0/I  & J104504.6-600303 & ... & ... &  \\ 
HH~1159  &  ...  &  ...  &  ...  &  ...  & ... & ... & ... &  \\
HH~1160  &  787  &  2.1 (1.3)  &  3.8 (1.3)  &  0/I  & J104509.4-600203 &  2.1 (1.7-2.5)  &  4.3 (1.4-6.4)  &  \\
HH~1161  & ... & ... & ... & ... & J104509.2-600220 & ... & ... &  \\ 
HH~1162  & ... & ... & ... & ... & J104513.0-600259 & ... & ... &  \\
HH~1163  &  803  &  1.3 (0.2)  &  1.8 (0.9)  &  A  & J104512.0-600310 & ... & ... &  \\
HH~1164  & 790 & 1.9 (0.6) & 3.2 (1.2) & 0/I & ... & ... & ... &  \\
HH~1156 &  986  &  1.4 (0.9)  &  1.4 (1.1)  &  0/I  & ... & ... & ... &   \\ 
\hline
\multicolumn{9}{l}{SBB2004 = \citet{smi04}; S2013 = \citet{shi13}; RS2013 = \citet{rei13}; RS2014 = \citet{rei14}; } \\ 
\multicolumn{9}{l}{R2015a = \citet{rei15a} } \\ 
\end{tabular}
\label{t:jets_ysos}
\end{table*}


\label{lastpage}

\end{document}